\documentclass[letterpaper, 12pt]{article}
\usepackage{natbib}
 \bibpunct[, ]{(}{)}{,}{a}{}{,}%
\usepackage[T1]{fontenc}
\usepackage[active]{srcltx}
\usepackage[bookmarks=true, colorlinks=true, linkcolor=blue, urlcolor=blue, citecolor=blue, breaklinks=true]{hyperref}
\hypersetup{colorlinks, urlcolor=blue, linkcolor=blue, citecolor=blue}
\usepackage{times}
\usepackage{graphicx}  							
\usepackage{amsthm}
\usepackage{amsfonts}
\usepackage{amsmath}
\usepackage{multirow}
\usepackage{mathrsfs}
\usepackage{mathtools}
\usepackage{algorithm}
\usepackage{algorithmicx}
\usepackage{algpseudocode}
\usepackage{bm}
\usepackage{booktabs}
\usepackage{setspace}

\usepackage{easyReview}
\usepackage{subfigure}
\usepackage{tikz}
\usepackage{pgfplots}
\pgfplotsset{compat=1.17}
\usepackage{pgf}

\makeatletter
\newenvironment{breakablealgorithm}
{
		\begin{center}
			\refstepcounter{algorithm}
			\hrule height.8pt depth0pt \kern2pt
			\renewcommand{\caption}[2][\relax]{
				{\raggedright\textbf{\ALG@name~\thealgorithm} ##2\par}%
				\ifx\relax##1\relax 
				\addcontentsline{loa}{algorithm}{\protect\numberline{\thealgorithm}##2}%
				\else 
				\addcontentsline{loa}{algorithm}{\protect\numberline{\thealgorithm}##1}%
				\fi
				\kern2pt\hrule\kern2pt
			}
		}{
		\kern2pt\hrule\relax
	\end{center}
}
\makeatother

\usepackage{endnotes}
\let\footnote=\endnote

\usepackage[normalem]{ulem}
\definecolor{mygreen}{rgb}{0,.5,0}

\usepackage{fullpage}
\usepackage[title,titletoc]{appendix}

\theoremstyle{plain}

\theoremstyle{plain}

\theoremstyle{plain}
\newtheorem*{assumption*}{\protect\assumptionname}
\theoremstyle{plain}

\theoremstyle{remark}
\newtheorem*{rem*}{\protect\remarkname}
\theoremstyle{plain}

\providecommand{\assumptionname}{Assumption}
\providecommand{\lemmaname}{Lemma}
\providecommand{\propositionname}{Proposition}
\providecommand{\remarkname}{Remark}
\providecommand{\theoremname}{Theorem}
\providecommand{\corollaryname}{Corollary}

\begin{document}

\title{A Risk Sensitive Contract-unified Reinforcement Learning Approach for Option Hedging}

\date{Nov 15, 2024}

\author{Xianhua Peng\thanks{HSBC Business School, Peking University, University Town, Nanshan District, Shenzhen, 518055, China. Email: xianhuapeng@pku.edu.cn.} \thanks{Sargent Institute of Quantitative Economics and Finance, HSBC Business School, Peking University, University Town, Nanshan District, Shenzhen, 518055, China.}  
\and Xiang Zhou\thanks{School of Data Science and Department of Mathematics, City University of Hong Kong, Tat Chee Ave., Kowloon, Hong Kong. Email: xiang.zhou@cityu.edu.hk.}
\and Bo Xiao \thanks{Department of Mathematics, City University of Hong Kong, Tat Chee Ave., Kowloon, Hong Kong. Email: boxiao3-c@my.cityu.edu.hk.}
\and Yi Wu \thanks{HSBC Business School, Peking University, University Town, Nanshan District, Shenzhen, 518055, China. Email: wuyi51@pku.edu.cn.}
}
\maketitle

\begin{abstract}

We propose a new risk sensitive reinforcement learning approach for the dynamic hedging of options. The approach focuses on the minimization of the tail risk of the final P\&L of the seller of an option. Different from most existing reinforcement learning approaches that require a parametric model of the underlying asset, our approach can learn the optimal hedging strategy directly from the historical market data without specifying a parametric model; in addition, the learned optimal hedging strategy is contract-unified, i.e., it applies to different options contracts with different initial underlying prices, strike prices, and maturities. Our approach extends existing reinforcement learning methods by learning the tail risk measures of the final hedging P\&L and the optimal hedging strategy at the same time. We carry out comprehensive empirical study to show that, in the out-of-sample tests, the proposed reinforcement learning hedging strategy can obtain statistically significantly lower tail risk and higher mean of the final P\&L than delta hedging methods. 


\end{abstract}

\section{Introduction}

Option hedging is a classic and important problem. A traditional hedging approach is delta hedging that is based on a parametric option pricing model. Under a delta hedging strategy, the hedging position at any given moment is equal to the partial derivative of the option pricing function with respect to the underlying stock price at that time. Although delta hedging is widely used in practice, it has a few shortcomings: (i) delta hedging assumes that when the underlying stock price changes, the other risk factors such as volatility do not change, which is inconsistent with empirical observations such as leverage effect; 
(ii) delta hedging assumes that the market is frictionless, has no transaction costs, and requires continuous adjustment of the hedging position;  
(iii) delta hedging is subject to model risk as it relies on a parametric option pricing model.

In addition to delta hedging, there is a large literature developing mean-variance hedging strategies based on minimization of a linear combination of mean and variance of the total hedging error upon the maturity of an option. 
\cite{Duffie1991} and \cite{Basak2012} studies mean-variance hedging of non-tradable assets by tradable assets under continuous parametric models. 
\cite{Schweizer1995} and \cite{Bertsimas2001} studies mean-variance hedging of options in discrete time under parametric models. 
In recent years, there are some papers applying model-based reinforcement learning to the mean-variance hedging of total hedging error; see, for example, \cite{Halperin2020, Kolm2019, Du2020, Cao2021}. In these paper, the cumulative reward of reinforcement learning 
is an approximation of linear combination of mean and variance of the total hedging error, and a parametric model of the underlying stock is needed for simulation of sample paths of the stock. 
\cite{Fecamp2021} proposes a deep learning approach to minimize the mean (asymmetric) squared error of total hedging error under a discrete time parametric model of the stock based on neural network representation of the hedging ratio. 
\cite{EvaLuetkebohmert2022} proposes a robust deep hedging approach 
by simulating 
sample paths of the underlying stock under generalized affine diffusion model that allows parameter uncertainty.

Another strand of literature studies the objective of minimizing the variance of the daily hedging error instead of the total hedging error. 
\cite{Hull2017, Nian2021, Ruf2022, Xia2023} assume that the minimum variance hedging ratio is a linear or nonlinear function of the option's Greeks and other variables of the option and use the market data of stock prices and options to fit the 
function. \cite{Mikkilae2023} proposes a model-free reinforcement learning approach that trains the model using using only S\&P 500 index and options data. The cumulative reward is defined as a rough approximation of the mean minus a multiple of the standard deviation of the \emph{episode} hedging error over five trading days.

However, hedging based on the first two moments of hedging error may not be adequate because it ignores the high-order moments and tail risk as measured by Value-at-Risk and Conditional Value-at-Risk (CVaR).
\cite{Buehler2019} propose a model-based deep hedging approach that minimizes an 
optimized certainty equivalent risk measure, e.g., CVaR, of the total hedging error of a particular option. 
\cite{Buehler2019} and their follow-up works \citep{Carbonneau2021,EvaLuetkebohmert2022,Limmer2023,Wu2023} has a few limitations: (i) 
their approaches need to 
train a different model for each different option with specific initial stock price, strike, or maturity; (ii)
their approaches 
need to specify a parametric model of the underlying stock in order to generate simulated sample paths of the stock for training the model, because one cannot find enough historical data of the stock starting from a particular initial stock price; (iii) their approaches cannot be used in a model-free way and hence cannot utilize the market data of stocks and options.

In this paper, we propose a new dynamic hedging strategy based on deep reinforcement learning, namely contract-unified reinforcement learning (CU-RL), to address the above challenges faced by existing options hedging approaches. In the CU-RL framework, we use deep reinforcement learning to train a contract-unified hedging strategy that can 
be used to 
hedge different options with different initial stock prices, strike prices, maturities, and other initial states of the hedger including initial cash and stock positions. The CU-RL approach 
minimizes the CVaR of the total hedging P\&L while maximizing its mean. 
The main difficulty of obtaining a contract-unified hedging strategy is that the distribution of the total hedging P\&L of an option depends on the contract parameters of the option and the initial states of the hedger of the option, so do the VaR and CVaR of the total hedging P\&L. To overcome the difficulty, our key innovation is to use a neural network to estimate the \emph{conditional VaR} of the total hedging P\&L of any option that is conditional on the 
contract parameters and initial states. The input of the neural network is the contract parameters and initial states; the output of the neural network is the conditional VaR of the total hedging P\&L conditional on the contract parameters and initial states. The output of the VaR network is then used to calculate the conditional CVaR of the total hedging P\&L of the option. 

The VaR network has two important implications. (i) It allows the training data to include any options with any contract parameters and initial states, so the a single unified reinforcement learning model can be trained and generates a unified hedging strategy for any options. (ii) In the existing reinforcement learning approaches, a parametric model is needed because the training data can only include the sample paths of the stock starting from a given initial price, but such sample paths are not available in the historical data. The VaR network allows the training data to include historical sample paths of the stock with any initial price and any options written on the stock, so there is no need for a parametric model of the underlying stock, and it allows utilization of all information of the historical data.

We also provide the algorithm of contract-specific reinforcement learning (CS-RL), a contract-specific version of CU-RL that finds the optimal hedging strategy for a specific option with fixed initial state. The VaR of the total hedging P\&L  is a number in the contract-specific setting, so a trainable parameter is sufficient to represent it. The CS-RL model has the limitation that it needs a parametric model of the underlying stock in order to simulate sample path of stocks that starts from the fixed initial stock price, and hence it cannot utilize the historical data of stock and option prices.

Specifically, the proposed CS-RL and CU-RL approaches are extensions of the Proximal Policy Optimization (PPO) approach that maximize a weighted sum of the expectation of total hedging P\&L and its negative CVaR. The CS-RL and CU-RL can also be easily extended to other the-state-of-art reinforcement learning algorithms other than PPO. The training process takes simulated or historical stock and option data trajectories combining other market information like interest rate as input. In each trajectory, the agent hedges an option until maturity at discrete time steps. At each step, the agent chooses underlying asset position and receives a reward. During the process, transaction costs can be incorporated. At each step before the last trading time step ahead of maturity, the reward is defined to be a chosen function of hedging error. At the last trading time step, the reward is defined as the weighted sum of total hedging P\&L and its negative CVaR. We propose to represent and estimate the VaR of total hedging P\&L for a particular option with a trainable parameter $\omega$ in the CS-RL approach, and to estimate the VaR of total hedging P\&L for any option with initial state $s_0$ by the output of a neural network in the CU-RL approach. 

The CS-RL and CU-RL extend the PPO algorithm by learning the optimal hedging strategy and the VaR of the total hedging P\&L simultaneously. In each training step, not only the value and policy networks in the original PPO, but $\omega$ or the VaR network will be updated. The loss of $\omega$ or VaR network is defined as the sample mean of the score function of quantile evaluating at the total hedging P\&L of each trajectory, so our algorithms do not need nested simulation to get or update risk measure estimation, which is the computational bottleneck in some risk-sensitive reinforcement learning algorithms such as \citet{Tamar2015} and \citet{Coache2023a}.

In summary, the proposed CU-RL has the following advantages. (i) It is contract-unified because it allows to train one single model to obtain the optimal hedging policy that applies to options with different strikes, maturity, and initial underlying stock price, etc. 
(ii) It is data-driven and does not require a parametric model of the underlying asset price, which can effectively avoid model misspecification and parameter estimation errors of parametric models. 
(iii) By proposing a new reward function, it introduces a risk measure of the total hedged P\&L in the objective function, which can effectively control the tail risk of the total hedged P\&L. 
(iv) It can accurately include market frictions like transaction costs. 

In the CU-RL approach, we minimize the static CVaR of the total hedging P\&L, i.e., the terminal P\&L realized at the maturity of an option, so the learned hedging policy is a precommitted strategy as in many other risk-sensitive reinforcement learning algorithms. Although the precommitted strategy is not time consistent, the risk objective has clear economic meaning and is desirable for an options dealer, who are concerned about monitoring the VaR and CVaR of his total hedging P\&L. 
Option hedging based on dynamic risk measures or recursive risk measure is studied in some recent papers; see, e.g., \cite{SaeedMarzban2023, Coache2023a, Coache2023, Buehler2023}. A dynamic risk measure has a recursive representation that defines the risk at current state by an one-step risk measure on the current loss plus the risk-to-go. 
Although hedging based on dynamic risk measure has the advantage of being time consistent, the economic meaning of the risk measurement of dynamic risk measure is less transparent and dynamic risk measure tends to be over conservative than static risk measure since risk is compounded in time \citep{Iancu2015}.

We also contribute to the literature of risk-sensitive reinforcement learning by formulating and solving a problem that measures the tail risk of the reward at the terminal time period; in contrast, most existing literature of risk-sensitive reinforcement learning applies risk measures to the cumulative or average reward.
For example, \cite{Borkar2001, Borkar2002} consider the risk-sensitive reinforcement learning problem with the expected exponential utility. \cite{Tamar2012} and \cite{Prashanth2013} consider variance-related risk measures. 
\cite{Morimurat2010}, \cite{Tamar2015}, and \cite{Chow2015} consider minimizing CVaR of the cumulative reward in reinforcement learning; 
these algorithms assume that the state transition dynamics is known or can be obtained by nested simulation, so they may be computationally inefficient and cannot be trained with real world data. 
\citet{Xu2023} propose a new episodic risk-sensitive reinforcement learning formulation and algorithm based on tabular
Markov decision processes with recursive optimized certainty equivalent risk measures. 
However, they all focus on recursive risk measures, and thus suffer from the problems mentioned above. Moreover, our framework can incorporate flexible reward design before the terminal step, which can provide the agent with additional guidance and thus accelerate the exploration.

	We validate CS-RL and CU-RL with stock and options data simulated under two parametric models, the Black-Scholes model and heavy-tail GARCH model proposed by \cite{Heston2000}, and find that our methods outperform the delta hedging methods in terms of tail-risk control and profit making. 
	For validating CS-RL, we simulate stock and options price trajectories of a single option with a fixed initial state including initial underlying stock price, strike, and maturity under BS model and heavy-tail GARCH model. We then use CS-RL to learn hedging strategies to minimize CVaR at 0.975 level 
	of final hedging P\&L in the case of zero and proportional transaction costs. 
	We compare the CS-RL with benchmark delta hedging methods, which are the BS delta for the case of BS model and the  BS delta and the GARCH delta
	for the case of GARCH model.
	We compare the out-of-sample final P\&L of CS-RL and benchmarks by their mean and tail risk measured by various risk measures including VaR, CVaR, and median shortfall (MS) at 0.95 and 0.975 level.
	CS-RL outperforms these benchmarks in the sense that in all the cases with different underlying dynamics and transaction costs, CS-RL obtains the significantly lowest CVaR of final P\&L at 0.975 level.
	Moreover, we find that for BS model with zero and proportional transaction costs and GARCH model with proportional transaction costs, (i) the final P\&L of CS-RL has significantly highest mean with P-value less than 0.001; (ii) CS-RL obtains significantly lowest tail risk measured by VaR, MS, and CVaR at 0.95 and 0.975 level.
	
	For validating CU-RL, we first simulate a ten-year daily S\&P 500 trajectory following BS and GARCH models, and list standard S\&P 500 call options contracts on a daily basis imitating CBOE's listing rules for specifying strikes and maturities of these products. The prices of these options are calculated under the assumed model. 
	The data of these options are divided into training set and test set according to their current dates and expiration dates to avoid information leakage. 
	We use 
	the CU-RL method to train a unified hedging model for options with any contract parameters and initial states. 
	For both underlying dynamics and transaction costs, CU-RL outperforms benchmark delta hedging methods in the out-of-sample test set: it obtains the significantly highest mean and significantly lowest tail risk among all the methods. The results show the effectiveness of proposed methods.
	
	To the best of our knowledge, our paper is the first to conduct comprehensive empirical study on the performance of different hedging strategies in terms of the mean and tail risk of the total hedging P\&L. 
	In a model-free setting, we test CU-RL using the market data of S\&P 500 index options from 01/01/2008 to 12/31/2017. Since for a single option with given initial state and contract parameters, there are rare options in the market data that have the same initial state as the given option, we cannot obtain enough market data of the option in order to train and test CS-RL in a model-free way.
	Similar to the parametric case, we divide the S\&P 500 option data set into training set, validation set, and test set.
	We use CU-RL method to train a unified model for each of eight groups of options: the group of all call (resp., put)  options, the group of all short-term near-the-money call (resp., put) options, and two groups of short-term deep-in-the-money call (resp., out-of-the-money put) options. 	
	We also check the robustness of CU-RL on all call/put options and short-term near-the-money groups with two alternative reward formulations: a reward that is zero before maturity and a relative return formation with respect to initial margin.
	In all the cases, we train and test CU-RL with both zero and proportional transaction costs.
	We use BS delta, local volatility function delta, and SABR delta as our benchmarks. 
	
	The advantage of CU-RL is statistically significant in most of the experiments, where CU-RL have significantly higher mean of final P\&L with P-value less than 0.001 and also significantly smaller tail risk measured by VaR and CVaR at 97.5\%-level, which is specified in the objective function of CU-RL, and VaR and CVaR at 95\% level. The under-performance of CU-RL only appears in the two short-term deep-in/out-of-the-money groups, which may be due to lack of enough training data.
	Moreover, we find that: (i) CU-RL performs much better than benchmarks in the settings with transaction costs, suggesting it can better incorporate market frictions into hedging strategies; (ii) CU-RL is robust even without using the information of hedging error in the intermediate reward;
	(iii) CU-RL can include margin requirement for selling uncovered options and optimize risk-controlled relative option return with respect to margin.

\subsection{Related Literature}

Traditional literature on option hedging in incomplete market starts with episode mean-variance criterion on the final hedging P\&L thanks to its tractability. 
\cite{Duffie1991} formulate a mean-variance hedging problem to use a correlated tradable asset to hedge a non-tradable asset continuously. \cite{Schweizer1992} generalizes \cite{Duffie1991} to allow the hedging object be a contingent claim depending on the non-tradable asset, such as an European call, and then extends mean-variance hedging to the discrete time \citep{Schweizer1995}. \cite{Bertsimas2001} apply stochastic optimal control to the mean-variance hedging problem in the Markov-state setting and provide tractable solutions for both continuous and discrete time cases, and is later developed by \cite{Cerny2004} to incorporate stochastic interest rate. \cite{Basak2012} address the time-consistency issue in the global mean-variance hedging and provide a tractable and dynamic optimal strategy. 
Beyond all these theoretical works on episode mean-variance hedging, there are papers to study daily mean-variance hedging. 
\cite{Bakshi1997} compare the empirical performance of several stochastic volatility models with jumps on S\&P 500 options, and \cite{Bakshi2000} compare the difference between short-term and long-term options. Since the market is incomplete due to stochastic volatility and jumps, they consider the daily minimum variance hedging that minimize the instant volatility of the hedged portfolio.
\cite{Hull2017} study the empirically optimal delta hedging in the sense that it minimizes the variance of daily hedging errors. They prove that the minimum variance delta is approximately equal to the Black-Scholes delta plus Vega multiplied by the sensitivity of expected implied volatility to the price, which can be obtained by regressing on real market data. They find that their method outperform several stochastic volatility models on a variety of real-world options.
Based on \cite{Hull2017}, \cite{Xia2023} propose a two-step approach utilizing the observation that the volatility-price elasticity is mean-reverting, and find their method gain marginal improvement over \cite{Hull2017} on index options.
Although mathematically favorable, mean-variance hedging has several shortcomings. (i) Minimizing variance also punish the gain over expectation, which means it also reject the chance to earn a profit from trading. (ii) All these works rely on parametric assumptions and estimation, thus cannot avoid the misspecification and estimation errors. (iii) Mean-variance objectives ignore the higher-order moments, which is crucial in practitioners' risk management.

The literature has recently started to use deep learning methods to solve option hedging problems.
\cite{Buehler2019} propose to use deep learning to solve the hedging problem of European options with an objective function defined as the terminal P\&L of the hedge measured by a convex risk metric. They construct  a series of neural networks with the underlying price and the holding position as input and trading position as output to learn a hedging policy by minimizing the risk of P\&L. The method use a different neural network at each time period so the model may be difficult to train for long-maturity options. It requires to train a different model for each option with a different initial stock price, strike or maturity. In addition, it requires a parametric model and can only use simulated data as the training data.
\cite{Zhang2021} empirically test the deep hedging method in \cite{Buehler2019}. 
\cite{Fecamp2021} propose an augmented LSTM algorithm by adding a feedforward network to the output of an LSTM cell to solve a discrete-time hedging problem in the incomplete market. The objective function is linear or quadratic function of the P\&L at maturity.
\cite{Carbonneau2021} use deep hedging in \cite{Buehler2019} for equal risk pricing of derivatives with complex convex risk measures like CVaR.  
\cite{EvaLuetkebohmert2022} consider robust deep hedging in the sense of parameter uncertainty of generalized affine diffusion model.
\cite{Limmer2023} propose a GAN structure to do robust deep hedging considering the uncertainty of underlying asset price dynamics, which is modelled by a neural stochastic differential equation.
\cite{Wu2023} consider the robust risk-aware option hedging problem by minimizing the maximum of hedging risk in a Wasserstein ball around the terminal P\&L distribution.
Since we can't exhaust this literature, we refer to \cite{Ruf2020} who provide a comprehensive review of using neural networks for option pricing and hedging.
However, the series of deep learning approaches proposed by \cite{Buehler2019} and their follow-up works have two drawbacks: (i) These approaches are parametric model-based approaches for stocks, which require generating simulated data for stocks based on the model and then obtaining the optimal hedging strategy based on the simulated data through deep learning. Therefore, these methods cannot avoid the risk of the model being misspecified. Although \cite{EvaLuetkebohmert2022, Limmer2023, Wu2023} consider parameter uncertainty or distribution uncertainty, their methods still rely on the choice of parameter distributions or reference models. (ii) These methods require training a model for each individual option, and when any of the underlying price, strike price, or expiration date of the option changes, a new model needs to be retrained, making it difficult to apply in practice. 

There are also machine learning methods for option hedging that are data-driven, and thus free of model misspecification.
To our best knowledge, \cite{Hutchinson1994} are the first in finance literature to use neural networks to price and hedge options. They use neural networks to approximate the option price both in simulated and historical data, and they choose the partial derivative of the pricing function with respect to moneyness as their hedging strategy. They provide early evidence that learning methods may complement parametric ones in terms of pricing and hedging. 
\cite{Chen2021} use deep neural networks to approximate the option pricing model by a surrogate function, which treats the state, parameters of model as well as the hidden state (e.g. volatility) as input. The hidden state variables and parameters can be estimated by minimizing the distance between the predicted price and real price. However, they use the partial derivative of pricing function as \cite{Hutchinson1994}, and share the same major drawback in the application of hedging: the partial derivative doesn't take transaction costs, discrete hedging time, and a variety of risk-sensitive objectives into consideration, which are crucial in a real-world hedging circumstance.

\cite{Nian2021} minimize the daily hedging error through an encoder-decoder structure of deep learning on option data in real market. 
\cite{Ruf2022} use neural networks and linear regression based on option Greeks to find the optimal delta hedging strategy to minimize daily hedging errors, and find that both data-driven methods outperform the model-based delta hedging. 
Although these methods are data-driven and do not require modeling the dynamic changes of the stock, they also share several drawbacks. (i)
Their objective functions are based on the first two moments of incremental hedging error during the hedging process, which does not reflect the total P\&L from option hedging. 
(ii) Both papers only valid their models on a sub-dataset with limit range of moneyness or time-to-maturity. 
(iii) Both papers only consider discrete hedging, and do not include market friction like transaction costs.

With the development of deep reinforcement learning, many researchers have proposed lots of popular approaches to solve decision-making problems, including deep Q-learning (DQL) \citep{Mnih2015}, trust region policy optimization (TRPO) \citep{Abbeel2015}, proximal policy optimization (PPO) \citep{Schulman2017}, deep deterministic policy gradient (DDPG) \citep{Lillicrap2019}, and Twin Delayed Deep Deterministic policy gradient (TD3) \citep{Fujimoto2018}. These reinforcement learning algorithms maximize the cumulative discounted reward.
There are several papers studying risk-sensitive reinforcement learning.
\cite{Borkar2001, Borkar2002} proposes Actor-critic and Q-learning algorithms to maximize expected exponential utility of stationary reward and prove the convergence of these algorithms. The stationary reward is defined as the log expectation of exponential of cumulative reward divided by the time length, and then let the time length goes to infinity. 
\cite{Tamar2012} propose a policy gradient algorithm to maximize mean-variance objective of the cumulative reward, and \cite{Prashanth2013} propose an actor-critic algorithm for both stationary and cumulative reward. 
For CVaR-based reinforcement learning, \cite{Morimurat2010} use particles to approximate the distribution of conditional expected cumulative reward for finite state Markov process. Then they use the quantile from particles as VaR. \cite{Tamar2015} gives the formula of the derivative of CVaR with respect to transition probability parameters for policy gradient algorithms. But their algorithm requires to do simulation from a fixed initial state. 
\cite{Chow2015} gives a contraction mapping for value function iteration based on the risk envelope of CVaR. The value function is dependent on the state and CVaR level. In the numerical algorithm, they linearly interpolate the CVaR level dimension, which inevitably introduces approximation error. \cite{Petrik2012} study the reinforcement learning with coherent risk measure on cumulative reward as the objective by risk envelope with known transition dynamics. 
From above, we can conclude that none of these papers are suitable for our task, i.e. to use real market data to train an agent that can hedge options and control the tail risk using reinforcement learning because they require the knowledge of market dynamics or the ability to simulate from a fixed state.

There are some reinforcement learning-based approaches to solving option hedging problems. 
\cite{Halperin2020} discretizes the stock price under a simple discrete-time Black-Scholes model and solves the hedging problem based on Q-learning using a mean-variance type reward function. 
\cite{Kolm2019} also solve the hedging problem based on the discrete-time Black-Scholes model and Q-learning with a reward function defined as a quadratic function of the incremental hedge P\&L to approximate the variance of the final P\&L of the hedge. 
\cite{Cao2021} consider mean-variance hedging final P\&L with deep Q-networks. In order to keep track of the standard deviation of the terminal P\&L, they use an additional Q-network to approximate the second moment of the hedging cost.
Thus the main drawbacks of these papers on option hedging based on reinforcement learning include that (i) The reward functions are always in the form of linear or quadratic. They only consider the mean or not strictly the variance of the hedging P\&L, but not the Value-at-Risk (VaR) or CVaR tail risk measures of the hedging P\&L. (ii) All these methods are equity-based parametric modeling methods and therefore, like the delta hedging methods, are not immune to parametric modeling errors. (iii) All these proposed models are dependent on parametric models and not adapted to different initial underlying prices and different options with varying maturity and moneyness. They require retraining different model parameters for different options and cannot learn a uniform model for options with different initial underlying prices, strike prices, and expiration dates.
\cite{Du2020} address the last issue by including the strike price in the state to hedge options with different strikes. But it is still based on parametric models and reward function is a quadratic function of the incremental hedge P\&L.
\cite{Mikkilae2023} test deep reinforcement learning approach (TD3) to hedge options with transaction costs using real world S\&P 500 hourly data. Their reward is defined as the hedging error minus a multiple of the absolute value of hedging error, which is only an approximation of the variance of the hedging error. They find that the agent trained with real-world data achieves higher reward than the one trained with synthetic data because the former can better capture the market dynamics. Although their method is data-driven, it still neglects the tail risk in their objective functions. In their experiments, the mean term is dominated by the variance term, so they model shows little mean improvement. Moreover, their model is not universal: they only consider hedging options for 5 sequential days and exclude any options that are close to maturity; nor do they include deep in/out-of-the-money options in their datasets.

There are several papers focusing on reinforcement learning with risk measures or utility functions and its applications in finance.
\cite{Murray2022} propose an actor-critic algorithm for hedging with exponential utility function, which is time-consistent. 
\cite{SaeedMarzban2023} propose an actor-critic algorithm for the hedging problem with recursive expectile risk measure, which they believe is an ideal choice because it's coherent, elicitable, and time-consistent. 
For general applications, \cite{Coache2023a} propose a modified actor-critic method for deep reinforcement learning with general recursive convex risk measures based on the risk envelope  representation of convex risk measures. Their method involves nested simulation at each visited state to estimate some value of interest, and thus is computationally expensive and cannot use real-world data to train.
Their follow-up work \citep{Coache2023} proposes a modified actor-critic method specifically for recursive conditionally elicitable risk measures like recursive CVaR. They use two neural networks to approximate the value of VaR and CVaR in each steps.
\cite{Buehler2023} solve the hedging problem with recursive optimized certainty equivalent (OCE) monetary utility by iterating value and policy functions with Bellman equations. They also propose to add a neural network to approximate the OCE value to the actor-critic framework.
All these methods utilize Bellman equations of recursive risk measure, and thus are not targeting for some risk measures wildly used in industries like CVaR, which are focusing on terminal P\&L and not time consistent. Moreover, all these papers above do not provide empirical performance of their methods trained and evaluated with real-world data.

The rest of paper is organized as follows. In Section~\ref{sec:formulation}, we present the formulation of the discrete-time option hedging problem. Section~\ref{sec:CU-RL algorithm} proposes the contract-unified reinforcement learning (CU-RL) approach for solving the discrete-time option hedging problem. In Section~\ref{sec:hedging_performance_parametric_models}, we study the hedging performance of the proposed CS-RL and CU-RL approaches when the hedging models are trained based on data simulated under parametric models of the underlying stock. In Section~\ref{sec:model_free_empirical_results}, we carry out comprehensive empirical study on the performance of the CU-RL approach for hedging S\&P 500 index options when the hedging model is trained by using the historical data of S\&P 500 index price and S\&P 500 index options. Section~\ref{sec:conclusion} concludes. 

\section{Formulation of the Dynamic Hedging Problem}\label{sec:formulation}

In this section, we formulate the dynamic hedging problem as a discrete-time stochastic control problem that achieves a tradeoff between minimizing the CVaR of the final P\&L and maximizing the expectation of the final P\&L at the maturity of the European option.

Suppose at time 0, an option dealer sells an European-style option on one share of a stock with strike $K$ and maturity $T$ periods. Let $S_t$ be the price of the stock underlying the option. The dealer hedges its short position on the option by dynamically trading the stock at time $t=0, 1, 2, \ldots, T-1$. Suppose before rebalancing at time $t$, the dealer holds $B_{t}$ amount of cash and $\delta_{t}$ shares of the stock in the hedging portfolio, and he or she observes the option price $Z_t$ from the market or calculated from some option pricing model. 

In particular, right before rebalancing at time $0$, the dealer holds cash $B_{0}$ and $\delta_{0}=0$ share of the stock, where $B_0$ is the amount of cash received by the seller due to the sale of the option. 

Since the dealer has a short position of the option, the dealer's total portfolio value right before rebalancing at time $t$ is 
\begin{equation}
	\label{wealth}
	W_{t} = B_{t} +\delta_{t}S_{t} - Z_{t}, \quad t = 0, 1, \cdots, T.
\end{equation}
Then, the dealer decides the number of shares $a_t$ that will be held in the hedging portfolio after the trading at time $t$, and then he buys $a_t-\delta_t$ shares of the stock at time $t$, which costs $(a_{t} - \delta_{t}) S_{t} + C_{t}(S_{t}, \delta_t, a_{t})$, where $C_{t}(S_{t}, \delta_t, a_{t})$ is the transaction cost. After rebalancing at time $t$, the dealer holds $B_{t} - (a_{t} - \delta_{t})S_{t} - C_{t}(S_{t}, \delta_t, a_{t})$ in cash. 

Hence, right before rebalancing at time $t+1$, the number of shares of stock held by the dealer is 
\begin{equation}\label{position_dyn}
	\delta_{t+1}  = a_{t}, 
\end{equation}
and the amount of cash held by the dealer is 
\begin{equation}\label{bank_dyn}
	B_{t+1} = (B_{t} - (a_{t} - \delta_{t})S_{t} - C_{t}(S_{t}, \delta_t, a_{t})) e^{r \Delta t}, 
\end{equation}
where $\Delta t$ is the time difference in year between time period $t$ and $t+1$, $r$ is the risk-free interest rate. 

The total P\&L of the dealer is realized at time $T$ and is equal to $W_T$. The tail risk of the future P\&L $X$ of a portfolio can be measured by VaR or CVaR. The VaR of $X$ at confidence level $\gamma$ is defined as 
$\text{VaR}_{\gamma}(X) := \inf \left\{x \in \mathbb{R}: F_{-X}(x)\geq \gamma \right\}$, where $F_{-X}(x):= \mathbb{P}(-X \leq x)$ is the probability distribution function of $-X$, which is the loss of the portfolio. The CVaR at level $\alpha$ of $X$ is defined as
\begin{equation}
	\label{cvar}
	\text{CVaR}_{\alpha}(X) := \frac{1}{1-\alpha} \int_{\alpha}^{1} \text{VaR}_{\gamma}(X) \, d\gamma.
\end{equation}

We assume that the dealer's objective is 
\begin{equation}\label{equ:dealer_objective}
	\max_{a_0,a_1,\ldots,a_{T-1}} - \lambda_{1} \text{CVaR}_{\alpha}(W_{T})+\lambda_{2} E[W_T],
\end{equation}
where $\lambda_{1}\geq 0$ and $\lambda_{2}\geq 0$ are two constant weights that specify the relative importance of controlling the tail risk of the final P\&L and maximizing the expectation of the final P\&L. 

\section{A Risk Sensitive Contract-unified  Reinforcement Learning Approach for Option Hedging}\label{sec:CU-RL algorithm}

There are two kinds of solutions to the problem \eqref{equ:dealer_objective}: contract-specific solution and contract-unified solution. The contract-specific solution solves the problem \eqref{equ:dealer_objective} for a specific option with given contract parameters, including
strike $K$, maturity $T$, initial stock price $S_0$, 
and some initial states of the dealer, including 
initial cash $B_0$ and initial stock position $\delta_0$. The contract-specific solution may also depend on the  initial option price $Z_0$ or  implied volatility $\sigma_0$. 

The contract-unified solution to the problem \eqref{equ:dealer_objective} is a \emph{unified} solution for an infinite number of options with any contract parameters (e.g., strike and maturity) and initial states (e.g., initial stock price, initial implied volatility, initial cash position, and initial hedging position). In practice, option dealers need to hedge a portfolio of options with different contract parameters, starting dates, and maturities, hence a contract-unified solution is desirable and important.

In Section \ref{subsec:hedging_a_single_option}, we first propose a contract-specific reinforcement learning algorithm for hedging a single option. In Section \ref{subsec:hedging_multiple_options}, we propose a contract-unified reinforcement learning algorithm. The algorithm learns a unified hedging policy that can be used to hedge different option contracts with different contract parameters and initial states of the dealer.

\subsection{A New Contract-specific Reinforcement Learning Algorithm for Hedging a Single Option}\label{subsec:hedging_a_single_option}

In this subsection, we propose a new contract-specific  reinforcement learning (CS-RL) approach for hedging a single option with fixed contract parameters including initial stock price, maturity, and strike, etc. The CS-RL approach can only be used when a parametric model is specified for pricing the option, because only in this case, we can simulate under the parametric model to obtain enough training data of sample paths of stock prices and option prices that all start from the same initial stock price and have the same contract parameters. The CS-RL approach cannot be applied in a model-free way, because we cannot obtain enough training data of historical prices of options with the same initial stock price and contract parameters as the option that is to be hedged. 

First, we propose a new formulation of the hedging problem \eqref{equ:dealer_objective} for a specific option contract in Section \ref{subsec:new_formulation_hedging_problem}.  Then, we propose a contract-specific reinforcement learning algorithm for solving the problem in Section \ref{subsubsec:contract_specific_rl_algo}.

\subsubsection{A New Formulation of the Hedging Problem \eqref{equ:dealer_objective}}\label{subsec:new_formulation_hedging_problem}
First, We reformulate the hedging problem \eqref{equ:dealer_objective} into a reinforcement learning problem. We use $\bm{s_{t}}$ to denote the state variable at time $t$. $ \bm{s_{t}}$
can include the information that the dealer observes at time $t$ right before the rebalancing at time $t$. 
$ \bm{s_{t}}$ include the stock price $S_t$, the dealer's cash position $B_t$ and stock position $\delta_t$ in the hedging portfolio right before the rebalancing at time $t$, the stochastic volatility $\sigma_t$, and the time to maturity $\tau_{t} = T-t$. The stochastic volatility $\sigma_t$ can be defined as the forecast next-time-period volatility in the parametric model of the stock price (e.g., in a GARCH model); it may also be 
defined as the implied volatility calculated from the option price $Z_t$ at time $t$, where $Z_t$ is the option price calculated based on the parametric option pricing model, such as a jump diffusion model. 
In summary, the state $\bm{s_{t}}$ can be represented by
\begin{equation}
	\label{formulation-state}
	\bm{s_{t}} := (B_{t}, \delta_{t}, S_{t},\sigma_{t}, \tau_{t}), t = 0, 1,\cdots, T.
\end{equation}

Having observed $\bm{s_{t}}$, the dealer decides the number of shares $a_t$ that will be held in the hedging portfolio after the rebalancing at time $t$, and carries out the trading. And then, the state transits to $\bm{s_{t+1}} = (B_{t+1}, \delta_{t+1}, S_{t+1},\sigma_{t+1}, \tau_{t+1})$ at the next time period $t+1$, where $B_{t+1}$ and $\delta_{t+1}$ are determined by \eqref{bank_dyn} and \eqref{position_dyn}, $S_{t+1}$ and $\sigma_{t+1}$ are determined by the parametric model of option pricing. 

The dealer's objective is to achieve a balance between minimizing the tail risk of the terminal P\&L and maximizing the expected terminal P\&L, as specified in \eqref{equ:dealer_objective}. The CVaR in \eqref{equ:dealer_objective}  has the representation \citep{Rockafellar2002}
\begin{align}
	\label{cvar-alter}
	\text{CVaR}_{\alpha}(W_T) = & \inf_{x \in \mathbb{R}} \left\{ x + \frac{1}{1- \alpha}\mathbb{E}\left[\max (-W_T-x, 0)\right]\right\}\notag\\
	= & \mathbb{E}\left[\omega + \frac{1}{1- \alpha}\max (-W_T-\omega, 0)\right],
\end{align} 
where 
\begin{equation}\label{equ:omega_star_one_contract}
	\omega = \text{VaR}_{\alpha}(W_T).
\end{equation}
Hence, we define the reward $R_T$ received by the dealer at the terminal time $T$ as 
\begin{equation}\label{equ:cvar_terminal_reward}
	R_{T}(\bm{s}_{T-1},a_{T-1},\bm{s}_{T}):= - \lambda_{1} \left[\omega + \frac{1}{1- \alpha}\max (-W_T-\omega, 0)\right]+\lambda_{2} W_T,	
\end{equation}
where $\lambda_{1}\geq 0$ and $\lambda_{2}\geq 0$ are the two weights in \eqref{equ:dealer_objective}.

At the intermediate time $t=0,1,\ldots, T-2$, the dealer takes the action $a_t$, and then right before rebalancing at time $t+1$,
the dealer's wealth becomes $W_{t+1}$. Although the dealer mainly focuses on the total P\&L realized at time $T$, 
$W_{t+1}$ provides information about the performance of the hedging action $a_t$, and can be used to defined the reward $R_{t+1}$ that the dealer receives at time $t+1$. 
The dealer may be concerned when negative hedging error occurs, i.e., when $W_{t+1}<0$, and may be indifferent to the value of the hedging error $W_{t+1}$ as long as $W_{t+1}\geq 0$. So the reward may be defined in an asymmetric way as 
\begin{equation}
	\label{R_relu}
	R_{t+1}(\bm{s}_{t},a_{t},\bm{s}_{t+1}) = -|W_{t+1}| \cdot 1_{ \{ W_{t+1}<0 \} }, t = 0, 1, \cdots, T-2.
\end{equation}	
This definition only penalizes negative hedging performance but does not provide award for positive hedging performance, so it does not encourage profit making. However, profit making is rewarded by the definition of the reward $R_T$ in \eqref{equ:cvar_terminal_reward}.

Suppose at each time $t=0,1,\ldots, T-1$, the dealer takes the action $a_{t}$ based on the state $\bm{s}_t$ by following the policy $\pi(a_t | \bm{s}_t)$, which can be deterministic or stochastic. 

The above formulation falls into the framework of reinforcement learning, except that there is an unknown parameter $\omega$ in the expression of $R_T$ in \eqref{equ:cvar_terminal_reward}. We will propose a new algorithm for solving the problem \eqref{equ:dealer_objective} that finds the optimal policy $\pi(a | \bm{s}_t)$ together with the unknown parameter $\omega$.

\subsubsection{The CS-RL Algorithm for Hedging a Single Option}\label{subsubsec:contract_specific_rl_algo}

In this subsection, we propose the CS-RL algorithm for hedging a \emph{single} options contract. 
This options contract has a fixed initial state $\bm{s}_0=(B_{0}, \delta_{0}, S_{0},\sigma_{0}, \tau_{0})$ and a given strike $K$. The reinforcement learning algorithm will learn the optimal hedging strategy that is optimal only for this particular options contract. In other words, the optimal hedging strategy cannot be used for hedging other options contracts. The CS-RL algorithm is an extension of the Proximal Policy Optimization (PPO) algorithm. 

Starting from the initial state $\bm{s}_{0}$, the agent takes actions following a policy $\pi$ until the target option matures, which result in a trajectory
\begin{equation}
	\label{trajectory}
	\mathcal{T} =(\bm{s}_{0}, a_{0}, R_{1}, \bm{s}_{1}, a_{1}, R_{2}, \cdots, \bm{s}_{T-1}, a_{T-1}, R_{T}, \bm{s}_{T}).
\end{equation}
Under the policy $\pi$, the state value function $V^{\pi}(\bm{s})$ is 
\begin{equation}
	\label{value_func}
	V^{\pi}(\bm{s}) =\mathbb{E}^{\pi} \left[ \sum \limits_{k=0}^{T-t-1} \gamma^{k} R_{t+1+k} \Bigg| \bm{s}_{t} = \bm{s} \right], t= 0, 1, \cdots, T-1,
\end{equation}
where $\gamma \in [0,1]$ is the discount rate. Since our problem is a finite-horizon problem, $\gamma$ is defined to be 1. 
The state-action value function under $\pi$ is 
\begin{equation}
	\label{q_func}
	Q^{\pi}(\bm{s},a) = \mathbb{E}^{\pi} \left[ \sum \limits_{k=0}^{T-t-1} \gamma^{k} R_{t+1+k} \Bigg| \bm{s}_{t} = \bm{s}, a_{t} = a \right], t=0, 1,\cdots, T-1.
\end{equation}
The advantage function under $\pi$ is 
\begin{equation}
	\label{ad_func}
	A^{\pi}(\bm{s},a) = Q^{\pi}(\bm{s},a) - V^{\pi}(\bm{s}).
\end{equation} 

We parameterize the policy $\pi(\cdot|\bm{s})$ by a stochastic policy $\pi_{\theta}(a | \bm{s})$ that follows a normal distribution $N(\mu (\bm{s};\phi), \psi)$, where $\mu (\bm{s};\phi)$ and $\psi$ are respectively the mean and variance of the normal distribution. 
The parameter $\theta$ is defined as $\theta:= (\phi, \psi)$. The mean $\mu (\bm{s};\phi)$ is represented by a neural network with $\bm{s}$ as input and $\phi$ as the network parameter, and $\psi$ is a free parameter that is not affected by the state. 
For an European call (resp., put) option, the activation function of the last layer of the $\mu$ network is defined to be the sigmoid (resp., negative sigmoid) function as the delta of option should be in the range $(0, 1)$ (resp., (-1, 0)).
The action $a_t$ is truncated to be within the interval $[-b, b]$ after it is sampled from the distribution $N(\mu (\bm{s}_t;\phi), \psi)$, where $b>0$ is a constant. For an option written on a single share of the underlying stock, $b$ is defined to be $1$. We parameterize the state-value function $V(\bm{s})$ by a neural network $V(\bm{s}; \xi)$ with parameter $\xi$.

However, different from the common formulation, the proposed rewards measured by the CVaR in \eqref{equ:cvar_terminal_reward} introduce an unknown parameter $\omega$. Therefore, we extend the PPO method to learn the additional parameter $\omega$ together with the optimal hedging policy at the same time.

For the policy network, the objective function $L^{P}(\bm{s}_{t}, a_{t}, \hat{A}^{\pi_{\theta_{\text{old}}}}(\bm{s}_t, a_t), \theta_{\text{old}}; \theta)$ is shown as follows.
\begin{equation}
	\label{obj_theta_ppo}
	L^{P}(\bm{s}_{t}, a_{t}, \hat{A}^{\pi_{\theta_{\text{old}}}}(\bm{s}_t, a_t), \theta_{\text{old}}; \theta) = \min \left(\frac{\pi(a_t|\bm{s}_{t}, \theta)}{\pi(a_t|\bm{s}_{t}, \theta_{old})} \hat{A}^{\pi_{\theta_{old}}}(\bm{s}_{t},a_t),~ g(\epsilon, \hat{A}^{\pi_{\theta_{old}}}(\bm{s}_{t},a_t)) \right),
\end{equation}
where $\hat{A}^{\pi_{\theta_{\text{old}}}}(\bm{s}_t, a_t)$ is an estimate of the advantage function in \eqref{ad_func} under the old policy $\pi_{\theta_{\text{old}}}$, and 
\begin{equation}
	g(\epsilon, A) = \begin{cases}
		(1+ \epsilon)A, & \text{if}\ A \geq 0, \\
		(1- \epsilon)A, & \text{else}.
	\end{cases}
\end{equation}
where the clip parameter $\epsilon \in[0, 1)$ is used to control how far away the updated policy can be allowed to move from the old policy, which avoids the instability caused by drastic policy change at one step.

The $1$-step TD residual of the value function is defined as 
\begin{equation*}
	\delta^{V}_{t}:= R_{t+1}+ \gamma V(\bm{s}_{t+1}) -V(\bm{s}_{t}), t = 0, \cdots, T-1.
\end{equation*}
In particular, $\delta^{V}_{T-1}= R_{T} - V(\bm{s}_{T-1})$ as $V(\bm{s}_{T})=0$. 
The generalized advantage estimator $\hat{A}(\bm{s}_t, a_t)$ is given by 
\begin{equation}
	\label{gae_ad}
	\hat{A}(\bm{s}_t, a_t) := \sum_{l=0}^{T-t-1}(\gamma \lambda^{gae})^{l} \delta_{t+l}^{V}.
\end{equation} 
where $\lambda^{gae} \in [0,1]$ is a parameter for controlling the tradeoff between bias and variance.
The generalized rewards-to-go at time $t$, i.e., the cumulative reward received from time $t+1$ to time $T$, is defined as 
\begin{equation}\label{equ:generalized_cumu_reward}
	G_t = \hat{A}(\bm{s}_t, a_t) + V(\bm{s}_t).
\end{equation}

The objective function $ L^{V}(\bm{s}_{t}, G_{t}; \xi)$ 
for learning the parameter $\xi$ of the value network is specified as
\begin{equation}
	\label{obj_value_ppo}
	L^{V}(\bm{s}_{t}, G_{t}; \xi) = \left(V(\bm{s}_{t}; \xi) - G_{t}\right)^{2},
\end{equation}
where $V(\bm{s}_{t}; \xi)$ is the state value predicted by the value network, 
$G_{t}$ is 
calculated 
based on the current parameter of the policy network and the value network.

We also need to learn the unknown parameter $\omega$ used by the terminal reward $R_T$ in \eqref{equ:cvar_terminal_reward}. By the definition of $\omega$ in \eqref{equ:omega_star_one_contract}, $\omega$ is the $\alpha$-level VaR of the total P\&L  $W_T$. Since $W_T$ depends on the contract specific information of the option including the initial state $\bm{s}_0$ and the strike $K$, $\omega$ also depends on these contract specific information. Since the option contract is fixed, the optimal $\omega$ is also a fixed value. 

By the result of quantile regression, the unknown parameter $\omega$ can be learned by minimizing the loss function
\begin{equation}
	\label{loss_omega1}
	L^{O}(\omega, W_{T})=
	\begin{cases}
		\alpha |\omega - (-W_{T})|, & -W_{T} \geq \omega, \\
		(1 - \alpha) |\omega - (-W_{T})|, & -W_{T} \leq \omega.
	\end{cases} 
\end{equation}

To encourage exploration, it is desirable to minimize the entropy loss for the policy $\pi_{\theta}(a | \bm{s})$
\begin{equation}
	\label{obj_entropy_ppo}
	L^{E}(\theta) = -\left[\frac{1}{2} + \frac{1}{2} \log(2\pi) + \log(\psi)\right].
\end{equation}

Overall, we propose to update the parameter from $(\theta_{\text{old}}, \xi_{\text{old}}, \omega_{\text{old}})$ to some new parameter value $(\theta_{\text{new}}, \xi_{\text{new}}, \omega_{\text{new}})$ by carrying out stochastic gradient descent updating for minimizing the objective function 
\begin{equation}
	\label{obj_newcvar1}
	L(\theta, \xi, \omega) = - \mathbb{E} \left[L^{P}(\bm{s}_{t}, a_{t}, \hat{A}^{\theta_{\text{old}}}_{t}(\bm{s}_{t}, a_{t}), \theta_{\text{old}}; \theta) + c_{0} L^{E}(\theta) + c_{1} L^{V}(\bm{s}_{t}, G_{t}; \xi) + c_{2} L^{O}(\omega, W_{T}) \right],
\end{equation}
where $c_{0}\geq 0, c_{1}>0$ and $c_{2}>0$ are constants.

Each parameter updating step for $\theta$ and $\xi$ is based on a minibatch dataset $\mathcal{B}$. In order to enhance the stability of the algorithm, we cap the $L^{2}$-norm of the gradient 
\begin{equation}\label{equ:gradient_norm_theta_xi}
\mathcal{N}_{\mathcal{B}}=\|\nabla_{\theta,\xi} \bar{L}_{\mathcal{B}}(\theta, \xi, \omega)\|_2	
\end{equation}
 by a maximal gradient norm $\mathcal{N}^{*}=0.5$, where 
$\bar{L}_{\mathcal{B}}(\theta, \xi, \omega)$ is the average value of the loss function in \eqref{obj_newcvar1} over the minibatch $\mathcal{B}$. 
Similarly, each parameter updating step for $\omega$ is based on a dataset $\mathcal{H}$. We cap the absolute value of the partial derivative 
\begin{equation}\label{equ:gradient_norm_omega}	
\mathcal{N}_{\mathcal{H}}=|\frac{\partial}{\partial \omega} \bar{L}_{\mathcal{H}}(\theta, \xi, \omega)|
\end{equation}
by an upper bound $\mathcal{N}^{*}=0.5$, where 
$\bar{L}_{\mathcal{H}}(\theta, \xi, \omega)$ is the average value of the loss function in \eqref{obj_newcvar1} over the dataset $\mathcal{H}$.

We summarize the new contract-specific reinforcement learning (CS-RL) algorithm for a single option in Algorithm~\ref{algorithm_ppo_omega}. Note that in the algorithm, 
all the training trajectories start from the same initial state $\bm{s}_0$ that corresponds to the single options contract. 

\begin{breakablealgorithm}
	\caption{The CS-RL algorithm for optimal hedging of a single option with a fixed initial state $\bm{s}_0$} \label{algorithm_ppo_omega}
	\begin{algorithmic}[0]
		\State Initialize parameters $\theta_{0}$, $\xi_{0}$, $\omega_0$. 
		\For{$k$ from $0$ to $\mathcal{K} - 1$}
		\State Initialize the buffer $\mathcal{D}_k=\emptyset$ and $\mathcal{H}_k=\emptyset$.
		\For{$i$ from $0$ to $N-1$}
		\State Generate a trajectory $\mathcal{T}_{i}$  by starting from $\bm{s}_0$ and following the policy $\pi_{\theta_{k}}$. Normalize states with the running mean and standard deviation.
		\State Calculate the generalized advantage estimator and rewards-to-go at time $T-1$ of the trajectory $\mathcal{T}_{i}$: 
		$$\hat{A}^{(i)}_{T-1} =R_{T}^{(i)}(\omega_{k}) - V(\bm{s}_{T-1}^{(i)}; \xi_{k}), \quad G^{(i)}_{T-1} =R_{T}^{(i)}(\omega_{k}).$$
		\For{$t$ from $T-2$ to 0}
		\State Compute generalized advantage estimator 
		\State $\hat{A}^{(i)}_{t} =(\lambda^{gae} \gamma) \cdot \hat{A}^{(i)}_{t+1} + R^{(i)}_{t+1} + \gamma V(\bm{s}^{(i)}_{t+1}; \xi_{k}) - V(\bm{s}^{(i)}_{t}; \xi_{k})$.
		\State Compute rewards-to-go: $G^{(i)}_{t} =\hat{A}^{(i)}_{t}+ V(\bm{s}^{(i)}_{t}; \xi_{k})$.
		\EndFor
		\State Add the tuples $\left\{ (\bm{s}_{t}^{(i)}, a_{t}^{(i)}, \pi_{\theta_{k}}(a_{t}^{(i)}|\bm{s}_{t}^{(i)}),G^{(i)}_{t}, \hat{A}^{(i)}_{t}) \right\}_{t=0}^{T-1}$  to the buffer $\mathcal{D}_{k}$. 
		\State Add the sample $W^{(i)}_T$ to the buffer $\mathcal{H}_k$.
		\EndFor
		\State Set $\theta = \theta_{k},~\xi=\xi_{k},~\omega=\omega_{k}$.
		\For{$m$ from $0$ to $M-1$}
		\State Shuffle the buffer $\mathcal{D}_{k}$.
		\For{i from $0$ to $\frac{|\mathcal{D}_{k}|}{n}-1$}
		\State Sequentially collect $n$ tuples from $\mathcal{D}_k$ as a minibatch: $\mathcal{B}=\{ (\bm{s}^{j}, a^{j}, \pi^{j}_{\theta_{k}}, G^{j}, \hat{A}^{j}) \}_{j=1}^{n}$.
		\State Calculate the $L^{2}$-norm $\mathcal{N}_{\mathcal{B}}$ defined in \eqref{equ:gradient_norm_theta_xi}.
		\State Update the policy network parameter $\theta$ by: 
		$$\theta \leftarrow \theta + \eta  \left[ \frac{1}{n} \sum_{j=1}^{n} \nabla_{\theta} L^{P}(\bm{s}^{j}, a^{j}, \hat{A}^{j}, \theta_{k}; \theta)- c_{0}  \nabla_{\theta} L^{E}(\theta) \right] \cdot \min \left(\frac{\mathcal{N}^{*}}{\mathcal{N}_{\mathcal{B}}}, 1\right);$$ 
		\State Update the value network parameter $\xi$ by:
		$$\xi \leftarrow \xi - c_{1} \eta  \left[ \frac{1}{n} \sum_{j=1}^{n} \nabla_{\xi} L^{V}(\bm{s}^{j}, G^{j}; \xi) \right] \cdot \min \left(\frac{\mathcal{N}^{*}}{\mathcal{N}_{\mathcal{B}}}, 1\right);$$ 
		\EndFor
		\State Use all the $N$ samples $\{W_{T}^{(j)}\}_{j=1}^{N}$ in $\mathcal{H}_{k}$ to update the parameter $\omega$:
		$$\omega \leftarrow \omega - c_{2} \eta \left[ \frac{1}{N} \sum_{j=1}^{N} \frac{\partial}{\partial \omega} L^{O}(\omega, W_{T}^{(j)}) \right] \cdot \min \left(\frac{\mathcal{N}^{*}}{\mathcal{N}_{\mathcal{H}_k}}, 1\right),$$
		where $\mathcal{N}_{\mathcal{H}_k}$ is defined in \eqref{equ:gradient_norm_omega} with $\mathcal{H}=\mathcal{H}_k$.
		\EndFor
		\State $\theta_{k+1} \leftarrow \theta;~\xi_{k+1} \leftarrow \xi;~\omega_{k+1} \leftarrow \omega.$ 
		\EndFor
	\end{algorithmic}
\end{breakablealgorithm}

\subsection{A Contract-Unified Reinforcement Learning Approach for Option Hedging}\label{subsec:hedging_multiple_options}

In this subsection, we propose a new contract-unified reinforcement learning (CU-RL) algorithm for learning the optimal hedging strategy for options with any contract parameters including initial stock price, maturity, and strike, and initial state of the dealer including initial cash and stock position. 

The CU-RL approach can be used in a model-free way, i.e., it can be used without a parametric model for option pricing. In contrast, the CS-RL approach can only be used under a parametric model for option pricing. 

We first extend the state variable to include all contract parameters that affect the option price, including initial stock price, maturity, strike, etc, so that the hedging position (i.e., the action) obtained from inputing the state variable to the policy network will depend on the parameter of the option contract. More precisely, we extend the definition of the state to be 
\begin{equation}\label{equ:cu_rl_state_def}
	\bm{s}_{t} = \left(B_{t}, \delta_{t}, S_{t}, \sigma_{t}, \tau_{t}, \frac{K}{S_{0}}, W_{t}, I_t\right),
\end{equation}
where $K$ is the strike of the option and $I_t$ represents market information that is relevant for the values of the stock or option. For example, $I_t$ can be defined as $I_t =(\Delta_{t}, \Gamma_{t}, \mathcal{V}_{t}, \Theta_t)$ that represent the delta, gamma, vega and theta, respectively.

However, the extension of state variables is far from enough to obtain hedging policy that is uniformly optimal for options with different initial states. In order for the hedging policy to be optimal for all options contracts, we will need to use training trajectories starting from different initial states. For an options contract with initial state $\bm{s}_0$, the distribution of the optimally hedging P\&L $W_T$ depends on $\bm{s}_0$; for example, the final hedging P\&L for an option with a maturity of five days and that for an option with a maturity of one year apparently have different probability distributions. Therefore, the $\alpha$-level VaR of the final hedging P\&L also depends on $\bm{s}_0$, i.e., $\omega$ in \eqref{equ:omega_star_one_contract} is a function of the $\bm{s}_0$ of the option contract. 

We propose to use a neural network $\omega(\bm{s}_{0}; \zeta)$ to represent the $\alpha$-level VaR of the final hedging P\&L $W_T$, where $\bm{s}_{0}$ is the input of the neural network and $\zeta$ is the network parameter. 
The output layer of the neural network is linear as the VaR can be positive or negative. 
Similar to the objective function in \eqref{loss_omega1}, the VaR network parameter $\zeta$ can be learned through minimizing the loss function for quantile regression
\begin{equation}
	\label{loss_zeta}
	L^{O}(\bm{s}_{0}, W_{T}; \zeta)=
	\begin{cases}
		\alpha |\omega(\bm{s}_{0}; \zeta) - (-W_{T})|, & -W_{T} \geq \omega(\bm{s}_{0}; \zeta), \\
		(1 - \alpha) |\omega(\bm{s}_{0}; \zeta) - (-W_{T})|, & -W_{T} \leq \omega(\bm{s}_{0}; \zeta).
	\end{cases} 
\end{equation}

Similar to the CS-RL algorithm, we use a policy network $\pi_{\theta}$ to represent the stochastic policy and a state value network $V(\cdot; \xi)$ to represent the state value function. The parameter $\theta$ is updated from $\theta_{\text{old}}$ by stochastic gradient descent updating for minimizing the policy objective function $L^{P}(\bm{s}_{t}, a_{t}, \hat{A}^{\theta_{\text{old}}}_{t}(\bm{s}_{t}, a_{t}), \theta_{\text{old}}; \theta)$, which is defined in \eqref{obj_theta_ppo}. The parameter $\xi$ is updated from minimizing the objective $L^{V}(\bm{s}_{t}, G_{t}; \xi)$ defined in \eqref{obj_value_ppo}. 

Overall, we propose to update the parameter from $(\theta_{\text{old}}, \xi_{\text{old}}, \zeta_{\text{old}})$ to some new parameter value $(\theta_{\text{new}}, \xi_{\text{new}}, \zeta_{\text{new}})$ by carrying out stochastic gradient descent updating for minimizing the objective function 
\begin{equation}
	\label{obj_newcvar_with_zeta}
	L(\theta, \xi, \zeta) = - \mathbb{E} \left[L^{P}(\bm{s}_{t}, a_{t}, \hat{A}^{\theta_{\text{old}}}_{t}(\bm{s}_{t}, a_{t}), \theta_{\text{old}}; \theta) + c_{0} L^{E}(\theta) + c_{1} L^{V}(\bm{s}_{t}, G_{t}; \xi) + c_{2} L^{O}(\bm{s}_0, W_{T}; \zeta) \right],
\end{equation}
where $c_{0}\geq 0, c_{1}>0$ and $c_{2}>0$ are constants, $L^{E}(\theta)$ is the entropy loss function defined in \eqref{obj_entropy_ppo}.

A big advantage of introducing the VaR network is that the CU-RL algorithm can be applied in a model-free way, because we can use the historical data of any option with any initial states and contract parameters as the training data, which is feasible only because the algorithm can consistently distinguish different option contracts by using the VaR network to compute the VaR of the hedging P\&L of different option contracts differently. In contrast, the CU-RL algorithm can not be used in a model free way, because there are no enough historical data of option prices that have the initial state as the particular option contract that is to be hedged.
 
Each parameter updating step for $\theta$ and $\xi$ is based on a minibatch dataset $\mathcal{B}$. In order to enhance the stability of the algorithm, we cap the $L^{2}$-norm of the gradient 
\begin{equation}\label{equ:gradient_norm_theta_xi_2}
	\mathcal{N}_{\mathcal{B}}=\|\nabla_{\theta,\xi} \bar{L}_{\mathcal{B}}(\theta, \xi, \zeta)\|_2	
\end{equation}
by a maximal gradient norm $\mathcal{N}^{*}=0.5$, where 
$\bar{L}_{\mathcal{B}}(\theta, \xi, \zeta)$ is the average value of the loss function in \eqref{obj_newcvar_with_zeta} over the minibatch $\mathcal{B}$. 
Similarly, each parameter updating step for $\zeta$ is based on a dataset $\mathcal{H}$. We cap the absolute value of the gradient 
\begin{equation}\label{equ:gradient_norm_zeta}	
	\mathcal{N}_{\mathcal{H}}=\|\nabla_{\zeta} \bar{L}_{\mathcal{H}}(\theta, \xi, \zeta)\|_2
\end{equation}
by an upper bound $\mathcal{N}^{*}=0.5$, where 
$\bar{L}_{\mathcal{H}}(\theta, \xi, \omega)$ is the average value of the loss function in \eqref{obj_newcvar_with_zeta} over the dataset $\mathcal{H}$.

We summarize the detailed CU-RL algorithm for optimal hedging of options with \textit{any} initial state $\bm{s}_0$ in Algorithm~\ref{algorithm_ppo_zeta}. 

\begin{breakablealgorithm}
	\caption{The CU-RL algorithm for optimal hedging of options with \textit{any} initial state $\bm{s}_0$} \label{algorithm_ppo_zeta}
	\begin{algorithmic}[0]
		\State Initialize policy network, value network, and VaR network parameter $\theta_{0}$, $\xi_{0}$, and $\zeta_{0}$, respectively.
		\For{$k$ from $0$ to $\mathcal{K} - 1$}
		\State Initialize buffer $\mathcal{D}_k=\emptyset$ and $\mathcal{H}_k=\emptyset$.
		\For{$i$ from $0$ to $N-1$}
		\State Generate a trajectory $\mathcal{T}_{i}$ by starting from some $\bm{s}^{(i)}_0$ and following the policy $\pi_{\theta_{k}}$ for an option with maturity $T^{(i)}$. Normalize states with the running mean and standard deviation.
		\State Calculate the generalized advantage estimator and rewards-to-go at time $T^{(i)}-1$ of the trajectory $\mathcal{T}_{i}$:
		$$\hat{A}^{(i)}_{T^{(i)}-1} =R_{T^{(i)}}^{(i)}(\omega(\bm{s}^{(i)}_{0}; \zeta_{k})) - V(\bm{s}_{T^{(i)}-1}^{(i)}; \xi_{k}), \quad G^{(i)}_{T^{(i)}-1} =R_{T^{(i)}}^{(i)}(\omega(\bm{s}^{(i)}_{0}; \zeta_{k})).$$
		\For{$t$ from $T^{(i)}-2$ to $0$}
		\State Compute generalized advantage estimator:
		\State $\hat{A}^{(i)}_{t} =(\lambda^{gae} \gamma) \cdot \hat{A}^{(i)}_{t+1} + R^{(i)}_{t+1} + \gamma V(\bm{s}^{(i)}_{t+1}; \xi_{k}) - V(\bm{s}^{(i)}_{t}; \xi_{k})$ 
		\State Compute rewards-to-go: $G^{(i)}_{t} =\hat{A}^{(i)}_{t}+ V(\bm{s}^{(i)}_{t}; \xi_{k})$. 
		\EndFor
		\State Add the tuples $\left\{ (\bm{s}_{t}^{(i)}, a_{t}^{(i)}, \pi_{\theta_{k}}(a_{t}^{(i)}|\bm{s}_{t}^{(i)}),G^{(i)}_{t}, \hat{A}^{(i)}_{t}) \right\}_{t=0}^{T^{(i)}-1}$  to the buffer $\mathcal{D}_{k}$.	
		\State Add the tuple $\left(\bm{s}_0^{(i)}, W_{T^{(i)}}^{(i)}\right)$ to the buffer $\mathcal{H}_k$.
		\EndFor
		\State Set $\theta = \theta_{k},~\xi=\xi_{k},~\zeta=\zeta_{k}$.
		\For{$m$ from $0$ to $M-1$}
		\State Shuffle the buffer $\mathcal{D}_{k}$.
			\For{$i$ from $0$ to $\frac{|\mathcal{D}_{k}|}{n}-1$}
			\State Sequentially collect $n$ tuples from $\mathcal{D}_k$ as a minibatch: $\mathcal{B}=\{ (\bm{s}^{j}, a^{j}, \pi^{j}_{\theta_{k}}, G^{j}, \hat{A}^{j}) \}_{j=1}^{n}$.
			\State Calculate the $L^{2}$-norm $\mathcal{N}_\mathcal{B}$ defined in \eqref{equ:gradient_norm_theta_xi_2}.  
			\State Update the policy parameter: 
			$$\theta \leftarrow \theta + \eta \left[ \frac{1}{n} \sum_{j=1}^{n} \nabla_{\theta}  L^{P}(\bm{s}^{j}, a^{j}, \hat{A}^{j}, \theta_{k}; \theta)- c_{0}  \nabla_{\theta} L^{E}(\theta) \right] \cdot \min \left(\frac{\mathcal{N}^{*}}{\mathcal{N}_\mathcal{B}}, 1\right);$$ 
			\State Update the value parameter:
			$$\xi \leftarrow \xi - c_{1} \eta \left[ \frac{1}{n} \sum_{j=1}^{n} \nabla_{\xi} L^{V}(\bm{s}^{j}, G^{j}; \xi) \right]\cdot \min \left(\frac{\mathcal{N}^{*}}{\mathcal{N}_\mathcal{B}}, 1\right);$$ 
			\EndFor
			\State Use all the $N$ tuples $\{(\bm{s}_{0}^{(j)}, W_{T^{(j)}}^{(j)})\}_{j=1}^{N}$ in $\mathcal{H}_k$ to update the VaR network parameter $\zeta$: 
			$$\zeta \leftarrow \zeta - c_{2} \eta \left[ \frac{1}{N} \sum_{j=1}^{N} \nabla_{\zeta} L^{O}(\bm{s}^{(j)}_{0}, W^{(j)}_{T^{(j)}}; \zeta)  \right] \cdot \min \left(\frac{\mathcal{N}^{*}}{\mathcal{N}_{\mathcal{H}_k}}, 1\right),$$
			where $\mathcal{N}_{\mathcal{H}_k}$ is defined in \eqref{equ:gradient_norm_zeta} with $\mathcal{H}=\mathcal{H}_k$.
			\EndFor
			\State $\theta_{k+1} \leftarrow \theta;~\xi_{k+1} \leftarrow \xi;~\zeta_{k+1} \leftarrow \zeta.$ 
			\EndFor
		\end{algorithmic}
	\end{breakablealgorithm}

	\section{Hedging Performance under Parametric Option Pricing Models}\label{sec:hedging_performance_parametric_models}
	
	In this section, we demonstrate the hedging performance of the CS-RL approach and the CU-RL approach based on simulated data generated by two classical parametric option pricing models, the Black-Scholes model in Section \ref{subsec:numerical_results_simulated_BS} and the GARCH model in Section \ref{subsec:numerical_results_simulated_GARCH}.
		
	Under each model, we compare the hedging performance of the CS-RL approach for hedging a single option with that of the delta hedging method. 
	We then compare the hedging performance of the CU-RL approach for hedging many different options with that of the delta hedging method. 
	
    In all numerical results based on simulated data in this section and the empirical data in Section \ref{sec:model_free_empirical_results}, we set the confidence level $\alpha=0.975$ for CVaR in the objective function \eqref{equ:dealer_objective}, as the CVaR at 97.5\% level is used for setting capital charge for the trading book under the Basel III accord \citep{Basel-2019} and is hence concerned in practice. 
    The data is in daily frequency and hedging is performed daily. 
    
	The hedging performance is measured by multiple criteria, including the mean and standard error of the final P\&L, the VaR, the CVaR, and the 
	median shortfall (MS) of the final P\&L at level 95\% and 97.5\%, respectively, as well as the 95\% confidence interval (CI) of each of the tail risk measures. The 
	MS at 97.5\% of the final P\&L is the median of the conditional tail distribution beyond the 97.5\% VaR; it is 	
	equal to VaR at 98.75\% and provides a robust alternative to  CVaR at 97.5\% level \citep{kou2016}.	
	We also calculate the P-value of the one-sided t-test for related samples that tests if the mean of P\&L of our method is higher than that of the benchmark method. In all experiments, we consider two cases: one is without transaction cost, the other is with $0.1\%$ proportional transaction cost. 
	
	\subsection{Hedging Options Based on Data Simulated under Black-Scholes (BS) Model}\label{subsec:numerical_results_simulated_BS}
	Under the Black-Scholes model, the dynamics of the underlying stock price is specified in \eqref{gbm}
	\begin{equation}
		\label{gbm}
		dS_{t} = \tilde{\mu} S_{t}dt + \tilde{\sigma} S_{t}dB_{t}^{S},
	\end{equation}
	where $B_{t}^{S}$ denotes the standard Brownian motion. Suppose the time unit in the model \eqref{gbm} is one day and 
	the parameters $\tilde{\mu}$ and $\tilde{\sigma}$ and the risk free rate $r$ is given by Table~\ref{table_gbm-params}. 
	\begin{table}[!htbp]
		\centering
		\caption{\textbf{Value of parameters for the BS model.} }
		\begin{tabular}{cccc}
			\toprule
			Parameter &$\tilde{\mu}$ & $\tilde{\sigma}$ & $r$ \\
			Value & $\frac{0.1}{252}$ & $\frac{0.2}{\sqrt{252}}$ & $\frac{0.03}{365}$ \\
			\bottomrule
		\end{tabular}
		\label{table_gbm-params}
	\end{table}
	
	\subsubsection{Hedging a Single Option Using the CS-RL Method}
	We consider hedging a single European call option with spot price $S_0=100$, maturity $T = 30$ days, and strike $K=105$. 
	In all simulated sample paths, the initial state $\bm{s}_{0}=(B_{0}, \delta_{0}, S_{0},\sigma_{0}, \tau_{0})$ is the same, where  $\sigma_{0} = \tilde{\sigma}$, $\tau_{0} = T$, $\delta_{0} = 0.0$, and $B_{0}= Z_{\text{BS}}(S_{0}, \sigma_{0}, \tau_{0})$; here $Z_{\text{BS}}$ denotes the option price under the BS model. 

	In the implementation of CS-RL method, the coefficients in \eqref{equ:dealer_objective} are $\lambda_1 = 1.0$, $\lambda_2 = 0.0$, i.e. the dealer only cares about the risk. 
	The coefficients in \eqref{obj_newcvar1} are $c_{0}=0$, $c_{1}=0.04$, and $c_{2}=0.08$. 
The policy network and value network are specified as fully connected feedforward neural networks with 3 hidden layers of 32 neurons. The nonlinear activation function for each hidden layer is the Swish function. Each hidden layer is followed by a batch normalization layer. 
In Algorithm~\ref{algorithm_ppo_omega}, the size of buffer $\mathcal{D}_k$ is 29988, the minibatch size $n=2048$, the number of epochs $\mathcal{K}=1000$, learning rate $\eta=0.0005$, and $M=5$. 

\begin{figure}[!htbp]
		\centering
		\subfigure[Cumulative reward for each training epoch]{
			\begin{minipage}[c]{0.95\linewidth}
				\centering
				\includegraphics[width=0.48\linewidth]{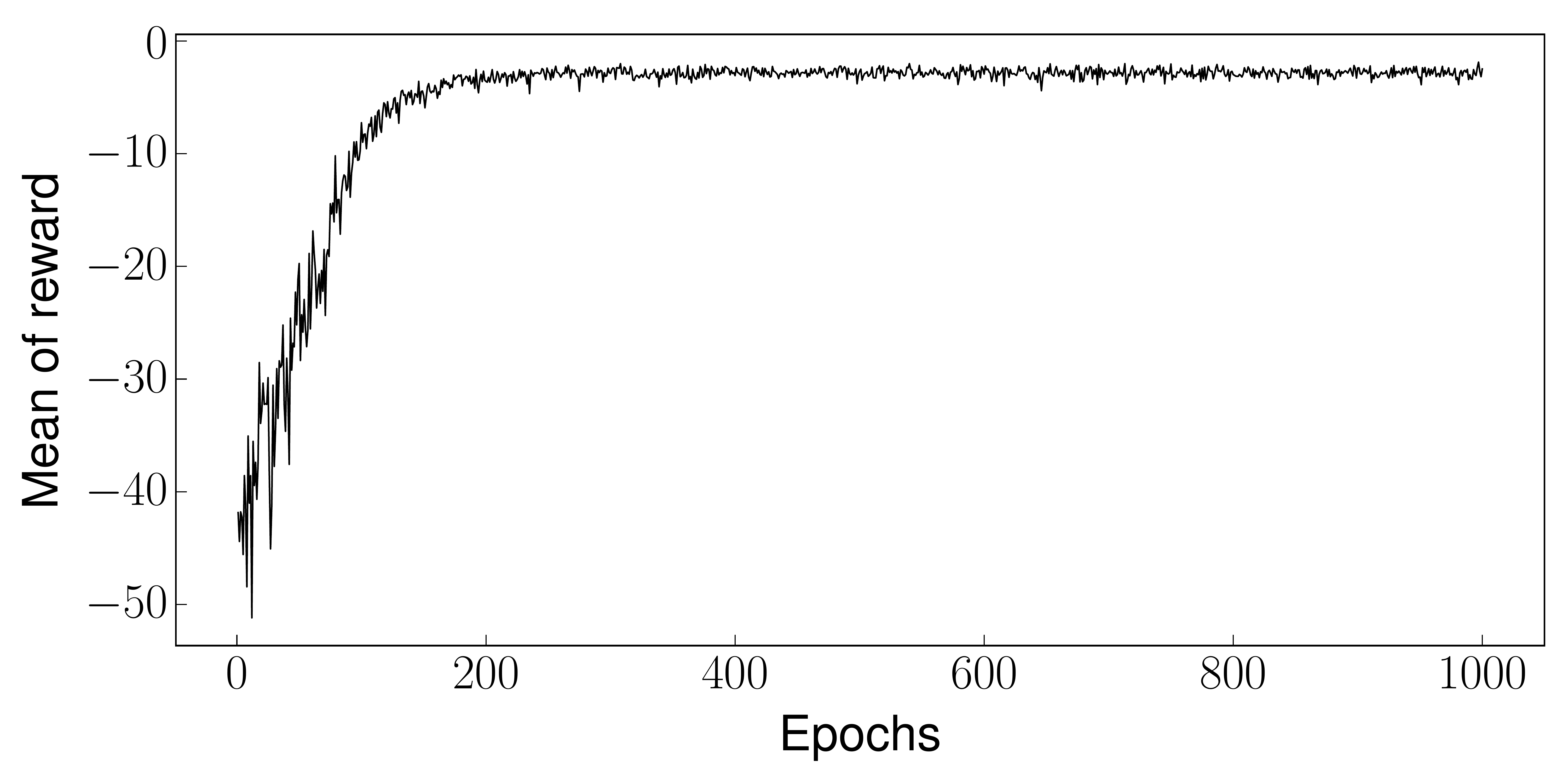}
			\end{minipage}
		}
		\subfigure[Loss of value network for each training epoch]{
			\begin{minipage}[c]{0.48\linewidth}
				\centering
				\includegraphics[width=0.95\linewidth]{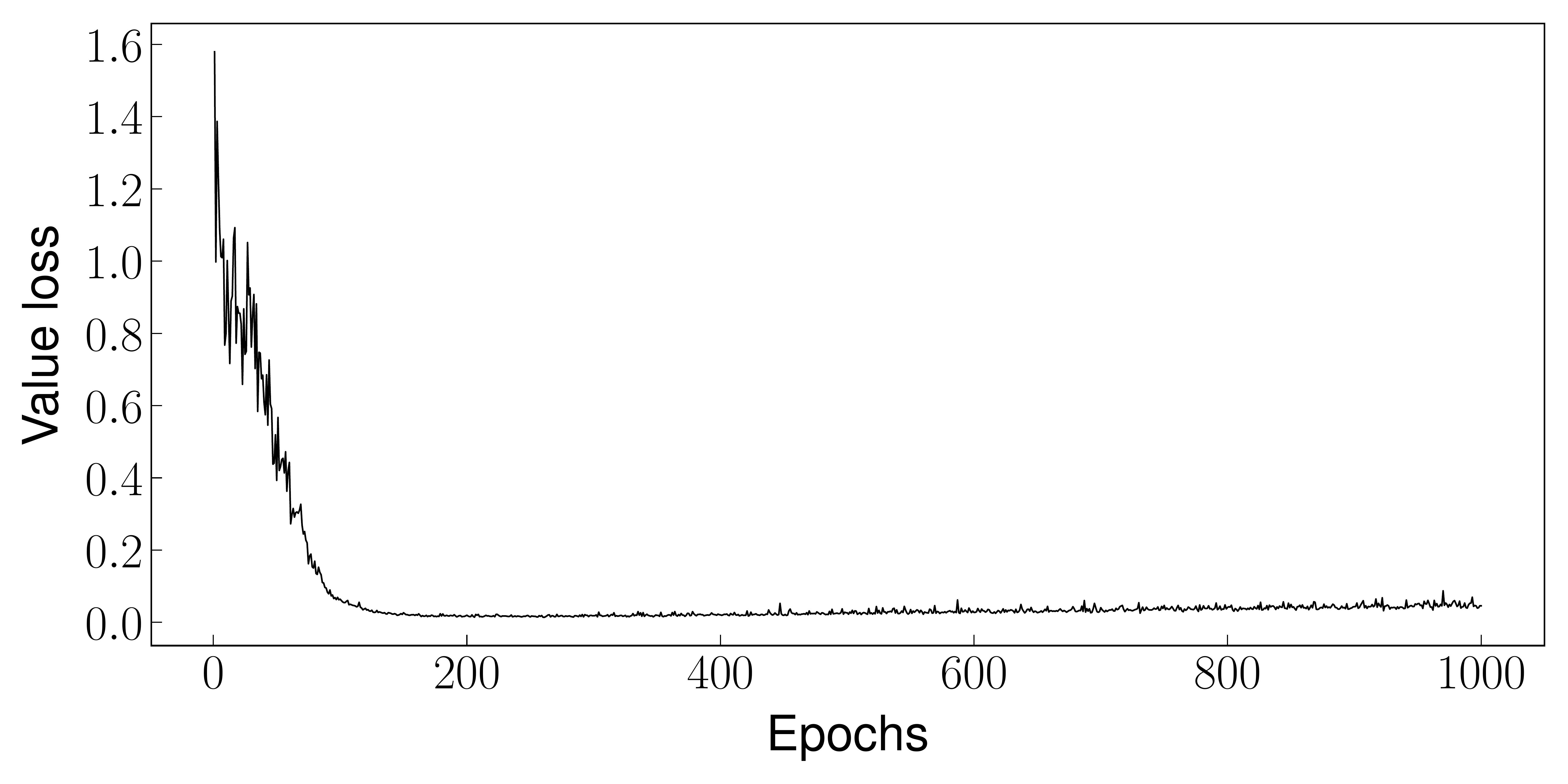}
			\end{minipage}
		}
		\subfigure[Loss of $\omega$ for each training epoch]{
			\begin{minipage}[c]{0.48\linewidth}
				\centering
				\includegraphics[width=0.95\linewidth]{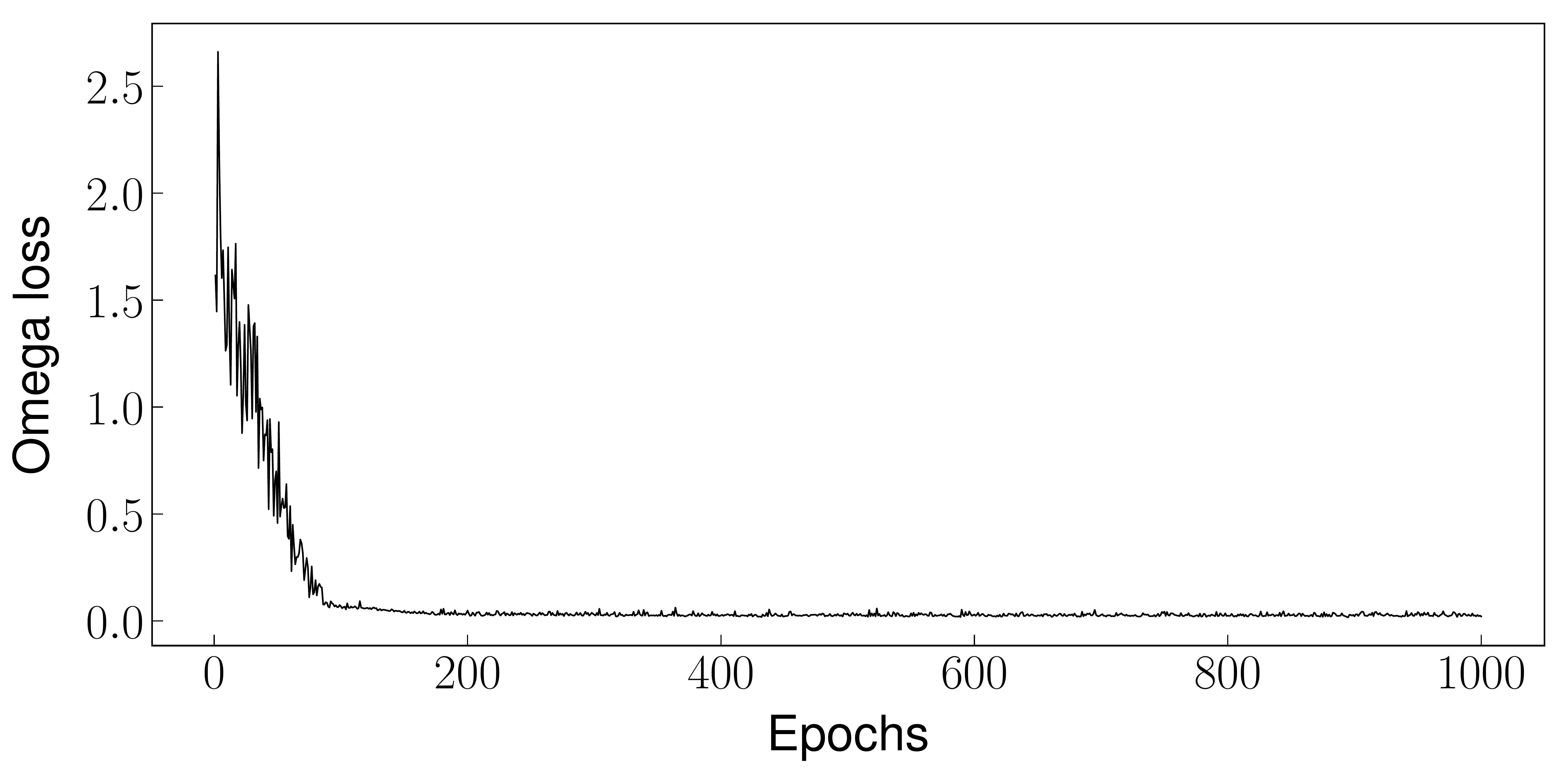}
			\end{minipage}
		}
		\caption{\textbf{Learning curves of the CS-RL method for hedging a single option under the Black-Scholes model.} The $x$-axis of all subfigures represent the training epoch. The $y$-axis represent the cumulative rewards, the loss of value network, and the loss of $\omega$, respectively for the subfigures (a), (b), and (c).}
		\label{fig:learning_curve_gbm}
	\end{figure}

Fig.~\ref{fig:learning_curve_gbm} shows the three learning curves of CS-RL method for the cumulative reward, the loss for value network, and the loss for variable $\omega$, respectively. The convergence of the CS-RL method seems to be stable.
	
We first train the model in the no-transaction cost case, and then we use the trained network parameters in the non-transaction cost case to initialize the parameters for the model in the case of $0.1\%$ transaction cost.

Table~\ref{table_gbm_single} shows the 		
		 performance of CS-RL method and the BS delta hedging method on a out-of-sample test dataset of 100000 trajectories in the no-transaction cost case and the 0.1\% transaction cost case.
	The panel A shows that: (i) CS-RL method obtains a significantly higher mean final P\&L than the benchmark method. Although the Black-Scholes delta hedging method is optimal in the continuously hedging case, CS-RL method obtains a higher mean P\&L due to discrete daily hedging. 	
	 (ii) CS-RL method obtains lower CVaR at 0.975 level, the designated risk measure in the total reward, lower MS at 0.975 level, and lower VaR at 0.975 level. (iii) CS-RL method obtains lower VaR and CVaR at 0.95 level, suggesting it can effectively minimize the tail risk measured at an alternative level. (iv) All the results with respect to tail risk measures are statistically significant, since all the confidence intervals of CS-RL method are lower and have no overlap with  those of BS delta hedging method. The panel B shows that similar conclusion holds when 0.1\% transaction costs are considered. Although the mean of final P\&L falls and the tail risk measures increase due to transaction costs, CS-RL method still obtains a significantly higher mean P\&L and lower tail risk than  the BS delta hedging method.

	\begin{table}[!htbp]
		\centering
		\caption{\textbf{The mean, standard error, and tail risk of the final P\&L of hedging a single option by the CS-RL method and the BS delta hedging method calculated from an out-of-sample test set.} The option has parameters $T = 30, K=105$. The test set is comprised of 100000 sample paths simulated under the Black-Scholes model. The p-values are calculated from a one-sided t-test for related samples that tests if the mean P\&L of our method is higher than that of the benchmark method. 0.95-CI means 95\% confidence interval. Scientific notation: $\text{1.23E-4} = 1.23\times 10^{-4}$. $\text{***}:p < 0.001$; $\text{**}:p < 0.01$; $*:p < 0.05$.}
	    \begin{tabular}{lll}
		\toprule
		& \multicolumn{1}{r}{CS-RL} & \multicolumn{1}{c}{Traditional Model} \\ 
		\cmidrule(lr){3-3}
		& & \multicolumn{1}{r}{BS delta} \\ 
		\midrule 
		\multicolumn{3}{c}{Panel A: without transaction cost} \\ 
		\midrule
		Mean & \multicolumn{1}{r}{1.4783E-02}& \multicolumn{1}{r}{5.9855E-05}\\
Std Err & \multicolumn{1}{r}{1.4534E-03}& \multicolumn{1}{r}{1.2621E-03}\\
P-Value & \multicolumn{1}{r}{}& \multicolumn{1}{r}{3.0907E-89$^{***}$}\\
0.95-VaR & \multicolumn{1}{r}{0.6481}& \multicolumn{1}{r}{0.6799}\\
0.95-CI of 0.95-VaR & \multicolumn{1}{r}{[0.6413, 0.6553]} & \multicolumn{1}{r}{[0.6727, 0.6898]} \\
0.975-VaR & \multicolumn{1}{r}{0.8307}& \multicolumn{1}{r}{0.8915}\\
0.95-CI of 0.975-VaR & \multicolumn{1}{r}{[0.8207, 0.8414]} & \multicolumn{1}{r}{[0.8784, 0.9038]} \\
0.975-MS & \multicolumn{1}{r}{1.0032}& \multicolumn{1}{r}{1.1053}\\
0.95-CI of 0.975-MS & \multicolumn{1}{r}{[0.9910, 1.0157]} & \multicolumn{1}{r}{[1.0871, 1.1227]} \\
0.95-CVaR & \multicolumn{1}{r}{0.8995}& \multicolumn{1}{r}{0.9862}\\
0.95-CI of 0.95-CVaR & \multicolumn{1}{r}{[0.8899, 0.9082]} & \multicolumn{1}{r}{[0.9739, 0.9982]} \\
0.975-CVaR & \multicolumn{1}{r}{1.0701}& \multicolumn{1}{r}{1.1977}\\
0.95-CI of 0.975-CVaR & \multicolumn{1}{r}{[1.0579, 1.0846]} & \multicolumn{1}{r}{[1.1806, 1.2141]} \\
		\midrule 
		\multicolumn{3}{c}{Panel B: with $0.1\%$ transaction cost} \\ 
		\midrule
		Mean & \multicolumn{1}{r}{-0.1646}& \multicolumn{1}{r}{-0.1725}\\
Std Err & \multicolumn{1}{r}{1.4701E-03}& \multicolumn{1}{r}{1.3216E-03}\\
P-Value & \multicolumn{1}{r}{}& \multicolumn{1}{r}{2.1201E-24$^{***}$}\\
0.95-VaR & \multicolumn{1}{r}{0.9017}& \multicolumn{1}{r}{0.9336}\\
0.95-CI of 0.95-VaR & \multicolumn{1}{r}{[0.8946, 0.9102]} & \multicolumn{1}{r}{[0.9246, 0.9433]} \\
0.975-VaR & \multicolumn{1}{r}{1.1010}& \multicolumn{1}{r}{1.1585}\\
0.95-CI of 0.975-VaR & \multicolumn{1}{r}{[1.0897, 1.1105]} & \multicolumn{1}{r}{[1.1466, 1.1704]} \\
0.975-MS & \multicolumn{1}{r}{1.2830}& \multicolumn{1}{r}{1.3951}\\
0.95-CI of 0.975-MS & \multicolumn{1}{r}{[1.2670, 1.2986]} & \multicolumn{1}{r}{[1.3777, 1.4133]} \\
0.95-CVaR & \multicolumn{1}{r}{1.1719}& \multicolumn{1}{r}{1.2610}\\
0.95-CI of 0.95-CVaR & \multicolumn{1}{r}{[1.1620, 1.1817]} & \multicolumn{1}{r}{[1.2490, 1.2739]} \\
0.975-CVaR & \multicolumn{1}{r}{1.3532}& \multicolumn{1}{r}{1.4873}\\
0.95-CI of 0.975-CVaR & \multicolumn{1}{r}{[1.3403, 1.3674]} & \multicolumn{1}{r}{[1.4697, 1.5054]} \\
		\bottomrule 
		\end{tabular}	
		\label{table_gbm_single} 
	\end{table}

	\subsubsection{Hedging Many Options Using the CU-RL Method}\label{subsubsec:bs_simulated_data_cu_rl}
	
	We consider hedging many options with different initial states by using the CU-RL method. We simulate the data set of stock index prices and option prices based on the BS model as follows. 
	
	 We first simulate a single sample path of stock index price $S_{t}$ from 01/02/2008 to 12/31/2017 under the BS model with parameters in  Table~\ref{table_gbm-params} and initial price $S_{0} = 1447.16$, which is the adjusted closing price of the S\&P 500 index on 01/02/2008.
	We then assume that call options on the stock index are listed at the exchange based on a set of rules similar to the strike listing rules of S\&P 500 (3rd Friday) non-quarterly index options, and that these call options are priced by the BS model using parameters in Table~\ref{table_gbm-params}. More precisely, 	
		on each day starting from 01/02/2008, call options with maturities of less than 12 months are listed by the following strike listing rules.
		\begin{itemize}
			\item
			Call options with strike prices being the at-the-money (ATM) price $\pm 10\%$ of exercise reference price (i.e., the closing price in the last trading day) in 5 index points intervals are listed. The option expiration date is roughly 1 week to 2 months from the current date and is the third Friday of a month.
			\item
			Call options with strike prices being the ATM price $\pm 20\%$ of exercise reference price in 10 index points intervals. The option expiration date is roughly 2 to 6 months from the current date and is the third Friday of a month.
			\item
			Call options with strike prices being the ATM price $\pm 50\%$ of exercise reference price in 25 index points intervals. The option expiration date is roughly 6 to 12 months from the current date and is the third Friday of the month.
		
			\item 			
			When the underlying price index is out of the current strike interval of each the above three rules, we list a new group of options following the respective rule. Meanwhile, we do not repeatedly list options with the same pair of maturity and strike.			
		\end{itemize}	
	
	We obtain 142987 options in total that expire before 01/02/2018.  We group these option paths into the training dataset and test dataset by the split date 12/31/2014. The options that mature before 12/31/2014 are classified as training data while those starting on or after 12/31/2014 are labeled as test data. The rest of options that cross the day are discarded. In this way, we ensure the ratio of size of training data to that of test data is roughly $5$ to $1$.
	The distribution of maturity and moneyness of training and test data are shown in Table~\ref{options_maturity_moneyness}.
	
	\begin{table}[!htbp]
		\centering
		\caption{\textbf{The number of paths of options in the training data and test data in different ranges of maturity and moneyness.} The maximum maturity of options in the training data and test data is 357 days. The moneyness of options in the training data is in the range of $[0.4681, 1.5689]$ and that of options in the test data is in the range of $[ 0.7963, 1.5573]$.}
		\begin{tabular}{rrrrrr}
			\toprule
			Maturity range &  Training set & Test set & Moneyness range  &  Training set & Test set \\
			\cmidrule(lr){2-3}\cmidrule(lr){5-6}
			$[7, 14]$ & 5698 & 1025 & $(0.1, 0.8]$ & 3254 & 3 \\
			$(14, 30]  $ & 11869 & 2309 & $(0.8, 0.9]$ & 3199 & 56 \\
			$(30, 90]$ & 28771 & 5584 & $(0.9, 1.1] $ & 41459 & 6472 \\
			$(90, 360]$ &  66809 & 12392 & $(1.1, 1.5]$ & 61595 & 13779 \\
			$(360, \infty]$ &  0 & 0 & $(1.5, 1.6]$ & 3640 & 1000 \\
			Total & 113147  & 21310 & Total & 113147 & 21310 \\
			\bottomrule
		\end{tabular}%
		\label{options_maturity_moneyness}%
	\end{table}%
	
	The state variable $\bm{s}_{t}$ in the CU-RL method is defined in \eqref{equ:cu_rl_state_def}, where 
	$\sigma_t=\tilde \sigma$ for all $t$, and $I_t =(\Delta_{t}, \Gamma_{t}, \mathcal{V}_{t}, \Theta_t)$, representing the delta, gamma, vega, and theta, respectively.	
	
	In the implementation of CU-RL method, the coefficients in \eqref{obj_newcvar_with_zeta} are $c_{0}=0$, $c_{1}=0.0004$, and $c_{2}=0.08$. 
	The policy, value, and VaR networks are specified as fully connected feedforward neural networks with 9 hidden layers of 32 neurons. The nonlinear activation function for each hidden layer is the Swish function. 
	Each hidden layer is followed by a batch normalization layer. 
	In Algorithm~\ref{algorithm_ppo_zeta}, the size of buffer $\mathcal{D}_k$ is 59990, the minibatch size $n=30000$, the number of epochs $\mathcal{K}=1000$, and $M=5$. 
	We apply a linearly decaying learning rate with the initial value 1e-7 and terminal value 1e-12.

	We use traditional supervised learning to pre-train an initializer based on the BS delta. The initializer is comprised of a policy network, a value network, and a VaR network, all of which have the same definition of input, output, and structure as the networks used in CU-RL. The policy network in the initializer is learned by minimizing the mean-squared loss of its output and the BS delta. For the value network, we simulate trajectories using option paths in the training set and BS delta as actions. Then we use Monte-Carlo method to estimate the value function, i.e. the estimator of the value function at the given state $s_t$ is equal to the sum of discounted realized reward:
	\begin{equation}
		\hat{V}^{init}(s_t)=\sum\limits_{k=t+1}^{T}\gamma^{k-t}R_k.
		\label{equ:initializer_value}
	\end{equation}
	The value network is trained to minimize the mean-square loss of $\hat{V}^{init}$ and its output. At the same time, we collect the initial states and final P\&L to train the VaR network to minimize \eqref{loss_zeta}. The initializer is only trained in non-transaction cost environment. 
	
	We train an initializer based on the Black-Scholes delta for 3000 epochs, and then train models for the non-transaction cost case and 0.1\% proportional transaction cost case respectively.
	
	Table~\ref{table_gbm_synreal} shows the 		
	performance of CU-RL method and the BS delta hedging method on a out-of-sample test dataset of 21310 options in the no-transaction cost case and the 0.1\% transaction cost case. 
	The panel A shows that: (i) CU-RL obtains a statistically significantly higher mean final P\&L than the benchmark method; (ii) CU-RL obtains lower CVaR at 0.975 level, the designated risk measure in the total reward of CU-RL, lower MS at 0.975 level, and lower VaR at 0.975 level; (iii) CU-RL obtains lower VaR and CVaR at 0.95 level, suggesting it can effectively minimize the tail risk measured at an alternative level; (iv) all the results with respect to tail risk measures are statistically significant, since all the confidence intervals of CU-RL are lower and have no overlap with those of BS model. The panel B shows that similar conclusion holds when transaction costs are considered, except that the confidence intervals of CU-RL for MS at 0.975 level overlaps with BS model's. 

	\begin{table}[!htbp]
		\centering
		\caption{\textbf{The mean, standard error, and tail risk of the final P\&L of hedging call options by the CU-RL method and the BS delta hedging method calculated from an out-of-sample test set.} The underlying prices of the  training data set and the test data set are simulated under the BS model and the option prices are calculated by the BS model. 
		The test set is comprised of 21310 call options with different initial states. 			
		The p-values are calculated from a one-sided t-test for related samples that tests if the mean P\&L of our method is higher than that of the benchmark method. 0.95-CI means 95\% confidence interval. Scientific notation: $\text{1.23E-4} = 1.23\times 10^{-4}$. $\text{***}:p < 0.001$; $\text{**}:p < 0.01$; $*:p < 0.05$.}
    \begin{tabular}{lll}
	\toprule
	& \multicolumn{1}{r}{CU-RL} & \multicolumn{1}{c}{Traditional Model} \\ 
	\cmidrule(lr){3-3}
	& & \multicolumn{1}{r}{BS delta} \\ 
	\midrule 
	\multicolumn{3}{c}{Panel A: without transaction cost} \\ 
	\midrule
	Mean & \multicolumn{1}{r}{19.7774}& \multicolumn{1}{r}{16.5946}\\
Std Err & \multicolumn{1}{r}{0.1276}& \multicolumn{1}{r}{0.1188}\\
P-Value & \multicolumn{1}{r}{}& \multicolumn{1}{r}{0.0000E+00$^{***}$}\\
0.95-VaR & \multicolumn{1}{r}{-1.9950}& \multicolumn{1}{r}{-0.7212}\\
0.95-CI of 0.95-VaR & \multicolumn{1}{r}{[-2.1807, -1.8808]} & \multicolumn{1}{r}{[-0.8254, -0.6255]} \\
0.975-VaR & \multicolumn{1}{r}{-0.5698}& \multicolumn{1}{r}{0.3742}\\
0.95-CI of 0.975-VaR & \multicolumn{1}{r}{[-0.6743, -0.4602]} & \multicolumn{1}{r}{[0.2738, 0.6007]} \\
0.975-MS & \multicolumn{1}{r}{0.4152}& \multicolumn{1}{r}{2.9188}\\
0.95-CI of 0.975-MS & \multicolumn{1}{r}{[0.1513, 0.9358]} & \multicolumn{1}{r}{[2.2079, 3.5900]} \\
0.95-CVaR & \multicolumn{1}{r}{0.3932}& \multicolumn{1}{r}{2.1178}\\
0.95-CI of 0.95-CVaR & \multicolumn{1}{r}{[0.1726, 0.6029]} & \multicolumn{1}{r}{[1.8551, 2.4543]} \\
0.975-CVaR & \multicolumn{1}{r}{2.1091}& \multicolumn{1}{r}{4.4515}\\
0.95-CI of 0.975-CVaR & \multicolumn{1}{r}{[1.7601, 2.4559]} & \multicolumn{1}{r}{[3.9563, 5.0340]}  \\
	\midrule 
	\multicolumn{3}{c}{Panel B: with $0.1\%$ transaction cost} \\ 
	\midrule
	Mean & \multicolumn{1}{r}{15.3985}& \multicolumn{1}{r}{12.2803}\\
Std Err & \multicolumn{1}{r}{0.1039}& \multicolumn{1}{r}{9.3143E-02}\\
P-Value & \multicolumn{1}{r}{}& \multicolumn{1}{r}{0.0000E+00$^{***}$}\\
0.95-VaR & \multicolumn{1}{r}{-0.9939}& \multicolumn{1}{r}{0.2501}\\
0.95-CI of 0.95-VaR & \multicolumn{1}{r}{[-1.1222, -0.8521]} & \multicolumn{1}{r}{[3.1943E-02, 0.4363]} \\
0.975-VaR & \multicolumn{1}{r}{0.3830}& \multicolumn{1}{r}{1.6582}\\
0.95-CI of 0.975-VaR & \multicolumn{1}{r}{[0.1113, 0.7369]} & \multicolumn{1}{r}{[1.4087, 1.8205]} \\
0.975-MS & \multicolumn{1}{r}{3.6044}& \multicolumn{1}{r}{4.1778}\\
0.95-CI of 0.975-MS & \multicolumn{1}{r}{[2.6998, 4.5668]} & \multicolumn{1}{r}{[3.5968, 5.2686]} \\
0.95-CVaR & \multicolumn{1}{r}{2.6001}& \multicolumn{1}{r}{3.9766}\\
0.95-CI of 0.95-CVaR & \multicolumn{1}{r}{[2.2324, 2.9707]} & \multicolumn{1}{r}{[3.5864, 4.4425]} \\
0.975-CVaR & \multicolumn{1}{r}{5.6087}& \multicolumn{1}{r}{7.0983}\\
0.95-CI of 0.975-CVaR & \multicolumn{1}{r}{[5.0622, 6.2609]} & \multicolumn{1}{r}{[6.4255, 7.9711]} \\
	\bottomrule 
	\end{tabular}
		\label{table_gbm_synreal} 
	\end{table}

		\subsection{Hedging Options Based on Data Simulated under GARCH Model}\label{subsec:numerical_results_simulated_GARCH}
	We consider a  GARCH(1,1) model  proposed in \cite{Heston2000}, in which the dynamics of the underlying stock price under the physical measure is specified as
		\begin{align}
			\log (\frac{S_{t}}{S_{t-1}}) & = r + \lambda \sigma_{t}^{2} + \sigma_{t} z_{t}, \notag\\
			\sigma_{t}^{2} &= \omega +\beta \sigma_{t-1}^{2} +  \alpha (z_{t - 1} - \gamma \sigma_{t-1})^{2},		\label{garch_model}
		\end{align}
	where $r$ is the risk-free interest rate, $\lambda$ is the risk premium parameter, and $z_{t} \mathop{\sim}\limits^{\text{i.i.d}} N(0,1)$. The parameters satisfy
	$\omega>0$, $\alpha > 0$, $\gamma>0$, $\beta > 0$, and the stationarity condition $\alpha \gamma^{2} + \beta < 1$.		
	 \cite{Heston2000} show that under the assumption that the value of a call option with one period to expiration obeys the BS option pricing formula, the dynamics of the GARCH process under the risk neutral measure is given by
	\begin{align}
		\log (\frac{S_{t}}{S_{t-1}}) & = r - \frac{1}{2} \sigma_{t}^{2} + \sigma_{t} z_{t}, \notag\\
		\sigma_{t}^{2} &= \omega +\beta \sigma_{t-1}^{2} +  \alpha (z_{t - 1} - \gamma^* \sigma_{t-1})^{2},\ \text{where}\ \gamma^*=\gamma+\frac{1}{2}+\lambda, 	\label{garch_model_heston2000}
	\end{align}
	which leads to the closed-form option pricing formula under the model. 
	\subsubsection{Hedging a single option using the CS-RL method}
	
	We consider hedging a single European call option with spot price $S_0=2585.64$, initial volatility $\sigma_1=0.6652\%$, maturity $T = 45$ days, and strike $K=S_0$. 
	We assume the stock price follows a GARCH(1, 1) model with parameters given in Table~\ref{table_garch_params}, which corresponds to the time unit of one day.\footnote{The parameters are obtained by fitting the GARCH(1, 1) model to the historical data of S\&P 500 index from 03/05/2013 to 11/16/2017 assuming $r=0.03/365$. $S_{0}$ is the adjusted closing price of S\&P 500 index on 11/16/2017, and the initial volatility $\sigma_{1}$ is computed from the data and model parameter.}

	\begin{table}[!htbp]
		\caption{\textbf{Value of parameters of GARCH(1, 1) model.}} \label{table_garch_params}
		\centering
		\begin{tabular}{cccccccc}
			\toprule
			Parameter & $\lambda$ & $\omega$ & $\alpha$ & $\beta$ & $\gamma$ & $\sigma_{1}$ & $r$ \\
			Value & 0.2981 & 3.4105e-07 & 9.6154e-06 & 0.8168 & 0.1497 & 0.6652\% & $0.03/365$\\
			\bottomrule
		\end{tabular}
	\end{table}
	
	In the implementation of CS-RL method for the GARCH model, the coefficients 
	$\lambda_{1} = 10.0, \lambda_{2} = 0.0$ in \eqref{equ:cvar_terminal_reward}. 
	The coefficients in \eqref{obj_newcvar1} are $c_{0}=0$, $c_{1}=0.004$, and $c_{2}=0.008$ for the non-transaction cost case and $c_{0}=0$, $c_{1}=0.004$, and $c_{2}=0.0004$ for the proportional transaction cost case.
	The policy network and value network are specified as fully connected feedforward neural networks with 3 hidden layers of 32 neurons; the nonlinear activation function for each hidden layer is the Swish function; each hidden layer is followed by a batch normalization layer.
	In Algorithm~\ref{algorithm_ppo_omega}, the size of buffer $\mathcal{D}_k$ is 29988, the minibatch size $n=6000$, the number of epochs $\mathcal{K}=1000$, and $M=5$. 
	We apply a linearly decaying learning rate with the initial value 0.001 and terminal value 1e-12.
		
	We compare the CS-RL with the BS delta hedging and GARCH delta hedging method. The BS delta at time $t$ is calculated using two definitions of volatility: one is the predicted volatility $\sigma_{t+1}$ of the GARCH model, and the other is the implied volatility derived from the option price under the GARCH model. The two BS delta hedging methods are denoted as ``BS delta (pred-vol)" and ``BS delta" in Table~\ref{table_garch}, respectively. The GARCH delta hedging position is the partial derivative of the option price under the GARCH model with respective to the underlying price. 

	We train the model for the non-transaction cost case and then use the trained parameters as an initializer to train a model the 0.1\% proportional transaction cost case. 

	Table~\ref{table_garch} shows the 		
	performance of CS-RL method, the two BS delta hedging methods, and the GARCH delta hedging method on a out-of-sample test dataset of 160000 sample paths in the no-transaction cost case and the 0.1\% transaction cost case.
	The panel A shows that: (i) CS-RL obtains the lowest CVaR and MS of the final P\&L at the chosen level 0.975, and its CVaR at 0.975 level is statistically significantly lower; (ii) BS delta method has insignificantly higher mean of the final P\&L than CS-RL, but it also has higher tail risk; (iii) GARCH delta method has significantly lower mean of final P\&L than that of CS-RL, 
	although the GARCH delta method has the lowest risk measured by 0.95-VaR, 0.975-VaR, and 0.95-CVaR; (iv) there is no benchmark method that has both higher profit and lower tail risk than CS-RL.
	The panel B shows the results when transaction costs are considered. In this case, CS-RL obtains the statistically significantly highest mean and lowest risk measure of the final P\&L, showing it can adapt to real market setting like transaction costs.
	
	\begin{table}[!htbp]
		\centering
		\caption{\textbf{The mean, standard error, and tail risk of the final P\&L of hedging a single option by the CS-RL method, the BS delta hedging method, and GARCH delta hedging method calculated from an out-of-sample test set.} 
		The test set is comprised of 160000 sample paths simulated under the GARCH model. The BS delta (pred-vol) means the BS delta calculate by using the predicted volatility in the GARCH model. The BS delta means BS delta calculated by using the implied volatility of the option price. The p-values are calculated from a one-sided t-test for related samples that tests if the mean P\&L of our method is higher than that of the benchmark method. 0.95-CI means 95\% confidence interval. Scientific notation: $\text{1.23E-4} = 1.23\times 10^{-4}$. $\text{***}:p < 0.001$; $\text{**}:p < 0.01$; $*:p < 0.05$.}
    \resizebox{\linewidth}{!}{
	\begin{tabular}{lllll}
		\toprule
		& \multicolumn{1}{r}{CS-RL} & \multicolumn{3}{c}{Traditional Model} \\ 
		\cmidrule(lr){3-5}
		& & \multicolumn{1}{r}{BS delta (pred-vol)} & \multicolumn{1}{r}{BS delta} & \multicolumn{1}{r}{GARCH delta} \\ 
		\midrule 
		\multicolumn{5}{c}{Panel A: without transaction cost} \\ 
		\midrule
		Mean & \multicolumn{1}{r}{-4.8162E-02}& \multicolumn{1}{r}{-3.5121E-03}& \multicolumn{1}{r}{-4.7628E-03}& \multicolumn{1}{r}{-0.1447}\\
Std Err & \multicolumn{1}{r}{3.2550E-02}& \multicolumn{1}{r}{2.9162E-02}& \multicolumn{1}{r}{2.8392E-02}& \multicolumn{1}{r}{2.6732E-02}\\
P-Value & \multicolumn{1}{r}{}& \multicolumn{1}{r}{0.9930$^{}$}& \multicolumn{1}{r}{0.9950$^{}$}& \multicolumn{1}{r}{1.0907E-02$^{*}$}\\
0.95-VaR & \multicolumn{1}{r}{20.3432}& \multicolumn{1}{r}{20.5685}& \multicolumn{1}{r}{19.7982}& \multicolumn{1}{r}{18.9447}\\
0.95-CI of 0.95-VaR & \multicolumn{1}{r}{[20.2019, 20.4794]} & \multicolumn{1}{r}{[20.4115, 20.7298]} & \multicolumn{1}{r}{[19.6100, 19.9565]} & \multicolumn{1}{r}{[18.7851, 19.1157]} \\
0.975-VaR & \multicolumn{1}{r}{24.9147}& \multicolumn{1}{r}{25.7011}& \multicolumn{1}{r}{25.2294}& \multicolumn{1}{r}{24.4295}\\
0.95-CI of 0.975-VaR & \multicolumn{1}{r}{[24.7285, 25.1431]} & \multicolumn{1}{r}{[25.4535, 25.9052]} & \multicolumn{1}{r}{[24.9864, 25.4727]} & \multicolumn{1}{r}{[24.1642, 24.6749]} \\
0.975-MS & \multicolumn{1}{r}{29.3582}& \multicolumn{1}{r}{30.7426}& \multicolumn{1}{r}{30.7524}& \multicolumn{1}{r}{29.9358}\\
0.95-CI of 0.975-MS & \multicolumn{1}{r}{[29.1231, 29.6659]} & \multicolumn{1}{r}{[30.3967, 31.0018]} & \multicolumn{1}{r}{[30.4376, 31.0781]} & \multicolumn{1}{r}{[29.5927, 30.3444]} \\
0.95-CVaR & \multicolumn{1}{r}{26.9283}& \multicolumn{1}{r}{27.8137}& \multicolumn{1}{r}{27.7320}& \multicolumn{1}{r}{26.8347}\\
0.95-CI of 0.95-CVaR & \multicolumn{1}{r}{[26.7402, 27.1381]} & \multicolumn{1}{r}{[27.5988, 28.0388]} & \multicolumn{1}{r}{[27.4949, 27.9838]} & \multicolumn{1}{r}{[26.5976, 27.0748]} \\
0.975-CVaR & \multicolumn{1}{r}{31.4645}& \multicolumn{1}{r}{32.7531}& \multicolumn{1}{r}{33.2446}& \multicolumn{1}{r}{32.3008}\\
0.95-CI of 0.975-CVaR & \multicolumn{1}{r}{[31.1943, 31.7834]} & \multicolumn{1}{r}{[32.4628, 33.0603]} & \multicolumn{1}{r}{[32.9099, 33.5935]} & \multicolumn{1}{r}{[31.9550, 32.6367]}  \\
		\midrule 
		\multicolumn{5}{c}{Panel B: with $0.1\%$ transaction cost} \\ 
		\midrule
		Mean & \multicolumn{1}{r}{-4.2708}& \multicolumn{1}{r}{-6.9340}& \multicolumn{1}{r}{-6.8413}& \multicolumn{1}{r}{-7.1186}\\
Std Err & \multicolumn{1}{r}{3.6128E-02}& \multicolumn{1}{r}{3.0048E-02}& \multicolumn{1}{r}{2.9897E-02}& \multicolumn{1}{r}{2.8512E-02}\\
P-Value & \multicolumn{1}{r}{}& \multicolumn{1}{r}{0.0000E+00$^{***}$}& \multicolumn{1}{r}{0.0000E+00$^{***}$}& \multicolumn{1}{r}{0.0000E+00$^{***}$}\\
0.95-VaR & \multicolumn{1}{r}{26.7733}& \multicolumn{1}{r}{28.4811}& \multicolumn{1}{r}{28.4687}& \multicolumn{1}{r}{27.9080}\\
0.95-CI of 0.95-VaR & \multicolumn{1}{r}{[26.6210, 26.9398]} & \multicolumn{1}{r}{[28.3106, 28.6669]} & \multicolumn{1}{r}{[28.2755, 28.6584]} & \multicolumn{1}{r}{[27.7263, 28.0909]} \\
0.975-VaR & \multicolumn{1}{r}{31.8456}& \multicolumn{1}{r}{33.9079}& \multicolumn{1}{r}{34.3516}& \multicolumn{1}{r}{34.0782}\\
0.95-CI of 0.975-VaR & \multicolumn{1}{r}{[31.6270, 32.0580]} & \multicolumn{1}{r}{[33.7069, 34.1424]} & \multicolumn{1}{r}{[34.1239, 34.6104]} & \multicolumn{1}{r}{[33.7949, 34.3876]} \\
0.975-MS & \multicolumn{1}{r}{36.9980}& \multicolumn{1}{r}{39.2778}& \multicolumn{1}{r}{40.4047}& \multicolumn{1}{r}{40.1944}\\
0.95-CI of 0.975-MS & \multicolumn{1}{r}{[36.6607, 37.3052]} & \multicolumn{1}{r}{[38.9173, 39.6481]} & \multicolumn{1}{r}{[40.0290, 40.8340]} & \multicolumn{1}{r}{[39.8498, 40.5740]} \\
0.95-CVaR & \multicolumn{1}{r}{34.2219}& \multicolumn{1}{r}{36.2107}& \multicolumn{1}{r}{37.1237}& \multicolumn{1}{r}{36.6094}\\
0.95-CI of 0.95-CVaR & \multicolumn{1}{r}{[33.9966, 34.4402]} & \multicolumn{1}{r}{[35.9716, 36.4342]} & \multicolumn{1}{r}{[36.8522, 37.3932]} & \multicolumn{1}{r}{[36.3524, 36.8867]} \\
0.975-CVaR & \multicolumn{1}{r}{39.4503}& \multicolumn{1}{r}{41.4951}& \multicolumn{1}{r}{43.1377}& \multicolumn{1}{r}{42.5885}\\
0.95-CI of 0.975-CVaR & \multicolumn{1}{r}{[39.1289, 39.7869]} & \multicolumn{1}{r}{[41.1799, 41.8056]} & \multicolumn{1}{r}{[42.7757, 43.5175]} & \multicolumn{1}{r}{[42.2150, 42.9299]} \\
		\bottomrule 
		\end{tabular}} 
		\label{table_garch} 
	\end{table}
	
	\subsubsection{Hedging Many Options Using the CU-RL Method}
	
	We consider hedging many options with different initial states by using the CU-RL method. Following the procedure in 
	Section~\ref{subsubsec:bs_simulated_data_cu_rl}, we first simulate a single sample path of stock index price $S_{t}$ from 01/02/2008 to 12/31/2017 under the GARCH model with parameters\footnote{The parameters are 
	obtained from 	
	fitting the GARCH model to the adjusted closing price of S\&P 500 index from 01/02/2008 to 12/31/2017, assuming $r=0.03/365$.} 
	in Table~\ref{table_garch_params_multi} and initial price $S_{0} = 1447.16$, which is the adjusted closing price of the S\&P 500 index on 01/02/2008. We then assume that call options on the stock index are listed at the exchange based on the same set of rules specified in Section~\ref{subsubsec:bs_simulated_data_cu_rl} and the option prices are calculated from the GARCH model.  
	
	\begin{table}[!htbp]
		\caption{\textbf{Value of parameters of GARCH(1, 1) model.}} \label{table_garch_params_multi}
		\centering
		\begin{tabular}{cccccccc}
			\toprule
			Parameter & $\lambda$ & $\omega$ & $\alpha$ & $\beta$ & $\gamma$ & $\sigma_1$ & $r$\\
			Value & 0.2886 & 1.6615e-09 & 1.8284e-05 & 0.8503 & 0.2993 & 0.01931 & $0.03/365$ \\
			\bottomrule
		\end{tabular}
	\end{table}
		
	We obtain 117427 options in total that expire before 12/31/2017. We group these option paths into the training dataset and test dataset by the split date of 12/31/2015. Options that mature before the day are classified as training data while those starting after the day are labeled as test data. The rest of options that cross the day are dropped. The distribution of maturity and moneyness of options in the training and test data are shown in Table~\ref{options_maturity_moneyness_garch}.
	
	\begin{table}[!htbp]
		\centering
		\caption{\textbf{The number of training paths and test paths in different ranges of maturity and moneyness.} The maximum maturity in both the training set and the test set is 357 days. The moneyness of options in the training set is in the range of $[0.4723, 1.5629]$ and that of options in the test set is in the range of $[0.9241, 1.5351]$.}
		\begin{tabular}{rrrrrr}
			\toprule
			Maturity range &  Training set & Test set & Moneyness range &  Training set & Test set \\ 
			\cmidrule(lr){2-3}\cmidrule(lr){5-6}
			$[7, 14]$ & 5211 & 865 & $(0.1, 0.9]$ & 6398 & 0 \\
			$(14, 30]  $ & 10519 & 1552 & $(0.9, 1.1] $ & 36059 & 5899 \\
			$(30, 90]$ & 27215 & 3971 & $(1.1, 1.5]$ & 55285 & 9426 \\
			$(90, 360]$ &  58851 & 9243 & $(1.5, 1.6]$ & 4054 & 306 \\
			Total & 101796  & 15631 & Total & 101796  & 15631 \\
			\bottomrule
		\end{tabular}%
		\label{options_maturity_moneyness_garch}%
	\end{table}%
	
	The state variable $\bm{s}_{t}$ in the CU-RL method is defined in \eqref{equ:cu_rl_state_def}, where 
		$\sigma_t$ is defined as the implied volatility calculated from the GARCH option price at time $t$, and $I_t =(\Delta_{t}, \Gamma_{t}, \mathcal{V}_{t}, \Theta_t)$ are the option's Greeks.

	We train two models for the non-transaction cost case and 0.1\% proportional transaction cost case, respectively. We still use the parameters in the pre-trained initializer based on the BS delta to initialize network parameters. The number of initializer training epochs is 2000, and the initial value of the linearly decaying learning rate is set to be 1e-6. Other hyperparameters and network structure are the same as those in Section \ref{subsubsec:bs_simulated_data_cu_rl}.

	\begin{table}[!htbp]
		\centering
		\caption{\textbf{The mean, standard error, and tail risk of the final P\&L of hedging  options by the CU-RL method, two BS delta hedging methods, and the GARCH delta hedging method calculated from an out-of-sample test set.} 
		The underlying prices of the  training data set and the test data set are simulated under the GARCH model and the option prices are calculated by the GARCH model. 
		The test set is comprised of 15631 call options with different initial states.				
			The BS delta (pred-vol) means the BS delta calculate by using the predicted volatility in the GARCH model. The BS delta means BS delta calculated by using the implied volatility of the option price.			
			The p-values are calculated from a one-sided t-test for related samples that tests if the mean P\&L of our method is higher than that of the benchmark method. 0.95-CI means 95\% confidence interval. Scientific notation: $\text{1.23E-4} = 1.23\times 10^{-4}$. $\text{***}:p < 0.001$; $\text{**}:p < 0.01$; $*:p < 0.05$.}
    \resizebox{\linewidth}{!}{
	\begin{tabular}{lllll}
		\toprule
		& \multicolumn{1}{r}{CU-RL} & \multicolumn{3}{c}{Traditional Model} \\ 
		\cmidrule(lr){3-5}
		& & \multicolumn{1}{r}{BS delta (pred-vol)} & \multicolumn{1}{r}{BS delta} & \multicolumn{1}{r}{GARCH delta} \\ 
		\midrule 
		\multicolumn{5}{c}{Panel A: without transaction cost} \\ 
		\midrule
		Mean & \multicolumn{1}{r}{11.4643}& \multicolumn{1}{r}{6.1651}& \multicolumn{1}{r}{6.1783}& \multicolumn{1}{r}{6.0164}\\
Std Err & \multicolumn{1}{r}{0.1496}& \multicolumn{1}{r}{0.1348}& \multicolumn{1}{r}{0.1335}& \multicolumn{1}{r}{0.1868}\\
P-Value & \multicolumn{1}{r}{}& \multicolumn{1}{r}{0.0000E+00$^{***}$}& \multicolumn{1}{r}{0.0000E+00$^{***}$}& \multicolumn{1}{r}{5.5937E-239$^{***}$}\\
0.95-VaR & \multicolumn{1}{r}{5.4522}& \multicolumn{1}{r}{10.2452}& \multicolumn{1}{r}{9.7802}& \multicolumn{1}{r}{18.8301}\\
0.95-CI of 0.95-VaR & \multicolumn{1}{r}{[5.1055, 5.7649]} & \multicolumn{1}{r}{[9.7737, 10.7511]} & \multicolumn{1}{r}{[9.3638, 10.2951]} & \multicolumn{1}{r}{[16.4881, 20.7133]} \\
0.975-VaR & \multicolumn{1}{r}{8.6245}& \multicolumn{1}{r}{17.9601}& \multicolumn{1}{r}{17.5868}& \multicolumn{1}{r}{41.2547}\\
0.95-CI of 0.975-VaR & \multicolumn{1}{r}{[8.2618, 9.0675]} & \multicolumn{1}{r}{[16.6295, 19.5196]} & \multicolumn{1}{r}{[15.8743, 19.3740]} & \multicolumn{1}{r}{[39.7569, 42.7711]} \\
0.975-MS & \multicolumn{1}{r}{16.9750}& \multicolumn{1}{r}{32.9698}& \multicolumn{1}{r}{32.9163}& \multicolumn{1}{r}{54.0249}\\
0.95-CI of 0.975-MS & \multicolumn{1}{r}{[13.2471, 20.3019]} & \multicolumn{1}{r}{[30.1767, 36.6814]} & \multicolumn{1}{r}{[30.1453, 36.4798]} & \multicolumn{1}{r}{[52.2357, 55.3731]} \\
0.95-CVaR & \multicolumn{1}{r}{13.8070}& \multicolumn{1}{r}{24.2084}& \multicolumn{1}{r}{23.8017}& \multicolumn{1}{r}{49.5249}\\
0.95-CI of 0.95-CVaR & \multicolumn{1}{r}{[12.7461, 15.1009]} & \multicolumn{1}{r}{[22.8256, 25.6482]} & \multicolumn{1}{r}{[22.3719, 25.2464]} & \multicolumn{1}{r}{[46.5163, 52.9706]} \\
0.975-CVaR & \multicolumn{1}{r}{20.7667}& \multicolumn{1}{r}{35.1175}& \multicolumn{1}{r}{35.0618}& \multicolumn{1}{r}{69.0765}\\
0.95-CI of 0.975-CVaR & \multicolumn{1}{r}{[18.9307, 23.2112]} & \multicolumn{1}{r}{[33.0889, 37.4025]} & \multicolumn{1}{r}{[32.9056, 37.3650]} & \multicolumn{1}{r}{[65.1642, 73.8666]}  \\
		\midrule 
		\multicolumn{5}{c}{Panel B: with $0.1\%$ transaction cost} \\ 
		\midrule
		Mean & \multicolumn{1}{r}{6.2717}& \multicolumn{1}{r}{1.9656}& \multicolumn{1}{r}{2.0434}& \multicolumn{1}{r}{0.8613}\\
Std Err & \multicolumn{1}{r}{0.1317}& \multicolumn{1}{r}{0.1254}& \multicolumn{1}{r}{0.1243}& \multicolumn{1}{r}{0.1849}\\
P-Value & \multicolumn{1}{r}{}& \multicolumn{1}{r}{0.0000E+00$^{***}$}& \multicolumn{1}{r}{0.0000E+00$^{***}$}& \multicolumn{1}{r}{6.0211E-247$^{***}$}\\
0.95-VaR & \multicolumn{1}{r}{8.7052}& \multicolumn{1}{r}{16.6240}& \multicolumn{1}{r}{16.0929}& \multicolumn{1}{r}{27.2466}\\
0.95-CI of 0.95-VaR & \multicolumn{1}{r}{[8.2856, 9.1425]} & \multicolumn{1}{r}{[15.9747, 17.2615]} & \multicolumn{1}{r}{[15.5966, 16.7680]} & \multicolumn{1}{r}{[24.6823, 30.2913]} \\
0.975-VaR & \multicolumn{1}{r}{14.8657}& \multicolumn{1}{r}{24.6764}& \multicolumn{1}{r}{24.5287}& \multicolumn{1}{r}{51.7348}\\
0.95-CI of 0.975-VaR & \multicolumn{1}{r}{[13.6176, 17.2352]} & \multicolumn{1}{r}{[23.3038, 26.3341]} & \multicolumn{1}{r}{[23.2227, 26.3301]} & \multicolumn{1}{r}{[50.9104, 54.9577]} \\
0.975-MS & \multicolumn{1}{r}{29.5319}& \multicolumn{1}{r}{40.8713}& \multicolumn{1}{r}{40.9004}& \multicolumn{1}{r}{68.8901}\\
0.95-CI of 0.975-MS & \multicolumn{1}{r}{[26.7955, 31.4739]} & \multicolumn{1}{r}{[37.3693, 44.6502]} & \multicolumn{1}{r}{[37.3048, 44.7364]} & \multicolumn{1}{r}{[67.2344, 70.0849]} \\
0.95-CVaR & \multicolumn{1}{r}{23.1402}& \multicolumn{1}{r}{32.1425}& \multicolumn{1}{r}{31.9942}& \multicolumn{1}{r}{61.0529}\\
0.95-CI of 0.95-CVaR & \multicolumn{1}{r}{[21.7837, 24.7895]} & \multicolumn{1}{r}{[30.4253, 33.5130]} & \multicolumn{1}{r}{[30.4297, 33.6202]} & \multicolumn{1}{r}{[58.0521, 64.7725]} \\
0.975-CVaR & \multicolumn{1}{r}{35.1989}& \multicolumn{1}{r}{44.7334}& \multicolumn{1}{r}{44.7098}& \multicolumn{1}{r}{82.5378}\\
0.95-CI of 0.975-CVaR & \multicolumn{1}{r}{[32.8350, 38.1706]} & \multicolumn{1}{r}{[42.2557, 47.4895]} & \multicolumn{1}{r}{[42.1666, 47.4908]} & \multicolumn{1}{r}{[78.7115, 87.7272]}  \\
		\bottomrule 
		\end{tabular}
		} 
		\label{table_garch_synreal} 
	\end{table} 

	Table~\ref{table_garch_synreal} shows the 		
	performance of CU-RL method, the BS delta hedging method, and the GARCH delta hedging method on a out-of-sample test dataset of 15631 call options.	
	The panel A shows that: (i) CU-RL obtains significantly higher mean final P\&L than all the benchmark methods; (ii) CU-RL obtains the lowest CVaR at 0.975 level, the designated risk measure in the total reward of CU-RL, the lowest MS at 0.975 level, and lowest VaR at 0.975 level; (iii) CU-RL obtains the lowest VaR and CVaR at 0.95 level, suggesting it can effectively minimize the tail risk measured at an alternative level; (iv) all the results with respect to tail risk measures are statistically significant, since all the confidence intervals of CU-RL are lower and have no overlap with those of other benchmark models. The panel B shows that similar conclusion holds when transaction costs are considered. 
	
\section{Hedging S\&P 500 Index Options in the Model-Free Setting} \label{sec:model_free_empirical_results}

The CU-RL method has the 
		unique advantage that it can be used in a model-free setting, i.e., it can use only historical data of the stock and option to learn the hedging strategy without specifying a parametric model for the underlying stock.
	In this section, we will train the CU-RL model using only the historical data of the S\&P 500 index and index options to demonstrate its empirical performance. 
	
Section~\ref{subsec:data_processing} describes the data of S\&P 500 index options and other market information variables, and explains the setting of empirical study. In Section~\ref{subsec:cu_rl_for_all_call_options} (resp., Section~\ref{subsec:cu_rl_for_all_put_options}), we use the data of all S\&P 500 index call (resp., put) options to train a unified model for hedging any call (resp., put) options. 
In Section~\ref{subsec:cu_rl_for_short_term_call_options} (resp., Section~\ref{subsec:cu_rl_for_short_term_put_options}), we use the data of short-term S\&P 500 index call (resp., put) options to train a unified model for hedging short-term call (resp., put) options. 
In Section~\ref{subsec:cu_rl_for_deep_itm_call_options} (resp., Section~\ref{subsec:cu_rl_for_short_term_deep_otm_put_options}), we use the data of deep-in-the-money (resp. deep-out-of-the-money) short-term S\&P 500 index call (resp., put) options to train a unified model for hedging deep-in-the-money (resp., deep-out-of-the-money) short-term call (resp., put) options. 
Section~\ref{subsec:cu_rl_zero_reward} presents the empirical hedging performance of the CU-RL approach when the reward at each time period $t=0, 1, \ldots, T-2$ is defined to be zero. Section~\ref{subsec:cu_rl_divided_by_margin_reward} presents the empirical hedging performance of the CU-RL approach when the reward at each time period is divided by the initial margin of the option. 
	
	\subsection{Data Description and Setting of Empirical Study}\label{subsec:data_processing}
	We use a data set of the historical daily market data of SPX options from 01/01/2008 to 12/31/2017 that is obtained from OptionsMetrics. We only retain options with expiration dates prior to 01/02/2018. 
	The data set include the SPX-Traditional (SPX), the SPX Weeklys (SPXW) and the SPX End Of Month (SPX EOM).
	
	To utilize the data more efficiently, we pre-process the data in the following way. For the historical data path of each option contract, we first specify any day before the option matures as 'today', then we can obtain an 'independent' data path from each 'today' to the expiration date. In this way, many paths of different lengths can be isolated from a single option, which greatly improves data utilization.  We divide all paths into three sets, the training set, validation set and test set. The training set comprises of all the paths that expire before 07/01/2017; the validation set comprises of all the paths that start after 07/01/2017 and expire before 10/01/2017; and the test set comprises of all the paths that start after 10/01/2017 and expires before 01/02/2018. We train CU-RL on the training set with different hyperparameters, including the learning rate $\eta$, minibatch size $n$, and replay buffer size $|\mathcal{D}_k|$, and then select the best model based on the performance on validation set. The results we present are based on the test set.
	
	The market data contain the best bid and ask price of options, the strike price, the implied volatility, the Greeks, and so on. We use the middle price of the best bid and the best ask price as the option price $Z_{t}, t= 0, \cdots, T-1$, which is utilized to compute the hedging error in \eqref{wealth}. The initial capital $B_{0}$ is specified to be $Z_0$. 
	The underlying price $S_{t}$ is the adjusted closing price of the S\&P 500 index. 
	We use the overnight London Interbank Offered Rate (LIBOR) $r_{t}$ as the risk-free interest rate. 
	
	The state variable $\bm{s}_{t}$ in the CU-RL method is defined in \eqref{equ:cu_rl_state_def}, where 
	$\sigma_t$ is the implied volatility on day $t$ provided in the market data, and $I_t =(r_t, \Delta_{t}, \Gamma_{t}, \mathcal{V}_{t}, \Theta_t)$, which include the LIBOR rate and the option's Greeks in the market data. 
	
	We compare the strategy learned by CU-RL with three benchmarks, the BS delta hedging, the local volatility function (LVF) and the SABR delta hedging. 
For the LVF, the strategy $\delta_{\text{lvf}} = \delta_{\text{BS}} + \nu \frac{\partial \sigma}{\partial K}$, where $\sigma$ is the implied volatility, and $\nu$ is the Vega ratio. Both $\sigma$ and $\nu$ are provided by market data. Therefore, we only need to estimate $\frac{\partial \sigma}{\partial K}$ for computing $\delta_{\text{lvf}}$. Assume that the implied volatility depends on the strike price quadratically, i.e. 
$\sigma = a_{1} K^{2} + a_{2} K + a_{3}$. We classify the option data by the date and expiration date. For each group of data, we fit a pair of $(a_1, a_2)$, and then get $\frac{\partial \sigma}{\partial K} = 2 a_{1} K + a_{2}$ as the estimated value of $\delta_{\text{lvf}}$ for the states in that group. If the sign of it is the same as the Black-Scholes' delta, we accept the calibrated value $\delta_{\text{lvf}}$ as the holding position. Otherwise, we maintain the delta value provided by the Black-Scholes model. Especially, if the group contains $n<4$ data points, we just adopt the value of $\delta_{\text{BS}}$ as the holding position rather than estimating the calibrated value $\delta_{\text{lvf}}$. For the SABR model, the strategy $\delta_{\text{SABR}}$ of test options can be computed as same as introduced in the pre-training part.

The SABR model is defined as
\begin{equation}
	\label{sabr}
	\begin{aligned}
		dF_{t} &= \alpha_{t} F_{t}^{\beta} dW_{t}^{(1)}, \\
		d\alpha_{t} &= \nu \alpha_{t} dW_{t}^{(2)}, 
	\end{aligned}
\end{equation}
where $W_{t}^{(1)}, W_{t}^{(2)}$ are standard Brownian motions and $dW_{t}^{(1)}dW_{t}^{(2)} = \rho dt$, $F_{t} := S_{t} \exp[(r-q)(T-t)]$ is the forward price. We first calibrate the SABR model with the option data. Test set data is never involved in the pre-training process. We classify option data in the training set by the date and expiration date. Data with the same date and expiration date is collected in a group. Therefore, for each group, the time-to-maturity $\tau$, which determined by the difference between the date and expiration date, is unique. Also, the date $t$ has the unique value in each group. 
Take one group as an example. Assume that there are $n$ data points in that group. Each data point includes the option price, the underlying price, time-to-maturity, implied volatility. For the SABR model shown in \eqref{sabr}, we can obtain the estimated value of $\alpha_{t}, \rho, \nu$ assuming $\beta=1$. We then compute the SABR delta value.  

	For hedging real options, the relative importance coefficients of final P\&L in \eqref{equ:dealer_objective} are fixed to be $\lambda_1=1.0$, $\lambda_{2}=0.0$. In the implementation of CU-RL method for hedging options in the real market, the coefficients in \eqref{obj_newcvar_with_zeta} are always $c_{0}=0$, $c_{1}=0.04$, and $c_{2}=0.08$. In Algorithm~\ref{algorithm_ppo_zeta}, we always set the number of epochs $\mathcal{K}=1000$, and $M=5$. We always apply a linearly decaying learning rate with terminal value 1e-12. 
	The policy, value, and VaR networks are specified as fully connected feedforward neural networks. The nonlinear activation function for each hidden layer is the Swish function. Each hidden layer is followed by a batch normalization layer. The policy network is appended by a Sigmoid layer before it outputs to scale the mean of action to $[0, 1]$.
	For put options, the output is additionally multiplied by -1 to scale the mean of action to $[-1, 0]$. The network structure is kept the same in all the empirical studies, and we only change the number of hidden layers and the number of neurons in each layer.
	Besides, we always pre-train an initializer using the corresponding training set. The training process of initializers is similar to the one based on Black-Scholes delta, except that we may replace the BS delta with SABR delta.

	\subsection{Performance of Hedging All Call Options by a Unified Model}\label{subsec:cu_rl_for_all_call_options}
	We train a unified model for all call options with any initial states by the CU-RL approach. The training data include the paths of all the call options in the training set. 
	In total, there are 513017 training paths, 113529 validation paths, and 114089 test paths. The distribution of maturity and moneyness of the training, validation, and test dataset of call options are shown in Table~\ref{table_real_call_all_data}.
	\begin{table}[H]
		\centering
		\caption{\textbf{The number of paths of call options in different maturity ranges and moneyness ranges.} The maximum maturity of options in the training, validation, and test data are 361 days, 88 days, and 88 days, respectively. The moneyness ranges of options in the training, validation, and test data are $[0.06426, 1.5816]$, $[0.1212, 1.3694]$, and $[0.1153,1.3839]$, respectively.}
		\resizebox{\textwidth}{!}{
		\begin{tabular}{cccccccc}
			\toprule
			Maturity range &  Training set & Valid set & Test set & Moneyness range &  Training set & Valid set & Test set \\ 
			\cmidrule(lr){2-4}\cmidrule(lr){6-8}
			$[0, 7]$ & 62231 & 18030 & 18979 & $(0.0, 0.1]$ & 129 & 0 & 0 \\
			$(7, 14]$ & 69967 & 19603 & 19616 & $(0.1, 0.5]$ & 11734 & 1374 & 591 \\
			$(14, 30]  $ & 151964 & 37632 & 40222 & $(0.5, 0.8]$ & 102802 & 11399 & 8084 \\
			$(30, 60]$ & 152372 & 31806 & 29836 & $(0.8, 0.9]$ &  104469 & 19486 & 20021 \\
			$(60, 90]$ & 60704 & 6458 & 5436 & $(0.9, 1.1] $ & 273336 & 77634 & 82542 \\
			$(90, 361]$ & 15779 & 0 & 0 & $(1.1, 1.5]$ & 20486 & 3636 & 2851 \\
			$(361, \infty]$ & 0 & 0 & 0 & $(1.5, 1.6]$ & 61 & 0 & 0 \\
			\bottomrule
		\end{tabular}%
		}
		\label{table_real_call_all_data}%
	\end{table}
	
	\begin{figure}[!htbp]
		\centering
		\subfigure[Cumulative reward for each training epoch]{
			\begin{minipage}[c]{0.95\linewidth}
				\centering
				\includegraphics[width=0.48\linewidth]{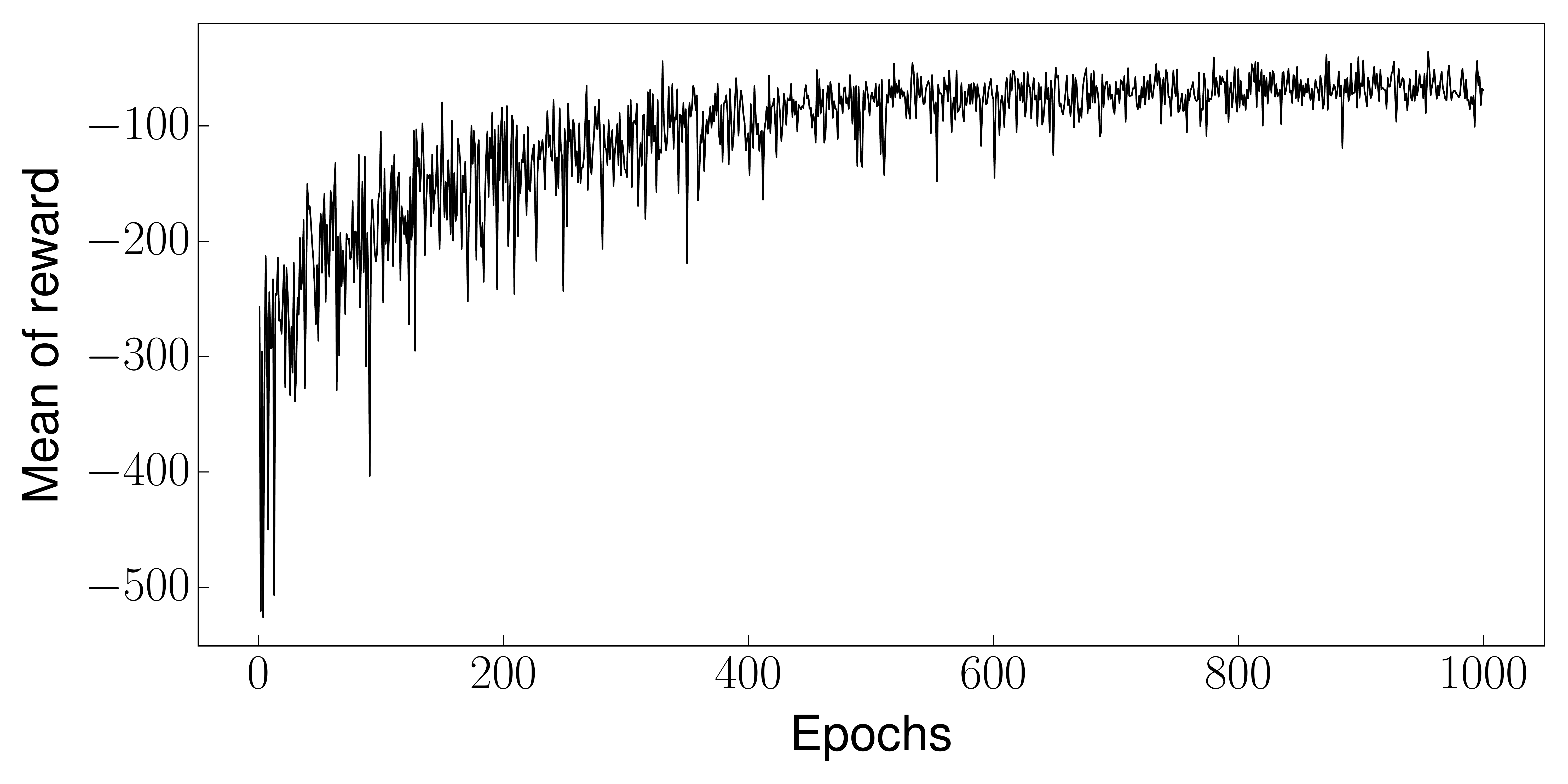}
			\end{minipage}
		}
		\subfigure[Loss of value network for each training epoch]{
			\begin{minipage}[c]{0.48\linewidth}
				\centering
				\includegraphics[width=0.95\linewidth]{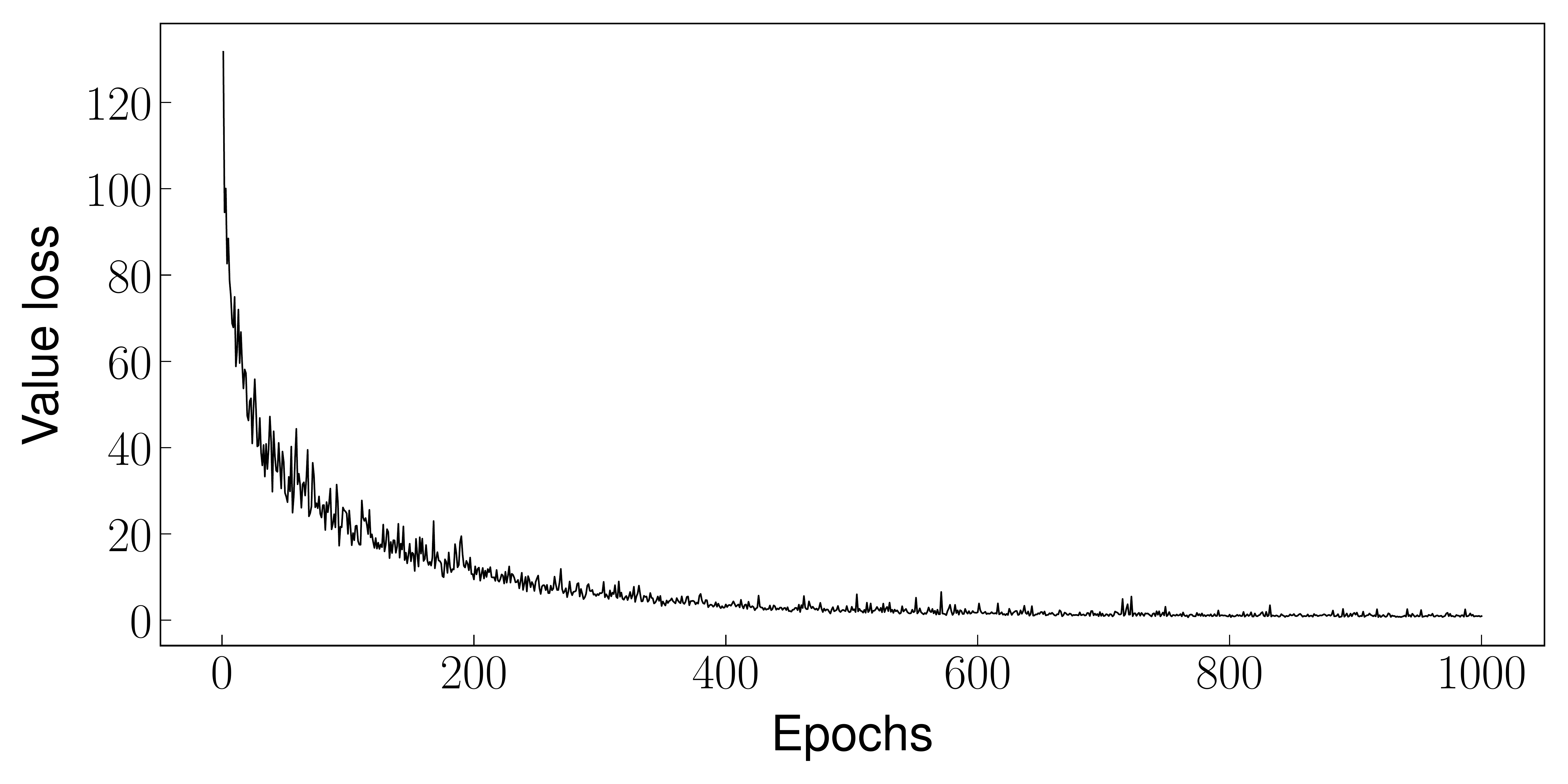}
			\end{minipage}
		}
		\subfigure[Loss of VaR network for each training epoch]{
			\begin{minipage}[c]{0.48\linewidth}
				\centering
				\includegraphics[width=0.95\linewidth]{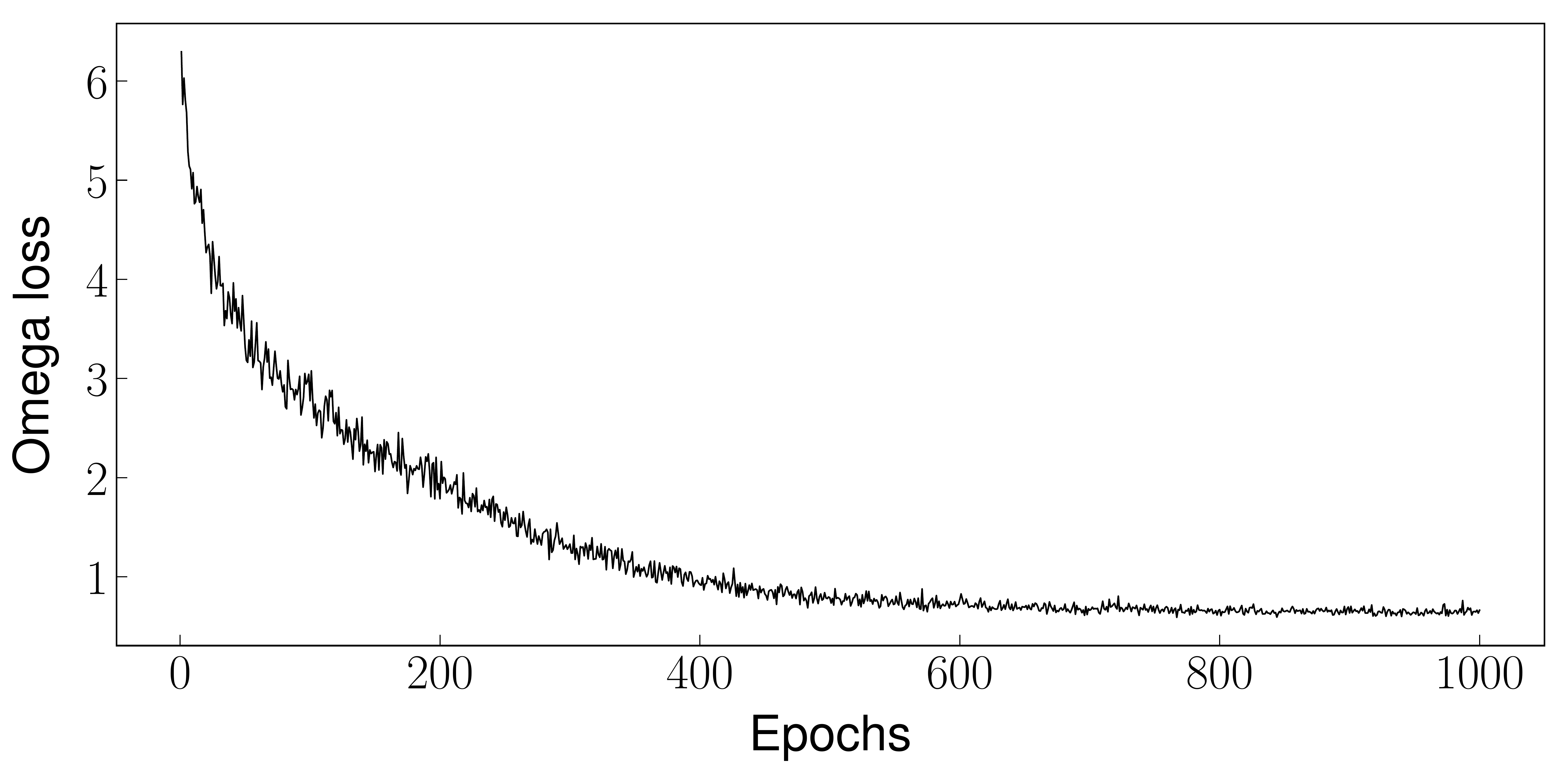}
			\end{minipage}
		}
		\caption{\textbf{Learning curves for the CU-RL method for hedging all S\&P 500 call options in the real market.} The $x$-axis of all subfigures represent the training epoch. The $y$-axis represent the cumulative rewards, the loss of value network, and the loss of VaR network, respectively for the subfigures (a), (b), and (c).}
		\label{fig:learning_curve_all}
	\end{figure}
	
	\begin{table}[!htbp]
		\centering
		\caption{\textbf{The mean, standard error, and tail risk of the final P\&L of hedging all call options by the CU-RL method, the Black-Scholes, local volatility function, and SABR delta hedging method calculated from an out-of-sample test set.} The test set include all S\&P 500 call options traded on or after 10/01/2017 and expired before 01/02/2018. The p-values are calculated from a one-sided t-test for related samples that tests if the mean P\&L of our method is higher than that of the benchmark method. 0.95-CI means 95\% confidence interval. Scientific notation: $\text{1.23E-4} = 1.23\times 10^{-4}$. $\text{***}:p < 0.001$; $\text{**}:p < 0.01$; $*:p < 0.05$.}
\resizebox{\linewidth}{!}{
	\begin{tabular}{lllll}
		\toprule
		& \multicolumn{1}{r}{CU-RL} & \multicolumn{3}{c}{Traditional Model} \\ 
		\cmidrule(lr){3-5}
		& & \multicolumn{1}{r}{BS Model} & \multicolumn{1}{r}{LVF Model} & \multicolumn{1}{r}{SABR Model} \\ 
		\midrule 
		\multicolumn{5}{c}{Panel A: without transaction cost} \\ 
		\midrule
		 Mean & \multicolumn{1}{r}{2.9249}& \multicolumn{1}{r}{-1.0406}& \multicolumn{1}{r}{-2.0096}& \multicolumn{1}{r}{0.2089}\\
 Std Err & \multicolumn{1}{r}{1.8068E-02}& \multicolumn{1}{r}{7.9599E-03}& \multicolumn{1}{r}{1.0556E-02}& \multicolumn{1}{r}{1.2193E-02}\\
 P-Value & \multicolumn{1}{r}{}& \multicolumn{1}{r}{0.0000E+00$^{***}$}& \multicolumn{1}{r}{0.0000E+00$^{***}$}& \multicolumn{1}{r}{0.0000E+00$^{***}$}\\
 0.95-VaR & \multicolumn{1}{r}{3.7794}& \multicolumn{1}{r}{5.8561}& \multicolumn{1}{r}{8.7729}& \multicolumn{1}{r}{5.2496}\\
 0.95-CI of 0.95-VaR & \multicolumn{1}{r}{[3.7359, 3.8328]} & \multicolumn{1}{r}{[5.8085, 5.9103]} & \multicolumn{1}{r}{[8.6952, 8.8562]} & \multicolumn{1}{r}{[5.1924, 5.3053]} \\
 0.975-VaR & \multicolumn{1}{r}{5.0189}& \multicolumn{1}{r}{7.1287}& \multicolumn{1}{r}{11.0766}& \multicolumn{1}{r}{6.7506}\\
 0.95-CI of 0.975-VaR & \multicolumn{1}{r}{[4.9602, 5.0801]} & \multicolumn{1}{r}{[7.0642, 7.1983]} & \multicolumn{1}{r}{[10.9690, 11.1896]} & \multicolumn{1}{r}{[6.6765, 6.8438]} \\
 0.975-MS & \multicolumn{1}{r}{6.1787}& \multicolumn{1}{r}{8.2424}& \multicolumn{1}{r}{12.4982}& \multicolumn{1}{r}{8.5785}\\
 0.95-CI of 0.975-MS & \multicolumn{1}{r}{[6.0986, 6.2639]} & \multicolumn{1}{r}{[8.1687, 8.3331]} & \multicolumn{1}{r}{[12.4096, 12.5886]} & \multicolumn{1}{r}{[8.4534, 8.7365]} \\
 0.95-CVaR & \multicolumn{1}{r}{5.4541}& \multicolumn{1}{r}{7.4568}& \multicolumn{1}{r}{11.2591}& \multicolumn{1}{r}{8.1723}\\
 0.95-CI of 0.95-CVaR & \multicolumn{1}{r}{[5.3940, 5.5081]} & \multicolumn{1}{r}{[7.4043, 7.5127]} & \multicolumn{1}{r}{[11.1849, 11.3346]} & \multicolumn{1}{r}{[8.0410, 8.2998]} \\
 0.975-CVaR & \multicolumn{1}{r}{6.5600}& \multicolumn{1}{r}{8.4783}& \multicolumn{1}{r}{12.7464}& \multicolumn{1}{r}{10.4350}\\
 0.95-CI of 0.975-CVaR & \multicolumn{1}{r}{[6.4845, 6.6419]} & \multicolumn{1}{r}{[8.4106, 8.5385]} & \multicolumn{1}{r}{[12.6687, 12.8228]} & \multicolumn{1}{r}{[10.2362, 10.6633]}  \\
		\midrule 
		\multicolumn{5}{c}{Panel B: with $0.1\%$ transaction cost} \\ 
		\midrule
		 Mean & \multicolumn{1}{r}{1.1755}& \multicolumn{1}{r}{-3.4508}& \multicolumn{1}{r}{-4.6257}& \multicolumn{1}{r}{-2.7045}\\
 Std Err & \multicolumn{1}{r}{2.3157E-02}& \multicolumn{1}{r}{1.0031E-02}& \multicolumn{1}{r}{1.3667E-02}& \multicolumn{1}{r}{1.3308E-02}\\
 P-Value & \multicolumn{1}{r}{}& \multicolumn{1}{r}{0.0000E+00$^{***}$}& \multicolumn{1}{r}{0.0000E+00$^{***}$}& \multicolumn{1}{r}{0.0000E+00$^{***}$}\\
 0.95-VaR & \multicolumn{1}{r}{6.4345}& \multicolumn{1}{r}{9.2068}& \multicolumn{1}{r}{13.0101}& \multicolumn{1}{r}{9.1282}\\
 0.95-CI of 0.95-VaR & \multicolumn{1}{r}{[6.3899, 6.4861]} & \multicolumn{1}{r}{[9.1579, 9.2491]} & \multicolumn{1}{r}{[12.9186, 13.0887]} & \multicolumn{1}{r}{[9.0536, 9.2073]} \\
 0.975-VaR & \multicolumn{1}{r}{7.6947}& \multicolumn{1}{r}{10.5362}& \multicolumn{1}{r}{15.6054}& \multicolumn{1}{r}{11.4791}\\
 0.95-CI of 0.975-VaR & \multicolumn{1}{r}{[7.6323, 7.7546]} & \multicolumn{1}{r}{[10.4645, 10.5985]} & \multicolumn{1}{r}{[15.4899, 15.7274]} & \multicolumn{1}{r}{[11.3261, 11.6428]} \\
 0.975-MS & \multicolumn{1}{r}{8.9001}& \multicolumn{1}{r}{11.6221}& \multicolumn{1}{r}{17.3305}& \multicolumn{1}{r}{14.7662}\\
 0.95-CI of 0.975-MS & \multicolumn{1}{r}{[8.7990, 8.9929]} & \multicolumn{1}{r}{[11.5416, 11.7006]} & \multicolumn{1}{r}{[17.2319, 17.4243]} & \multicolumn{1}{r}{[14.4888, 15.0669]} \\
 0.95-CVaR & \multicolumn{1}{r}{8.1483}& \multicolumn{1}{r}{10.8269}& \multicolumn{1}{r}{15.8995}& \multicolumn{1}{r}{14.0771}\\
 0.95-CI of 0.95-CVaR & \multicolumn{1}{r}{[8.0870, 8.2041]} & \multicolumn{1}{r}{[10.7770, 10.8816]} & \multicolumn{1}{r}{[15.8114, 15.9940]} & \multicolumn{1}{r}{[13.8530, 14.2852]} \\
 0.975-CVaR & \multicolumn{1}{r}{9.2947}& \multicolumn{1}{r}{11.8595}& \multicolumn{1}{r}{17.6550}& \multicolumn{1}{r}{18.0190}\\
 0.95-CI of 0.975-CVaR & \multicolumn{1}{r}{[9.2246, 9.3839]} & \multicolumn{1}{r}{[11.7940, 11.9177]} & \multicolumn{1}{r}{[17.5565, 17.7463]} & \multicolumn{1}{r}{[17.6693, 18.4025]}  \\
		\bottomrule 
		\end{tabular}} 
		\label{table_real_call_all_asym} 
	\end{table} 

	In the implementation of CU-RL method for hedging all call options in the real market, the number of hidden layers is 9 and the number of neurons in each layer is 32. In Algorithm~\ref{algorithm_ppo_zeta}, the size of buffer $\mathcal{D}_k$ is 59990, the minibatch size $n=30000$. The initial value for the linearly decaying learning rate is 1e-4 for the non-transaction cost case and 1e-5 for the 0.1\% proportional transaction cost case. We pre-train the initializer for all call options based on the SABR delta data. 
   
Fig. \ref{fig:learning_curve_all} shows the three learning curves of CU-RL method for the cumulative reward, the loss for value network, and the loss for VaR network, respectively. The convergence of the CU-RL method seems to be stable.
The out-of-sample performance of CU-RL on the test data in both non-transaction cost and proportional transaction cost cases  is shown in Table~\ref{table_real_call_all_asym}.
	
	The panel A of Table~\ref{table_real_call_all_asym} shows that: (i) CU-RL obtains the significantly highest mean final P\&L among all the methods; (ii) CU-RL obtains the lowest CVaR at 0.975 level, the designated risk measure in the total reward of CU-RL, the lowest MS at 0.975 level, and lowest VaR at 0.975 level; (iii) CU-RL obtains the lowest VaR and CVaR at 0.95 level, suggesting it can effectively minimize the tail risk measured at an alternative level; (iv) all the results with respect to tail risk measures are statistically significant, since all the confidence intervals of CU-RL are lower and have no overlap with those of other benchmark models.
		
	The panel B of Table~\ref{table_real_call_all_asym} shows that similar conclusion holds when transaction costs are considered. Although the mean of final P\&L falls and the risk measure rises due to transaction costs, CU-RL still obtains the significantly highest mean and lowest risk of P\&L.

	\subsection{Performance of Hedging Short-Term Near-the-Money Call Options by a Unified Model}\label{subsec:cu_rl_for_short_term_call_options}
	In this subsection, we train a unified model for all the short-term call options with maturity $T \in [5, 30]$ and moneyness $\frac{K}{S_{0}} \in [0.9, 1.1]$. The training set, validation set, and test set contains the data of all such options in the corresponding sets in Section \ref{subsec:data_processing}.	
	There are 160102 training paths, 39537 validation paths, and 43228 test paths. The distribution of maturity and moneyness in the training dataset and test dataset is shown in Table~\ref{table_real_call_short_data}.
	\begin{table}[!htbp]
		\centering
		\caption{\textbf{The number of paths of short-term close-to-the-money call options ($T \in [5, 30], \frac{K}{S_{0}} \in [0.9, 1.1]$) in different maturity ranges and moneyness ranges.}}
		\resizebox{\textwidth}{!}{
		\begin{tabular}{cccccccc}
			\toprule
			Maturity range &  Training set & Valid set & Test set & Moneyness range &  Training set & Valid set & Test set \\ 
			\cmidrule(lr){2-4}\cmidrule(lr){6-8}
			$[5, 15]$ & 55608 & 14703 & 16236 & $[0.9, 0.95]$ & 40098 & 8517 & 9253 \\
			$[16, 21]$ & 42787 & 10466 & 11791 & $(0.95, 1.0]$ & 43386 & 12276 & 13476 \\
			$[22, 30]$ & 61707 & 14368 & 15201 & $(1.0, 1.05]$ & 47942 & 13235 & 14715 \\
			- & - & - & - &  $(1.05, 1.1]$ & 28676 & 5509 & 5784 \\
			\bottomrule
		\end{tabular}}
		\label{table_real_call_short_data}
	\end{table} 

	In the implementation of CU-RL method for short-term call options, the network structure are the same as those in the all call option case, i.e. the three networks are specified as fully connected feedforward neural networks with 9 hidden layers of 32 neurons.
	In Algorithm~\ref{algorithm_ppo_zeta}, the size of buffer $\mathcal{D}_k$ is 29988, the minibatch size $n=20000$, and the initial value of the linearly decaying learning rate is 1e-6.
	
\begin{figure}[!htbp]
		\centering
		\subfigure[Cumulative reward for each training epoch]{
			\begin{minipage}[c]{0.95\linewidth}
				\centering
				\includegraphics[width=0.48\linewidth]{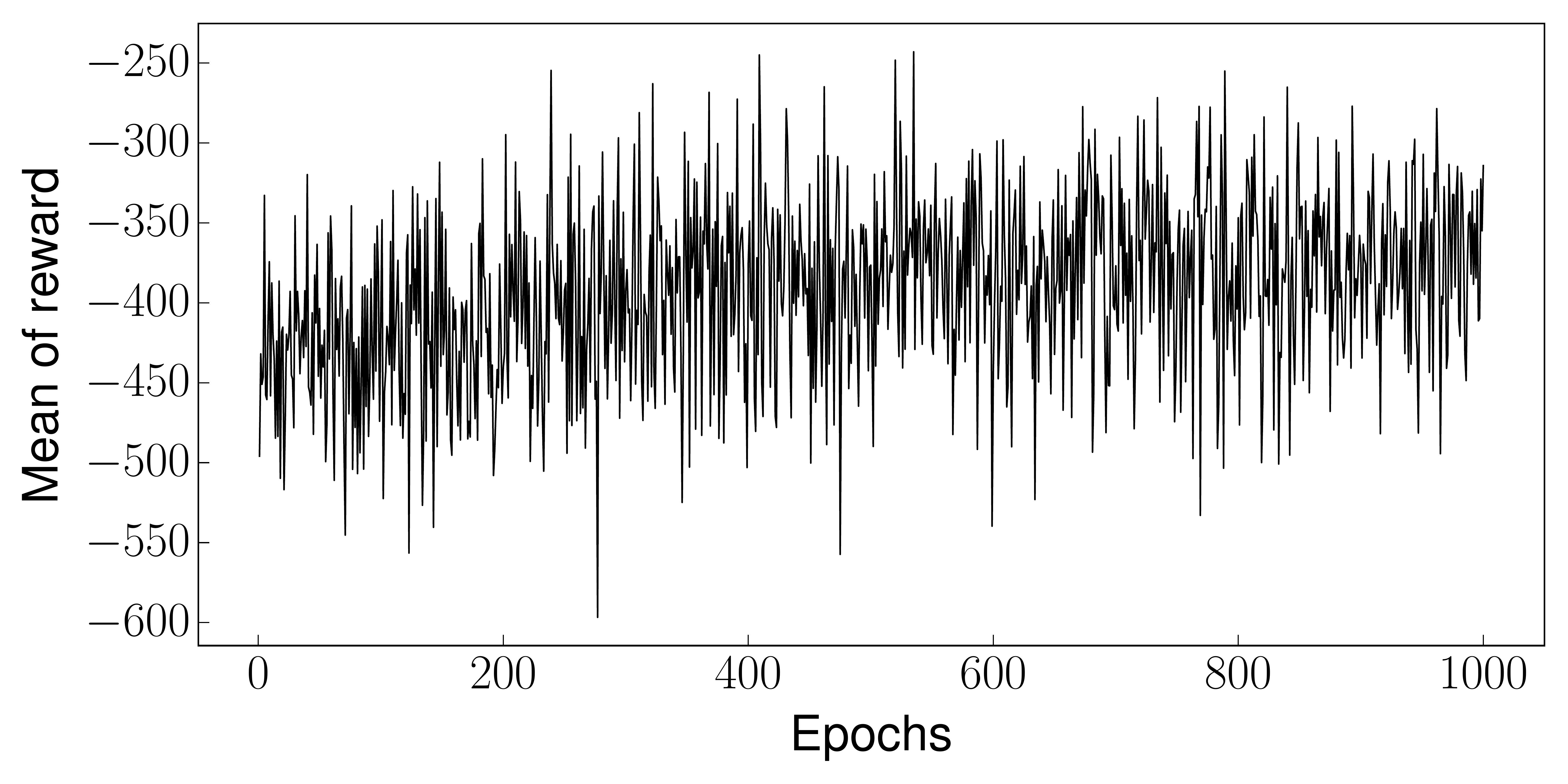}
			\end{minipage}
		}
		\subfigure[Loss of value network for each training epoch]{
			\begin{minipage}[c]{0.48\linewidth}
				\centering
				\includegraphics[width=0.95\linewidth]{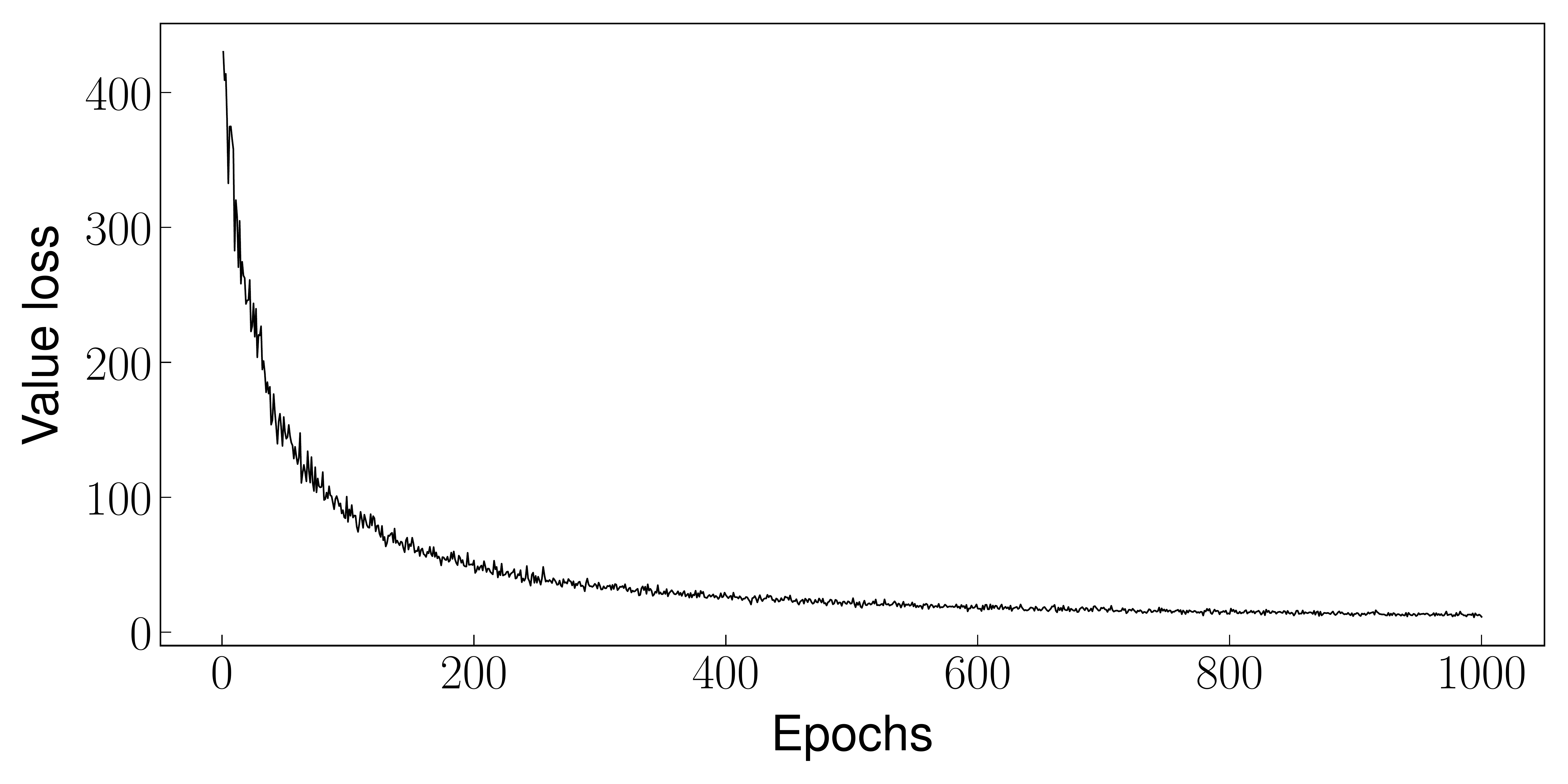}
			\end{minipage}
		}
		\subfigure[Loss of VaR network for each training epoch]{
			\begin{minipage}[c]{0.48\linewidth}
				\centering
				\includegraphics[width=0.95\linewidth]{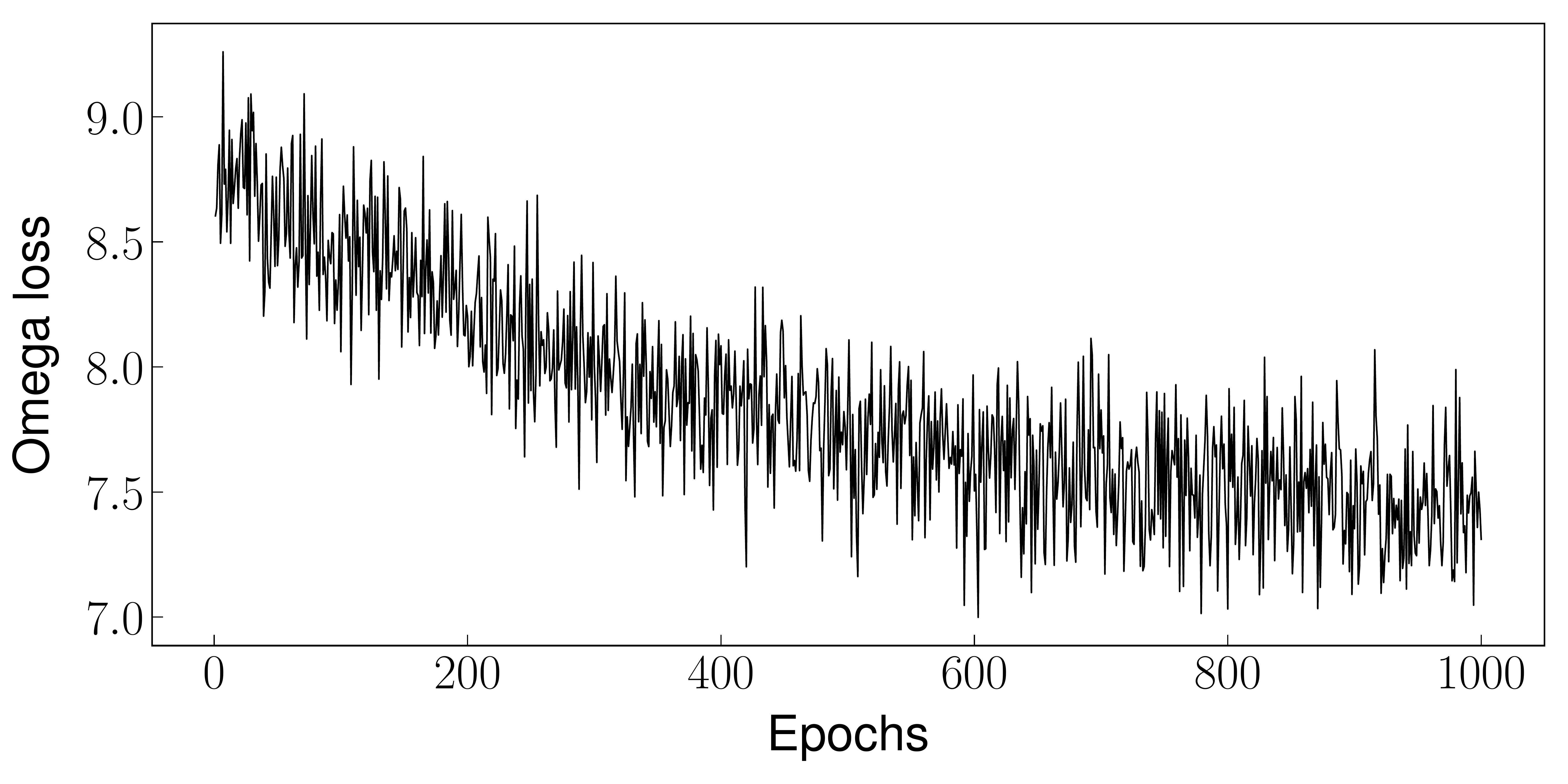}
			\end{minipage}
		}
		\caption{\textbf{Learning curves for the CU-RL method for hedging S\&P 500 short-term call options in the real market.} The $x$-axis of all subfigures represent the training epoch. The $y$-axis represent the cumulative rewards, the loss of value network, and the loss of VaR network, respectively for the subfigures (a), (b), and (c).}
		\label{fig:learning_curve_short}
	\end{figure}

	Fig. \ref{fig:learning_curve_short} shows the three learning curves of CU-RL method for the cumulative reward, the loss for value network, and the loss for VaR network, respectively. The convergence of the CU-RL method seems to be stable.
	The out-of-sample performance of CU-RL on the test data in both non-transaction cost and proportional transaction cost cases  is shown in Table~\ref{table_real_call_short_asym}.

	\begin{table}[!htbp]
		\centering
		\caption{\textbf{The mean, standard error, and tail risk of the final P\&L of hedging short-term call options by the CU-RL method, the Black-Scholes, local volatility function, and SABR delta hedging method calculated from an out-of-sample test set.} The test set include all S\&P 500 short-term call options with $T \in [5,30], K/S_{0} \in[0.9,1.1]$ traded on or after 10/01/2017 and expired before 01/02/2018. The p-values are calculated from a one-sided t-test for related samples that tests if the mean P\&L of our method is higher than that of the benchmark method. 0.95-CI means 95\% confidence interval. Scientific notation: $\text{1.23E-4} = 1.23\times 10^{-4}$. $\text{***}:p < 0.001$; $\text{**}:p < 0.01$; $*:p < 0.05$.}
\resizebox{\linewidth}{!}{
	\begin{tabular}{lllll}
		\toprule
		& \multicolumn{1}{r}{CU-RL} & \multicolumn{3}{c}{Traditional Model} \\ 
		\cmidrule(lr){3-5}
		& & \multicolumn{1}{r}{BS Model} & \multicolumn{1}{r}{LVF Model} & \multicolumn{1}{r}{SABR Model} \\ 
		\midrule 
		\multicolumn{5}{c}{Panel A: without transaction cost} \\ 
		\midrule
		 Mean & \multicolumn{1}{r}{2.4479}& \multicolumn{1}{r}{-0.3079}& \multicolumn{1}{r}{-1.0298}& \multicolumn{1}{r}{0.7770}\\
 Std Err & \multicolumn{1}{r}{2.2538E-02}& \multicolumn{1}{r}{1.1242E-02}& \multicolumn{1}{r}{1.4010E-02}& \multicolumn{1}{r}{1.8853E-02}\\
 P-Value & \multicolumn{1}{r}{}& \multicolumn{1}{r}{0.0000E+00$^{***}$}& \multicolumn{1}{r}{0.0000E+00$^{***}$}& \multicolumn{1}{r}{0.0000E+00$^{***}$}\\
 0.95-VaR & \multicolumn{1}{r}{1.5239}& \multicolumn{1}{r}{4.0753}& \multicolumn{1}{r}{5.9742}& \multicolumn{1}{r}{4.4919}\\
 0.95-CI of 0.95-VaR & \multicolumn{1}{r}{[1.4919, 1.5491]} & \multicolumn{1}{r}{[4.0187, 4.1357]} & \multicolumn{1}{r}{[5.8757, 6.0803]} & \multicolumn{1}{r}{[4.3136, 4.6473]} \\
 0.975-VaR & \multicolumn{1}{r}{2.0097}& \multicolumn{1}{r}{4.9829}& \multicolumn{1}{r}{7.5718}& \multicolumn{1}{r}{6.5233}\\
 0.95-CI of 0.975-VaR & \multicolumn{1}{r}{[1.9556, 2.0593]} & \multicolumn{1}{r}{[4.9061, 5.0849]} & \multicolumn{1}{r}{[7.4473, 7.6647]} & \multicolumn{1}{r}{[6.3532, 6.7066]} \\
 0.975-MS & \multicolumn{1}{r}{2.6801}& \multicolumn{1}{r}{6.5480}& \multicolumn{1}{r}{9.2724}& \multicolumn{1}{r}{8.8325}\\
 0.95-CI of 0.975-MS & \multicolumn{1}{r}{[2.6021, 2.7812]} & \multicolumn{1}{r}{[6.3322, 6.7633]} & \multicolumn{1}{r}{[9.0487, 9.5064]} & \multicolumn{1}{r}{[8.5744, 9.0669]} \\
 0.95-CVaR & \multicolumn{1}{r}{2.4143}& \multicolumn{1}{r}{5.6511}& \multicolumn{1}{r}{8.2637}& \multicolumn{1}{r}{7.7263}\\
 0.95-CI of 0.95-CVaR & \multicolumn{1}{r}{[2.3499, 2.4751]} & \multicolumn{1}{r}{[5.5673, 5.7563]} & \multicolumn{1}{r}{[8.1390, 8.3906]} & \multicolumn{1}{r}{[7.5117, 7.9199]} \\
 0.975-CVaR & \multicolumn{1}{r}{3.1050}& \multicolumn{1}{r}{6.8327}& \multicolumn{1}{r}{9.8537}& \multicolumn{1}{r}{10.0016}\\
 0.95-CI of 0.975-CVaR & \multicolumn{1}{r}{[3.0135, 3.2055]} & \multicolumn{1}{r}{[6.7043, 6.9722]} & \multicolumn{1}{r}{[9.6701, 10.0390]} & \multicolumn{1}{r}{[9.7116, 10.3233]}  \\
		\midrule 
		\multicolumn{5}{c}{Panel B: with $0.1\%$ transaction cost} \\ 
		\midrule
		 Mean & \multicolumn{1}{r}{1.2380}& \multicolumn{1}{r}{-2.3988}& \multicolumn{1}{r}{-3.2881}& \multicolumn{1}{r}{-2.1633}\\
 Std Err & \multicolumn{1}{r}{2.9677E-02}& \multicolumn{1}{r}{1.4498E-02}& \multicolumn{1}{r}{1.8475E-02}& \multicolumn{1}{r}{2.1290E-02}\\
 P-Value & \multicolumn{1}{r}{}& \multicolumn{1}{r}{0.0000E+00$^{***}$}& \multicolumn{1}{r}{0.0000E+00$^{***}$}& \multicolumn{1}{r}{0.0000E+00$^{***}$}\\
 0.95-VaR & \multicolumn{1}{r}{4.5229}& \multicolumn{1}{r}{7.3841}& \multicolumn{1}{r}{9.8077}& \multicolumn{1}{r}{9.5096}\\
 0.95-CI of 0.95-VaR & \multicolumn{1}{r}{[4.4763, 4.5772]} & \multicolumn{1}{r}{[7.3077, 7.4348]} & \multicolumn{1}{r}{[9.7235, 9.9045]} & \multicolumn{1}{r}{[9.3464, 9.6470]} \\
 0.975-VaR & \multicolumn{1}{r}{5.4778}& \multicolumn{1}{r}{8.3330}& \multicolumn{1}{r}{11.2466}& \multicolumn{1}{r}{12.4072}\\
 0.95-CI of 0.975-VaR & \multicolumn{1}{r}{[5.3950, 5.5769]} & \multicolumn{1}{r}{[8.2405, 8.4352]} & \multicolumn{1}{r}{[11.1512, 11.3675]} & \multicolumn{1}{r}{[12.1395, 12.5717]} \\
 0.975-MS & \multicolumn{1}{r}{6.5112}& \multicolumn{1}{r}{9.8437}& \multicolumn{1}{r}{13.0815}& \multicolumn{1}{r}{15.0653}\\
 0.95-CI of 0.975-MS & \multicolumn{1}{r}{[6.3929, 6.6523]} & \multicolumn{1}{r}{[9.6109, 10.0783]} & \multicolumn{1}{r}{[12.7899, 13.3427]} & \multicolumn{1}{r}{[14.7583, 15.4513]} \\
 0.95-CVaR & \multicolumn{1}{r}{6.0319}& \multicolumn{1}{r}{8.9888}& \multicolumn{1}{r}{12.0782}& \multicolumn{1}{r}{13.9373}\\
 0.95-CI of 0.95-CVaR & \multicolumn{1}{r}{[5.9339, 6.1235]} & \multicolumn{1}{r}{[8.9041, 9.0956]} & \multicolumn{1}{r}{[11.9504, 12.1997]} & \multicolumn{1}{r}{[13.6502, 14.1959]} \\
 0.975-CVaR & \multicolumn{1}{r}{7.1246}& \multicolumn{1}{r}{10.1843}& \multicolumn{1}{r}{13.6890}& \multicolumn{1}{r}{17.1279}\\
 0.95-CI of 0.975-CVaR & \multicolumn{1}{r}{[6.9963, 7.2752]} & \multicolumn{1}{r}{[10.0535, 10.3302]} & \multicolumn{1}{r}{[13.4952, 13.8857]} & \multicolumn{1}{r}{[16.7323, 17.5958]}  \\
		\bottomrule 
		\end{tabular}} 
		\label{table_real_call_short_asym} 
	\end{table} 
	
	The panel A of Table~\ref{table_real_call_short_asym} shows that: (i) CU-RL obtains the significantly highest mean final P\&L among all the methods; (ii) CU-RL obtains the lowest CVaR at 0.975 level, the designated risk measure in the total reward of CU-RL, the lowest MS at 0.975 level, and lowest VaR at 0.975 level; (iii) CU-RL obtains the lowest VaR and CVaR at 0.95 level, suggesting it can effectively minimize the tail risk measured at an alternative level; (iv) all the results with respect to tail risk measures are statistically significant, since all the confidence intervals of CU-RL are lower and have no overlap with those of other benchmark models.
		
	The panel B of Table~\ref{table_real_call_short_asym} shows that similar conclusion holds when transaction costs are considered. Although the mean of final P\&L falls and the risk measure rises due to transaction costs, CU-RL still obtains the significantly highest mean and lowest risk of P\&L.	
	
	\subsection{Performance of Hedging Short-Term Deep In-the-money (ITM) Call Options by a Unified Model}\label{subsec:cu_rl_for_deep_itm_call_options}

	We then train two unified models, one for all the short-term deep ITM call options with $K/S_{0} \in (0.1, 0.8]$ and $T \in [5,30]$, and the other for all the short-term deep ITM call options with $K/S_{0} \in (0.8, 0.9]$ and $T \in [5,30]$. The distribution of maturity in each dataset is shown in Table~\ref{table_real_data_call_deepin}. 
	We do not train a separate model for short-term deep out-of-the-money (OTM) call options since there are not enough such option data.
	
	\begin{table}[!htbp]
		\centering
		\caption{\textbf{The number of paths of short-term deep ITM call options ($T \in [5, 30], \frac{K}{S_{0}} \in [0.1, 0.9]$) in different maturity ranges and moneyness ranges.}}
		\begin{tabular}{ccccccc}
			\toprule
			Maturity Range & \multicolumn{3}{c}{$K/S_{0} \in (0.1, 0.8]$} & \multicolumn{3}{c}{$K/S_{0} \in (0.8, 0.9]$} \\
			\cmidrule(lr){2-4}\cmidrule(lr){5-7}
			 & Training set  & Validation set & Test set  & Training set & Validation set & Test set\\
			\midrule
			$ [5, 15]$ & 11213 & 1177 & 1302 & 20710 & 2405 & 2549 \\
			$ (15, 21]$ & 5862 & 1222 & 1137 & 10567 & 2244 & 2343 \\
			$(21, 30]$ & 9405 & 1797 & 1260 & 16619 & 2812 & 2942 \\
			Total & 26480 & 4196 & 3699 & 47896 & 7461 & 7838 \\
			\bottomrule
		\end{tabular}
		\label{table_real_data_call_deepin}
	\end{table}
	
	In the implementation of CU-RL method for short-term deep ITM call options, the number of hidden layers is 3 and the number of neurons in each layer is 64.
	In Algorithm~\ref{algorithm_ppo_zeta}, the size of buffer $\mathcal{D}_k$ is 29988, and the initial value of the linearly decaying learning rate is 1e-5. The minibatch size $n$ is 10000 for the group with $K/S_{0} \in (0.1, 0.8]$ and 20000 for $K/S_{0} \in (0.8, 0.9]$.
	
	The out-of-sample performance of CU-RL on the test data in both non-transaction cost and proportional transaction cost cases  for both two groups is shown in Table~\ref{table_real_call_deepin0} and Table~\ref{table_real_call_deepin1}.
	
	\begin{table}[!htbp]
		\centering
		\caption{\textbf{The mean, standard error, and tail risk of the final P\&L of hedging short-term deep ITM call options by the CU-RL method, the Black-Scholes, local volatility function, and SABR delta hedging method calculated from an out-of-sample test set.} The test set include all S\&P 500 short-term deep ITM call options with $K/S_{0} \in (0.1, 0.8], T \in [5, 30]$ traded on or after 10/01/2017 and expired before 01/02/2018. The p-values are calculated from a one-sided t-test for related samples that tests if the mean P\&L of our method is higher than that of the benchmark method. 0.95-CI means 95\% confidence interval. Scientific notation: $\text{1.23E-4} = 1.23\times 10^{-4}$. $\text{***}:p < 0.001$; $\text{**}:p < 0.01$; $*:p < 0.05$.}
	\resizebox{\linewidth}{!}{
	\begin{tabular}{lllll}
		\toprule
		& \multicolumn{1}{r}{CU-RL} & \multicolumn{3}{c}{Traditional Model} \\ 
		\cmidrule(lr){3-5}
		& & \multicolumn{1}{r}{BS Model} & \multicolumn{1}{r}{LVF Model} & \multicolumn{1}{r}{SABR Model} \\ 
		\midrule 
		\multicolumn{5}{c}{Panel A: without transaction cost} \\ 
		\midrule
		 Mean & \multicolumn{1}{r}{-0.9170}& \multicolumn{1}{r}{-1.2021}& \multicolumn{1}{r}{-1.6346}& \multicolumn{1}{r}{-0.9364}\\
 Std Err & \multicolumn{1}{r}{2.1436E-02}& \multicolumn{1}{r}{2.1928E-02}& \multicolumn{1}{r}{2.4261E-02}& \multicolumn{1}{r}{2.1641E-02}\\
 P-Value & \multicolumn{1}{r}{}& \multicolumn{1}{r}{0.0000E+00$^{***}$}& \multicolumn{1}{r}{0.0000E+00$^{***}$}& \multicolumn{1}{r}{4.6431E-256$^{***}$}\\
 0.95-VaR & \multicolumn{1}{r}{2.8302}& \multicolumn{1}{r}{3.0917}& \multicolumn{1}{r}{3.7633}& \multicolumn{1}{r}{2.8610}\\
 0.95-CI of 0.95-VaR & \multicolumn{1}{r}{[2.7706, 2.8746]} & \multicolumn{1}{r}{[3.0538, 3.1576]} & \multicolumn{1}{r}{[3.7121, 3.8255]} & \multicolumn{1}{r}{[2.8118, 2.9167]} \\
 0.975-VaR & \multicolumn{1}{r}{3.0144}& \multicolumn{1}{r}{3.3047}& \multicolumn{1}{r}{3.9797}& \multicolumn{1}{r}{3.0619}\\
 0.95-CI of 0.975-VaR & \multicolumn{1}{r}{[2.9753, 3.0791]} & \multicolumn{1}{r}{[3.2469, 3.3639]} & \multicolumn{1}{r}{[3.9329, 4.0338]} & \multicolumn{1}{r}{[3.0211, 3.1152]} \\
 0.975-MS & \multicolumn{1}{r}{3.1749}& \multicolumn{1}{r}{3.4509}& \multicolumn{1}{r}{4.1427}& \multicolumn{1}{r}{3.2031}\\
 0.95-CI of 0.975-MS & \multicolumn{1}{r}{[3.1188, 3.2183]} & \multicolumn{1}{r}{[3.3891, 3.5506]} & \multicolumn{1}{r}{[4.0693, 4.2222]} & \multicolumn{1}{r}{[3.1652, 3.2768]} \\
 0.95-CVaR & \multicolumn{1}{r}{3.0646}& \multicolumn{1}{r}{3.3813}& \multicolumn{1}{r}{4.0893}& \multicolumn{1}{r}{3.1075}\\
 0.95-CI of 0.95-CVaR & \multicolumn{1}{r}{[3.0231, 3.1066]} & \multicolumn{1}{r}{[3.3316, 3.4375]} & \multicolumn{1}{r}{[4.0218, 4.1599]} & \multicolumn{1}{r}{[3.0671, 3.1512]} \\
 0.975-CVaR & \multicolumn{1}{r}{3.2070}& \multicolumn{1}{r}{3.5660}& \multicolumn{1}{r}{4.3072}& \multicolumn{1}{r}{3.2514}\\
 0.95-CI of 0.975-CVaR & \multicolumn{1}{r}{[3.1572, 3.2570]} & \multicolumn{1}{r}{[3.5008, 3.6467]} & \multicolumn{1}{r}{[4.2222, 4.4325]} & \multicolumn{1}{r}{[3.2078, 3.3083]}  \\
		\midrule 
		\multicolumn{5}{c}{Panel B: with $0.1\%$ transaction cost} \\ 
		\midrule
		 Mean & \multicolumn{1}{r}{-3.5174}& \multicolumn{1}{r}{-3.8672}& \multicolumn{1}{r}{-4.4077}& \multicolumn{1}{r}{-3.5424}\\
 Std Err & \multicolumn{1}{r}{2.0953E-02}& \multicolumn{1}{r}{2.2049E-02}& \multicolumn{1}{r}{2.5460E-02}& \multicolumn{1}{r}{2.1144E-02}\\
 P-Value & \multicolumn{1}{r}{}& \multicolumn{1}{r}{0.0000E+00$^{***}$}& \multicolumn{1}{r}{0.0000E+00$^{***}$}& \multicolumn{1}{r}{0.0000E+00$^{***}$}\\
 0.95-VaR & \multicolumn{1}{r}{5.3888}& \multicolumn{1}{r}{5.7928}& \multicolumn{1}{r}{6.6989}& \multicolumn{1}{r}{5.4325}\\
 0.95-CI of 0.95-VaR & \multicolumn{1}{r}{[5.3424, 5.4451]} & \multicolumn{1}{r}{[5.7327, 5.8421]} & \multicolumn{1}{r}{[6.6379, 6.7802]} & \multicolumn{1}{r}{[5.3925, 5.4883]} \\
 0.975-VaR & \multicolumn{1}{r}{5.5728}& \multicolumn{1}{r}{5.9834}& \multicolumn{1}{r}{7.0165}& \multicolumn{1}{r}{5.6369}\\
 0.95-CI of 0.975-VaR & \multicolumn{1}{r}{[5.5505, 5.6389]} & \multicolumn{1}{r}{[5.9418, 6.0233]} & \multicolumn{1}{r}{[6.9419, 7.0831]} & \multicolumn{1}{r}{[5.5884, 5.6918]} \\
 0.975-MS & \multicolumn{1}{r}{5.7304}& \multicolumn{1}{r}{6.1250}& \multicolumn{1}{r}{7.1544}& \multicolumn{1}{r}{5.7720}\\
 0.95-CI of 0.975-MS & \multicolumn{1}{r}{[5.6917, 5.8040]} & \multicolumn{1}{r}{[6.0641, 6.2563]} & \multicolumn{1}{r}{[7.1279, 7.2956]} & \multicolumn{1}{r}{[5.7349, 5.8545]} \\
 0.95-CVaR & \multicolumn{1}{r}{5.6276}& \multicolumn{1}{r}{6.0792}& \multicolumn{1}{r}{7.1095}& \multicolumn{1}{r}{5.6768}\\
 0.95-CI of 0.95-CVaR & \multicolumn{1}{r}{[5.5858, 5.6699]} & \multicolumn{1}{r}{[6.0262, 6.1429]} & \multicolumn{1}{r}{[7.0303, 7.1944]} & \multicolumn{1}{r}{[5.6371, 5.7197]} \\
 0.975-CVaR & \multicolumn{1}{r}{5.7674}& \multicolumn{1}{r}{6.2688}& \multicolumn{1}{r}{7.3699}& \multicolumn{1}{r}{5.8197}\\
 0.95-CI of 0.975-CVaR & \multicolumn{1}{r}{[5.7185, 5.8146]} & \multicolumn{1}{r}{[6.1990, 6.3626]} & \multicolumn{1}{r}{[7.2737, 7.5052]} & \multicolumn{1}{r}{[5.7792, 5.8737]}  \\
		\bottomrule 
		\end{tabular}} 
		\label{table_real_call_deepin0} 
	\end{table} 
	
	\begin{table}[!htbp]
		\centering
		\caption{\textbf{The mean, standard error, and tail risk of the final P\&L of hedging short-term deep ITM call options by the CU-RL method, the Black-Scholes, local volatility function, and SABR delta hedging method calculated from an out-of-sample test set.} The test set include all S\&P 500 short-term deep ITM call options with $K/S_{0} \in (0.8, 0.9], T \in [5, 30]$ traded on or after 10/01/2017 and expired before 01/02/2018. The p-values are calculated from a one-sided t-test for related samples that tests if the mean P\&L of our method is higher than that of the benchmark method. 0.95-CI means 95\% confidence interval. Scientific notation: $\text{1.23E-4} = 1.23\times 10^{-4}$. $\text{***}:p < 0.001$; $\text{**}:p < 0.01$; $*:p < 0.05$.}
	\resizebox{\linewidth}{!}{
	\begin{tabular}{lllll}
		\toprule
		& \multicolumn{1}{r}{CU-RL} & \multicolumn{3}{c}{Traditional Model} \\ 
		\cmidrule(lr){3-5}
		& & \multicolumn{1}{r}{BS Model} & \multicolumn{1}{r}{LVF Model} & \multicolumn{1}{r}{SABR Model} \\ 
		\midrule 
		\multicolumn{5}{c}{Panel A: without transaction cost} \\ 
		\midrule
		 Mean & \multicolumn{1}{r}{-0.8550}& \multicolumn{1}{r}{-1.4554}& \multicolumn{1}{r}{-2.0387}& \multicolumn{1}{r}{-0.9186}\\
 Std Err & \multicolumn{1}{r}{1.4531E-02}& \multicolumn{1}{r}{1.5305E-02}& \multicolumn{1}{r}{1.7549E-02}& \multicolumn{1}{r}{1.4812E-02}\\
 P-Value & \multicolumn{1}{r}{}& \multicolumn{1}{r}{0.0000E+00$^{***}$}& \multicolumn{1}{r}{0.0000E+00$^{***}$}& \multicolumn{1}{r}{5.7469E-198$^{***}$}\\
 0.95-VaR & \multicolumn{1}{r}{2.6239}& \multicolumn{1}{r}{3.2628}& \multicolumn{1}{r}{4.2091}& \multicolumn{1}{r}{2.7167}\\
 0.95-CI of 0.95-VaR & \multicolumn{1}{r}{[2.5869, 2.6587]} & \multicolumn{1}{r}{[3.1856, 3.3580]} & \multicolumn{1}{r}{[4.1432, 4.2826]} & \multicolumn{1}{r}{[2.6884, 2.7490]} \\
 0.975-VaR & \multicolumn{1}{r}{2.8483}& \multicolumn{1}{r}{3.8865}& \multicolumn{1}{r}{5.1523}& \multicolumn{1}{r}{2.9877}\\
 0.95-CI of 0.975-VaR & \multicolumn{1}{r}{[2.7983, 2.8948]} & \multicolumn{1}{r}{[3.8236, 3.9404]} & \multicolumn{1}{r}{[4.9532, 5.3184]} & \multicolumn{1}{r}{[2.9419, 3.0411]} \\
 0.975-MS & \multicolumn{1}{r}{3.0496}& \multicolumn{1}{r}{4.1914}& \multicolumn{1}{r}{5.6199}& \multicolumn{1}{r}{3.1872}\\
 0.95-CI of 0.975-MS & \multicolumn{1}{r}{[3.0010, 3.1116]} & \multicolumn{1}{r}{[4.1139, 4.2689]} & \multicolumn{1}{r}{[5.5426, 5.6980]} & \multicolumn{1}{r}{[3.1380, 3.2274]} \\
 0.95-CVaR & \multicolumn{1}{r}{2.9071}& \multicolumn{1}{r}{3.8957}& \multicolumn{1}{r}{5.1113}& \multicolumn{1}{r}{3.0371}\\
 0.95-CI of 0.95-CVaR & \multicolumn{1}{r}{[2.8764, 2.9435]} & \multicolumn{1}{r}{[3.8255, 3.9633]} & \multicolumn{1}{r}{[5.0121, 5.2318]} & \multicolumn{1}{r}{[3.0034, 3.0838]} \\
 0.975-CVaR & \multicolumn{1}{r}{3.0915}& \multicolumn{1}{r}{4.2133}& \multicolumn{1}{r}{5.6696}& \multicolumn{1}{r}{3.2431}\\
 0.95-CI of 0.975-CVaR & \multicolumn{1}{r}{[3.0497, 3.1354]} & \multicolumn{1}{r}{[4.1593, 4.2675]} & \multicolumn{1}{r}{[5.5843, 5.7540]} & \multicolumn{1}{r}{[3.1991, 3.2952]}  \\
		\midrule 
		\multicolumn{5}{c}{Panel B: with $0.1\%$ transaction cost} \\ 
		\midrule
		 Mean & \multicolumn{1}{r}{-3.4388}& \multicolumn{1}{r}{-4.1924}& \multicolumn{1}{r}{-4.9259}& \multicolumn{1}{r}{-3.5600}\\
 Std Err & \multicolumn{1}{r}{1.4215E-02}& \multicolumn{1}{r}{1.5822E-02}& \multicolumn{1}{r}{1.9034E-02}& \multicolumn{1}{r}{1.4545E-02}\\
 P-Value & \multicolumn{1}{r}{}& \multicolumn{1}{r}{0.0000E+00$^{***}$}& \multicolumn{1}{r}{0.0000E+00$^{***}$}& \multicolumn{1}{r}{0.0000E+00$^{***}$}\\
 0.95-VaR & \multicolumn{1}{r}{5.1834}& \multicolumn{1}{r}{6.0773}& \multicolumn{1}{r}{7.3344}& \multicolumn{1}{r}{5.3393}\\
 0.95-CI of 0.95-VaR & \multicolumn{1}{r}{[5.1475, 5.2109]} & \multicolumn{1}{r}{[6.0188, 6.1789]} & \multicolumn{1}{r}{[7.2631, 7.4307]} & \multicolumn{1}{r}{[5.3076, 5.3721]} \\
 0.975-VaR & \multicolumn{1}{r}{5.4114}& \multicolumn{1}{r}{6.7948}& \multicolumn{1}{r}{8.4621}& \multicolumn{1}{r}{5.6069}\\
 0.95-CI of 0.975-VaR & \multicolumn{1}{r}{[5.3542, 5.4631]} & \multicolumn{1}{r}{[6.6689, 6.9024]} & \multicolumn{1}{r}{[8.2142, 8.6605]} & \multicolumn{1}{r}{[5.5608, 5.6578]} \\
 0.975-MS & \multicolumn{1}{r}{5.6097}& \multicolumn{1}{r}{7.1441}& \multicolumn{1}{r}{9.0274}& \multicolumn{1}{r}{5.7983}\\
 0.95-CI of 0.975-MS & \multicolumn{1}{r}{[5.5637, 5.6687]} & \multicolumn{1}{r}{[7.0661, 7.2487]} & \multicolumn{1}{r}{[8.9277, 9.1204]} & \multicolumn{1}{r}{[5.7632, 5.8362]} \\
 0.95-CVaR & \multicolumn{1}{r}{5.4643}& \multicolumn{1}{r}{6.7904}& \multicolumn{1}{r}{8.4009}& \multicolumn{1}{r}{5.6594}\\
 0.95-CI of 0.95-CVaR & \multicolumn{1}{r}{[5.4343, 5.5007]} & \multicolumn{1}{r}{[6.7072, 6.8683]} & \multicolumn{1}{r}{[8.2838, 8.5346]} & \multicolumn{1}{r}{[5.6249, 5.7089]} \\
 0.975-CVaR & \multicolumn{1}{r}{5.6483}& \multicolumn{1}{r}{7.1859}& \multicolumn{1}{r}{9.0514}& \multicolumn{1}{r}{5.8666}\\
 0.95-CI of 0.975-CVaR & \multicolumn{1}{r}{[5.6084, 5.6914]} & \multicolumn{1}{r}{[7.1192, 7.2462]} & \multicolumn{1}{r}{[8.9523, 9.1409]} & \multicolumn{1}{r}{[5.8234, 5.9222]}  \\
		\bottomrule 
\end{tabular}} 
		\label{table_real_call_deepin1} 
	\end{table} 

	The panel A of Table~\ref{table_real_call_deepin0} and Table~\ref{table_real_call_deepin1} share some similar results: (i) CU-RL obtains the significantly highest mean final P\&L among all the models; (ii) CU-RL obtains the lowest CVaR at 0.975 level, the designated risk measure in the total reward of CU-RL, the lowest MS at 0.975 level, and the lowest VaR at 0.975 level; (iii) CU-RL obtains the lowest VaR and CVaR at 0.95 level, suggesting it can effectively minimize the tail risk measured at an alternative level; (iv) the risk measures of CU-RL are significantly lower than all benchmark models for the group with $K/S_0\in[0.8, 0.9]$, and significantly lower than those of the BS and LVF models for the group with $K/S_0\in[0.1, 0.8]$, and lower (but not significant) than those of the SABR model with $K/S_0\in[0.1, 0.8]$. In summary, CU-RL performs better than all benchmark models in both profit-making and risk-control in both two short-term deep ITM call groups.
		
	The panel B of Table~\ref{table_real_call_deepin0} and Table~\ref{table_real_call_deepin1} show the results when transaction costs are considered: (i) CU-RL still obtains higher mean and lower risk measure than all benchmark models; (ii) CU-RL's tail risk is significantly lower than BS and LVF models in the two groups and SABR model in the group with $K/S_0\in[0.8, 0.9]$, ans is lower (but not significant) than the SABR model in the group with $K/S_0\in[0.8, 0.9]$.

	\subsection{Performance of Hedging All Put Options by a Unified Model}\label{subsec:cu_rl_for_all_put_options}
	We train a unified model for all put options with any maturity and moneyness. The numbers of training, validation, and test paths in different ranges of maturity and moneyness is shown in Table~\ref{table_real_put_all_data}. Overall, we have 483564 training paths, 125772 validation paths, and 150037 test paths. 
	\begin{table}[!htbp]
		\centering
		\caption{\textbf{The number of paths of put options in different maturity ranges and different moneyness ranges.} The maximum maturity of options in the training, validation, and test data are 550 days, 88 days, and 88 days, respectively. The moneyness ranges of options in the training, validation, and test data are $[0.1059, 2.2423]$, $[0.1201, 1.3490]$, and $[0.03855, 1.3839]$, respectively.}
		\resizebox{\textwidth}{!}{
		\begin{tabular}{cccccccc}
			\toprule
			Maturity range &  Training Set & Valid set &Test Set & Moneyness range &  Training Set & Valid set & Test Set \\ 
			\cmidrule(lr){2-4}\cmidrule(lr){6-8}
			$(0, 7]$ & 74943 & 22439 & 25726 & $(0.1, 0.5]$ &5004 & 3449 & 4005 \\
			$(7, 14]$ & 76774 & 21236 & 26327 & $(0.5, 0.8]$ & 69732 & 21429 & 25725 \\
			$(14, 30]  $ & 153615 & 40687 & 49900 & $(0.8, 0.9]$ & 103702 & 28651 & 35149 \\
			$(30, 60]$ & 131998 & 34878 & 40951 & $(0.9, 1.1] $ & 280256 & 70787 & 83136 \\
			$(60, 90]$ &  38877 & 6532 & 7133 & $(1.1, 1.5]$ & 24718 & 1456 & 1887 \\
			$(90, 550]$ & 7357 & 0 & 0 & $(1.5, 2.3]$ & 152 & 0 & 0 \\
			\bottomrule
		\end{tabular}}
		\label{table_real_put_all_data}%
	\end{table}%
	
	In the implementation of CU-RL method, 
	the three networks are specified the same as those in the all call option case.
	In Algorithm~\ref{algorithm_ppo_zeta}, the size of buffer $\mathcal{D}_k$ is 59990, the minibatch size $n=30000$, and the initial value of the linearly decaying learning rate is 1e-4 for the non-transaction cost case and 5e-4 for the 0.1\% proportional transaction cost case.
	
	The out-of-sample  performance of CU-RL in both non-transaction cost and proportional transaction cost cases on the test data is shown in Table~\ref{table_real_put_all_asym}.
	
	\begin{table}[!htbp]
		\centering
		\caption{\textbf{The mean, standard error, and tail risk of the final P\&L of hedging all put options by the CU-RL method, the Black-Scholes, local volatility function, and SABR delta hedging method calculated from an out-of-sample test set.} The test set include all S\&P 500 put options traded on or after 10/01/2017 and expired before 01/02/2018. The p-values are calculated from a one-sided t-test for related samples that tests if the mean P\&L of our method is higher than that of the benchmark method. 0.95-CI means 95\% confidence interval. Scientific notation: $\text{1.23E-4} = 1.23\times 10^{-4}$. $\text{***}:p < 0.001$; $\text{**}:p < 0.01$; $*:p < 0.05$.}
\resizebox{\linewidth}{!}{
	\begin{tabular}{lllll}
		\toprule
		& \multicolumn{1}{r}{CU-RL} & \multicolumn{3}{c}{Traditional Model} \\ 
		\cmidrule(lr){3-5}
		& & \multicolumn{1}{r}{BS Model} & \multicolumn{1}{r}{LVF Model} & \multicolumn{1}{r}{SABR Model} \\ 
		\midrule 
		\multicolumn{5}{c}{Panel A: without transaction cost} \\ 
		\midrule
		 Mean & \multicolumn{1}{r}{3.3132}& \multicolumn{1}{r}{1.7698}& \multicolumn{1}{r}{1.0325}& \multicolumn{1}{r}{2.5043}\\
 Std Err & \multicolumn{1}{r}{1.2835E-02}& \multicolumn{1}{r}{6.6780E-03}& \multicolumn{1}{r}{6.4837E-03}& \multicolumn{1}{r}{9.9172E-03}\\
 P-Value & \multicolumn{1}{r}{}& \multicolumn{1}{r}{0.0000E+00$^{***}$}& \multicolumn{1}{r}{0.0000E+00$^{***}$}& \multicolumn{1}{r}{0.0000E+00$^{***}$}\\
 0.95-VaR & \multicolumn{1}{r}{1.8135E-02}& \multicolumn{1}{r}{-1.4677E-02}& \multicolumn{1}{r}{1.0552}& \multicolumn{1}{r}{-2.4172E-02}\\
 0.95-CI of 0.95-VaR & \multicolumn{1}{r}{[1.5388E-02, 2.2266E-02]} & \multicolumn{1}{r}{[-1.5031E-02, -1.4325E-02]} & \multicolumn{1}{r}{[1.0077, 1.1090]} & \multicolumn{1}{r}{[-2.4224E-02, -2.4111E-02]} \\
 0.975-VaR & \multicolumn{1}{r}{0.1511}& \multicolumn{1}{r}{0.1280}& \multicolumn{1}{r}{2.9708}& \multicolumn{1}{r}{0.2625}\\
 0.95-CI of 0.975-VaR & \multicolumn{1}{r}{[0.1455, 0.1579]} & \multicolumn{1}{r}{[0.1032, 0.1551]} & \multicolumn{1}{r}{[2.8788, 3.0718]} & \multicolumn{1}{r}{[0.2065, 0.3192]} \\
 0.975-MS & \multicolumn{1}{r}{0.5738}& \multicolumn{1}{r}{1.1783}& \multicolumn{1}{r}{5.1877}& \multicolumn{1}{r}{2.3654}\\
 0.95-CI of 0.975-MS & \multicolumn{1}{r}{[0.3860, 0.7930]} & \multicolumn{1}{r}{[1.0880, 1.2697]} & \multicolumn{1}{r}{[5.0258, 5.3361]} & \multicolumn{1}{r}{[2.1944, 2.5420]} \\
 0.95-CVaR & \multicolumn{1}{r}{1.1961}& \multicolumn{1}{r}{0.9016}& \multicolumn{1}{r}{3.8585}& \multicolumn{1}{r}{1.8332}\\
 0.95-CI of 0.95-CVaR & \multicolumn{1}{r}{[1.1414, 1.2622]} & \multicolumn{1}{r}{[0.8550, 0.9415]} & \multicolumn{1}{r}{[3.7798, 3.9456]} & \multicolumn{1}{r}{[1.7549, 1.9240]} \\
 0.975-CVaR & \multicolumn{1}{r}{2.3126}& \multicolumn{1}{r}{1.7958}& \multicolumn{1}{r}{5.8443}& \multicolumn{1}{r}{3.6586}\\
 0.95-CI of 0.975-CVaR & \multicolumn{1}{r}{[2.1870, 2.4377]} & \multicolumn{1}{r}{[1.7203, 1.8731]} & \multicolumn{1}{r}{[5.7304, 5.9616]} & \multicolumn{1}{r}{[3.5077, 3.8106]}  \\
		\midrule 
		\multicolumn{5}{c}{Panel B: with $0.1\%$ transaction cost} \\ 
		\midrule
		 Mean & \multicolumn{1}{r}{3.2330}& \multicolumn{1}{r}{0.6398}& \multicolumn{1}{r}{-0.2402}& \multicolumn{1}{r}{1.3500}\\
 Std Err & \multicolumn{1}{r}{1.2606E-02}& \multicolumn{1}{r}{4.7636E-03}& \multicolumn{1}{r}{6.3944E-03}& \multicolumn{1}{r}{9.0985E-03}\\
 P-Value & \multicolumn{1}{r}{}& \multicolumn{1}{r}{0.0000E+00$^{***}$}& \multicolumn{1}{r}{0.0000E+00$^{***}$}& \multicolumn{1}{r}{0.0000E+00$^{***}$}\\
 0.95-VaR & \multicolumn{1}{r}{-2.4865E-02}& \multicolumn{1}{r}{1.4115}& \multicolumn{1}{r}{4.4806}& \multicolumn{1}{r}{1.6581}\\
 0.95-CI of 0.95-VaR & \multicolumn{1}{r}{[-2.4877E-02, -2.4851E-02]} & \multicolumn{1}{r}{[1.3590, 1.4621]} & \multicolumn{1}{r}{[4.3757, 4.5848]} & \multicolumn{1}{r}{[1.5897, 1.7050]} \\
 0.975-VaR & \multicolumn{1}{r}{0.6123}& \multicolumn{1}{r}{3.0216}& \multicolumn{1}{r}{7.8318}& \multicolumn{1}{r}{5.1990}\\
 0.95-CI of 0.975-VaR & \multicolumn{1}{r}{[0.5303, 0.6885]} & \multicolumn{1}{r}{[2.9267, 3.1172]} & \multicolumn{1}{r}{[7.6875, 7.9599]} & \multicolumn{1}{r}{[5.0026, 5.3900]} \\
 0.975-MS & \multicolumn{1}{r}{2.1933}& \multicolumn{1}{r}{5.2327}& \multicolumn{1}{r}{10.6417}& \multicolumn{1}{r}{9.4210}\\
 0.95-CI of 0.975-MS & \multicolumn{1}{r}{[2.0984, 2.3027]} & \multicolumn{1}{r}{[5.0877, 5.3966]} & \multicolumn{1}{r}{[10.4548, 10.7951]} & \multicolumn{1}{r}{[9.1458, 9.7027]} \\
 0.95-CVaR & \multicolumn{1}{r}{1.8365}& \multicolumn{1}{r}{3.9472}& \multicolumn{1}{r}{8.6063}& \multicolumn{1}{r}{6.8241}\\
 0.95-CI of 0.95-CVaR & \multicolumn{1}{r}{[1.7487, 1.9270]} & \multicolumn{1}{r}{[3.8594, 4.0215]} & \multicolumn{1}{r}{[8.4893, 8.7219]} & \multicolumn{1}{r}{[6.6475, 6.9837]} \\
 0.975-CVaR & \multicolumn{1}{r}{3.6154}& \multicolumn{1}{r}{5.8264}& \multicolumn{1}{r}{11.2287}& \multicolumn{1}{r}{10.7077}\\
 0.95-CI of 0.975-CVaR & \multicolumn{1}{r}{[3.4637, 3.7859]} & \multicolumn{1}{r}{[5.7068, 5.9436]} & \multicolumn{1}{r}{[11.0766, 11.3523]} & \multicolumn{1}{r}{[10.4892, 10.9313]}  \\
		\bottomrule 
		\end{tabular}} 
		\label{table_real_put_all_asym} 
	\end{table} 
	
	The panel A of Table~\ref{table_real_put_all_asym} shows that: (i) CU-RL obtains the significantly highest mean final P\&L among  all the methods; (ii) CU-RL obtains the second lowest CVaR at 0.95 and 0.975 level and VaR at 0.975 level, and obtains the third lowest VaR at 0.95 level; 
	(iii) CU-RL obtains the significantly lowest MS at 0.975 level among all the methods.
	
	The panel B of Table~\ref{table_real_put_all_asym} shows the results when transaction costs are considered. CU-RL obtains significantly higher mean and lower risk than all the benchmark models, suggesting it can efficiently adapt to real market settings like transaction costs.		
	
	\subsection{Performance of Hedging Short-Term Near-the-Money Put Options by a Unified Model}\label{subsec:cu_rl_for_short_term_put_options}
	We train a unified model for all the short-term put options with maturity $T \in [5,30]$ and moneyness $K/S_{0} \in [0.9, 1.1]$. The distribution of data is shown in Table~\ref{table_real_put_short_data}. We have 153433 training paths, 36011 validation paths, and 43560 test paths in total.  
	\begin{table}[!htbp]
		\centering
		\caption{\textbf{The number of paths of short-term close-to-the-money put options ($T \in [5,30], K/S_{0} \in [0.9, 1.1]$) in different maturity ranges and moneyness ranges.}}
		\resizebox{\textwidth}{!}{
		\begin{tabular}{cccccccc}
			\toprule
			Maturity range &  Training set & Valid set & Test set & Moneyness range &  Training set & Valid set & Test set \\ 
			\cmidrule(lr){2-4}\cmidrule(lr){6-8}
			$[5, 15]$ & 57322 & 14031 & 16927 & $[0.9, 0.95]$ & 49857 & 12756 & 14791 \\
			$[16, 21]$ & 39671 & 9168 & 11634 & $(0.95, 1.0]$ & 47056 & 13790 & 15410 \\
			$[22, 30]$ & 56440 & 12812 & 14999 & $(1.0, 1.05]$ & 39701 & 7801 & 10436 \\
			- & - & - & - & $(1.05, 1.1]$ & 16819 & 1664 & 2923 \\
			\bottomrule
		\end{tabular}}
		\label{table_real_put_short_data}
	\end{table} 

	In the implementation of CU-RL method for short-term put options, 
	the three networks are specified the same as those in the short-term call option case.
	In Algorithm~\ref{algorithm_ppo_zeta}, the size of buffer $\mathcal{D}_k$ is 29988, the minibatch size $n=20000$, and the initial value of the linearly decaying learning rate is 1e-7.
	The out-of-sample performance of CU-RL on the test data in both non-transaction cost and proportional transaction cost cases  is shown in Table~\ref{table_real_put_short_asym}.

	\begin{table}[!htbp]
		\centering
		\caption{\textbf{The mean, standard error, and tail risk of the final P\&L of hedging short-term put options by the CU-RL method, the Black-Scholes, local volatility function, and SABR delta hedging method calculated from an out-of-sample test set.} The test set include all S\&P 500 short-term put options with $T \in [5,30], K/S_{0} \in[0.9,1.1]$ traded on or after 10/01/2017 and expired before 01/02/2018. The p-values are calculated from a one-sided t-test for related samples that tests if the mean P\&L of our method is higher than that of the benchmark method. 0.95-CI means 95\% confidence interval. Scientific notation: $\text{1.23E-4} = 1.23\times 10^{-4}$. $\text{***}:p < 0.001$; $\text{**}:p < 0.01$; $*:p < 0.05$.}
\resizebox{\linewidth}{!}{
	\begin{tabular}{lllll}
		\toprule
		& \multicolumn{1}{r}{CU-RL} & \multicolumn{3}{c}{Traditional Model} \\ 
		\cmidrule(lr){3-5}
		& & \multicolumn{1}{r}{BS Model} & \multicolumn{1}{r}{LVF Model} & \multicolumn{1}{r}{SABR Model} \\ 
		\midrule 
		\multicolumn{5}{c}{Panel A: without transaction cost} \\ 
		\midrule
		 Mean & \multicolumn{1}{r}{3.6391}& \multicolumn{1}{r}{2.1169}& \multicolumn{1}{r}{1.2713}& \multicolumn{1}{r}{2.8813}\\
 Std Err & \multicolumn{1}{r}{1.4954E-02}& \multicolumn{1}{r}{1.1516E-02}& \multicolumn{1}{r}{1.3597E-02}& \multicolumn{1}{r}{1.5848E-02}\\
 P-Value & \multicolumn{1}{r}{}& \multicolumn{1}{r}{0.0000E+00$^{***}$}& \multicolumn{1}{r}{0.0000E+00$^{***}$}& \multicolumn{1}{r}{0.0000E+00$^{***}$}\\
 0.95-VaR & \multicolumn{1}{r}{-0.3462}& \multicolumn{1}{r}{0.2967}& \multicolumn{1}{r}{2.8337}& \multicolumn{1}{r}{0.1565}\\
 0.95-CI of 0.95-VaR & \multicolumn{1}{r}{[-0.3491, -0.3422]} & \multicolumn{1}{r}{[0.2501, 0.3555]} & \multicolumn{1}{r}{[2.7215, 2.9420]} & \multicolumn{1}{r}{[8.2989E-02, 0.2311]} \\
 0.975-VaR & \multicolumn{1}{r}{-0.2232}& \multicolumn{1}{r}{1.6854}& \multicolumn{1}{r}{5.0178}& \multicolumn{1}{r}{2.4963}\\
 0.95-CI of 0.975-VaR & \multicolumn{1}{r}{[-0.2381, -0.2108]} & \multicolumn{1}{r}{[1.5671, 1.8278]} & \multicolumn{1}{r}{[4.7887, 5.2264]} & \multicolumn{1}{r}{[2.2016, 2.7177]} \\
 0.975-MS & \multicolumn{1}{r}{-3.0624E-02}& \multicolumn{1}{r}{3.0683}& \multicolumn{1}{r}{7.4019}& \multicolumn{1}{r}{5.8128}\\
 0.95-CI of 0.975-MS & \multicolumn{1}{r}{[-0.1380, 0.2908]} & \multicolumn{1}{r}{[2.9316, 3.2132]} & \multicolumn{1}{r}{[7.1735, 7.6450]} & \multicolumn{1}{r}{[5.5032, 6.1951]} \\
 0.95-CVaR & \multicolumn{1}{r}{0.4815}& \multicolumn{1}{r}{2.2504}& \multicolumn{1}{r}{5.7857}& \multicolumn{1}{r}{3.9512}\\
 0.95-CI of 0.95-CVaR & \multicolumn{1}{r}{[0.4086, 0.5624]} & \multicolumn{1}{r}{[2.1449, 2.3651]} & \multicolumn{1}{r}{[5.6220, 5.9502]} & \multicolumn{1}{r}{[3.7296, 4.1748]} \\
 0.975-CVaR & \multicolumn{1}{r}{1.2492}& \multicolumn{1}{r}{3.6460}& \multicolumn{1}{r}{7.8232}& \multicolumn{1}{r}{6.8865}\\
 0.95-CI of 0.975-CVaR & \multicolumn{1}{r}{[1.1070, 1.4046]} & \multicolumn{1}{r}{[3.4992, 3.8027]} & \multicolumn{1}{r}{[7.6186, 8.0282]} & \multicolumn{1}{r}{[6.5428, 7.2275]}  \\
		\midrule 
		\multicolumn{5}{c}{Panel B: with $0.1\%$ transaction cost} \\ 
		\midrule
		 Mean & \multicolumn{1}{r}{2.4949}& \multicolumn{1}{r}{0.4693}& \multicolumn{1}{r}{-0.5903}& \multicolumn{1}{r}{1.1193}\\
 Std Err & \multicolumn{1}{r}{1.4396E-02}& \multicolumn{1}{r}{1.0408E-02}& \multicolumn{1}{r}{1.4632E-02}& \multicolumn{1}{r}{1.7650E-02}\\
 P-Value & \multicolumn{1}{r}{}& \multicolumn{1}{r}{0.0000E+00$^{***}$}& \multicolumn{1}{r}{0.0000E+00$^{***}$}& \multicolumn{1}{r}{0.0000E+00$^{***}$}\\
 0.95-VaR & \multicolumn{1}{r}{0.6014}& \multicolumn{1}{r}{3.3338}& \multicolumn{1}{r}{7.0026}& \multicolumn{1}{r}{4.8350}\\
 0.95-CI of 0.95-VaR & \multicolumn{1}{r}{[0.5210, 0.6527]} & \multicolumn{1}{r}{[3.1943, 3.4723]} & \multicolumn{1}{r}{[6.8557, 7.1900]} & \multicolumn{1}{r}{[4.6088, 5.0868]} \\
 0.975-VaR & \multicolumn{1}{r}{2.3465}& \multicolumn{1}{r}{5.6212}& \multicolumn{1}{r}{9.7243}& \multicolumn{1}{r}{8.8841}\\
 0.95-CI of 0.975-VaR & \multicolumn{1}{r}{[2.0971, 2.5911]} & \multicolumn{1}{r}{[5.4690, 5.8033]} & \multicolumn{1}{r}{[9.4652, 9.9380]} & \multicolumn{1}{r}{[8.5287, 9.1957]} \\
 0.975-MS & \multicolumn{1}{r}{5.2083}& \multicolumn{1}{r}{7.6481}& \multicolumn{1}{r}{12.2785}& \multicolumn{1}{r}{13.3470}\\
 0.95-CI of 0.975-MS & \multicolumn{1}{r}{[4.9034, 5.5251]} & \multicolumn{1}{r}{[7.3692, 7.9166]} & \multicolumn{1}{r}{[12.0182, 12.5933]} & \multicolumn{1}{r}{[12.8422, 13.7619]} \\
 0.95-CVaR & \multicolumn{1}{r}{3.5991}& \multicolumn{1}{r}{6.2739}& \multicolumn{1}{r}{10.5583}& \multicolumn{1}{r}{10.5216}\\
 0.95-CI of 0.95-CVaR & \multicolumn{1}{r}{[3.4086, 3.7708]} & \multicolumn{1}{r}{[6.1303, 6.4243]} & \multicolumn{1}{r}{[10.3660, 10.7509]} & \multicolumn{1}{r}{[10.2091, 10.8206]} \\
 0.975-CVaR & \multicolumn{1}{r}{5.9896}& \multicolumn{1}{r}{8.1932}& \multicolumn{1}{r}{12.9010}& \multicolumn{1}{r}{14.4257}\\
 0.95-CI of 0.975-CVaR & \multicolumn{1}{r}{[5.7212, 6.2719]} & \multicolumn{1}{r}{[8.0094, 8.3915]} & \multicolumn{1}{r}{[12.6551, 13.1424]} & \multicolumn{1}{r}{[14.0018, 14.8322]}  \\
		\bottomrule 
\end{tabular}} 
		\label{table_real_put_short_asym} 
	\end{table} 
	
	The panel A of Table~\ref{table_real_put_short_asym} shows that: (i) CU-RL obtains the significantly highest mean final P\&L among all the methods; (ii) CU-RL obtains the lowest CVaR at 0.975 level, the designated risk measure in the total reward of CU-RL, the lowest MS at 0.975 level, and the lowest VaR at 0.975 level; (iii) CU-RL obtains the lowest VaR and CVaR at 0.95 level, suggesting it can effectively minimize the tail risk measured at an alternative level; (iv) all the results with respect to tail risk measures are statistically significant, since all the confidence intervals of CU-RL are lower and have no overlap with those of other benchmark models.
	
	The panel B of Table~\ref{table_real_put_short_asym} shows that similar conclusion holds when transaction costs are considered. Although the mean of final P\&L falls and the risk measure rises due to transaction costs, CU-RL still obtains the significantly highest mean P\&L and lowest tail risk among all the models.		
	
	\subsection{Performance of Hedging Short-Term Deep OTM Put Options by a Unified Model}\label{subsec:cu_rl_for_short_term_deep_otm_put_options}
	We train two unified models for two groups of short-term deep OTM put options, one for options with $K/S_{0} \in (0.1, 0.8]$  and $T \in [5, 30]$, and another for options with $K/S_{0} \in (0.8, 0.9]$ and $T \in [5, 30]$. The number of paths in these two groups distributed in each range of maturity is listed in Table~\ref{table_real_put_deepout_data}.
	\begin{table}[!htbp]
		\centering
		\caption{\textbf{The number of paths of short-term deep OTM put options ($T \in [5, 30], \frac{K}{S_{0}} \in [0.1, 0.9]$) in different maturity ranges and moneyness ranges.}}
		\begin{tabular}{ccccccc}
			\toprule
			Maturity range &\multicolumn{3}{c}{$K/S_{0} \in (0.1, 0.8]$} &\multicolumn{3}{c}{$K/S_{0} \in (0.8, 0.9]$} \\
			\cmidrule(lr){2-4}\cmidrule(lr){5-7}
			 & Training set  & Validation set & Test set & Training set & Validation set & Test set  \\
			\midrule
			$ [5, 15]$ & 12992 & 4662 & 5997 & 30464 & 6691 & 8224 \\
			$ (15, 21]$ & 6317 & 2842 & 3573 & 13345 & 3597 & 4701 \\
			$(21, 30]$ & 9481 & 3956 & 4582 & 18086 & 4634 & 5750 \\
			Total & 28790 & 11460 & 10830 & 61895 & 14922 & 18675 \\
			\bottomrule
		\end{tabular}
		\label{table_real_put_deepout_data}
	\end{table}
	For other groups of put options such as deep ITM put options, we do not train a separate model for them due to the small number of sample paths for these groups of options. 

	In the implementation of CU-RL method for short-term deep OTM put options, 
	for the group of options with $K/S_{0} \in (0.1, 0.8], T \in [5, 30]$, the number of hidden layers is 3 and the number of neurons in each layer is 64; for the group of options with $K/S_{0} \in (0.8, 0.9], T \in [5, 30]$, the number of hidden layers is 9 and the number of neurons in each layer is 32. 
	In Algorithm~\ref{algorithm_ppo_zeta}, the size of buffer $\mathcal{D}_k$ is 29988. The minibatch size $n=10000$ for the group of options with $K/S_{0} \in (0.1, 0.8], T \in [5, 30]$ and $n=20000$ for the group of options with $K/S_{0} \in (0.8, 0.9], T \in [5, 30]$.
	The initial value of the linearly decaying learning rate is 1e-5 and 1e-6 for the two groups, respectively.
	The out-of-sample performance of CU-RL on the test data for these two groups in both non-transaction cost and proportional transaction cost cases is shown in Table~\ref{table_real_put_deepout0} and Table~\ref{table_real_put_deepout1}.
	
	\begin{table}[!htbp]
		\centering
		\caption{\textbf{The mean, standard error, and tail risk of the final P\&L of hedging short-term deep OTM put options by the CU-RL method, the Black-Scholes, local volatility function, and SABR delta hedging method calculated from an out-of-sample test set.} The test set include all S\&P 500 short-term deep OTM put options with $K/S_{0} \in (0.1, 0.8], T \in [5, 30]$ traded on or after 10/01/2017 and expired before 01/02/2018. The p-values are calculated from a one-sided t-test for related samples that tests if the mean P\&L of our method is higher than that of the benchmark method. 0.95-CI means 95\% confidence interval. Scientific notation: $\text{1.23E-4} = 1.23\times 10^{-4}$. $\text{***}:p < 0.001$; $\text{**}:p < 0.01$; $*:p < 0.05$.}
\resizebox{\linewidth}{!}{
	\begin{tabular}{lllll}
		\toprule
		& \multicolumn{1}{r}{CU-RL} & \multicolumn{3}{c}{Traditional Model} \\ 
		\cmidrule(lr){3-5}
		& & \multicolumn{1}{r}{BS Model} & \multicolumn{1}{r}{LVF Model} & \multicolumn{1}{r}{SABR Model} \\ 
		\midrule 
		\multicolumn{5}{c}{Panel A: without transaction cost} \\ 
		\midrule
		 Mean & \multicolumn{1}{r}{8.5582E-02}& \multicolumn{1}{r}{7.2430E-02}& \multicolumn{1}{r}{4.6072E-02}& \multicolumn{1}{r}{8.9578E-02}\\
 Std Err & \multicolumn{1}{r}{7.6920E-04}& \multicolumn{1}{r}{6.6749E-04}& \multicolumn{1}{r}{5.5973E-04}& \multicolumn{1}{r}{8.1035E-04}\\
 P-Value & \multicolumn{1}{r}{}& \multicolumn{1}{r}{0.0000E+00$^{***}$}& \multicolumn{1}{r}{0.0000E+00$^{***}$}& \multicolumn{1}{r}{1.0000$^{}$}\\
 0.95-VaR & \multicolumn{1}{r}{-1.8503E-02}& \multicolumn{1}{r}{-1.3678E-02}& \multicolumn{1}{r}{1.7931E-02}& \multicolumn{1}{r}{-2.3806E-02}\\
 0.95-CI of 0.95-VaR & \multicolumn{1}{r}{[-1.9034E-02, -1.8108E-02]} & \multicolumn{1}{r}{[-1.4009E-02, -1.3326E-02]} & \multicolumn{1}{r}{[1.7032E-02, 1.8527E-02]} & \multicolumn{1}{r}{[-2.3857E-02, -2.3760E-02]} \\
 0.975-VaR & \multicolumn{1}{r}{-1.3243E-02}& \multicolumn{1}{r}{-1.0464E-02}& \multicolumn{1}{r}{2.5295E-02}& \multicolumn{1}{r}{-2.3239E-02}\\
 0.95-CI of 0.975-VaR & \multicolumn{1}{r}{[-1.4181E-02, -1.1892E-02]} & \multicolumn{1}{r}{[-1.0937E-02, -9.9875E-03]} & \multicolumn{1}{r}{[2.4158E-02, 2.6366E-02]} & \multicolumn{1}{r}{[-2.3329E-02, -2.3132E-02]} \\
 0.975-MS & \multicolumn{1}{r}{-3.5757E-03}& \multicolumn{1}{r}{-7.4580E-03}& \multicolumn{1}{r}{3.3576E-02}& \multicolumn{1}{r}{-2.2592E-02}\\
 0.95-CI of 0.975-MS & \multicolumn{1}{r}{[-5.4143E-03, -2.0991E-03]} & \multicolumn{1}{r}{[-7.9426E-03, -6.5728E-03]} & \multicolumn{1}{r}{[3.1568E-02, 3.6246E-02]} & \multicolumn{1}{r}{[-2.2753E-02, -2.2429E-02]} \\
 0.95-CVaR & \multicolumn{1}{r}{-8.6098E-03}& \multicolumn{1}{r}{-8.9487E-03}& \multicolumn{1}{r}{3.0199E-02}& \multicolumn{1}{r}{-2.2796E-02}\\
 0.95-CI of 0.95-CVaR & \multicolumn{1}{r}{[-9.8438E-03, -6.6239E-03]} & \multicolumn{1}{r}{[-9.4146E-03, -8.5024E-03]} & \multicolumn{1}{r}{[2.8950E-02, 3.1760E-02]} & \multicolumn{1}{r}{[-2.2916E-02, -2.2669E-02]} \\
 0.975-CVaR & \multicolumn{1}{r}{-8.7025E-04}& \multicolumn{1}{r}{-5.7024E-03}& \multicolumn{1}{r}{3.9368E-02}& \multicolumn{1}{r}{-2.2035E-02}\\
 0.95-CI of 0.975-CVaR & \multicolumn{1}{r}{[-3.0705E-03, 3.2885E-03]} & \multicolumn{1}{r}{[-6.3221E-03, -4.9219E-03]} & \multicolumn{1}{r}{[3.7535E-02, 4.1911E-02]} & \multicolumn{1}{r}{[-2.2214E-02, -2.1780E-02]}  \\
		\midrule 
		\multicolumn{5}{c}{Panel B: with $0.1\%$ transaction cost} \\ 
		\midrule
		 Mean & \multicolumn{1}{r}{8.9664E-02}& \multicolumn{1}{r}{6.6259E-02}& \multicolumn{1}{r}{3.1905E-02}& \multicolumn{1}{r}{8.7802E-02}\\
 Std Err & \multicolumn{1}{r}{8.3541E-04}& \multicolumn{1}{r}{6.0998E-04}& \multicolumn{1}{r}{4.7966E-04}& \multicolumn{1}{r}{7.9536E-04}\\
 P-Value & \multicolumn{1}{r}{}& \multicolumn{1}{r}{0.0000E+00$^{***}$}& \multicolumn{1}{r}{0.0000E+00$^{***}$}& \multicolumn{1}{r}{5.1955E-264$^{***}$}\\
 0.95-VaR & \multicolumn{1}{r}{-2.0142E-02}& \multicolumn{1}{r}{-1.2569E-02}& \multicolumn{1}{r}{2.8042E-02}& \multicolumn{1}{r}{-2.3095E-02}\\
 0.95-CI of 0.95-VaR & \multicolumn{1}{r}{[-2.0407E-02, -1.9873E-02]} & \multicolumn{1}{r}{[-1.2889E-02, -1.2190E-02]} & \multicolumn{1}{r}{[2.6859E-02, 2.9196E-02]} & \multicolumn{1}{r}{[-2.3161E-02, -2.3009E-02]} \\
 0.975-VaR & \multicolumn{1}{r}{-1.7304E-02}& \multicolumn{1}{r}{-8.9517E-03}& \multicolumn{1}{r}{3.9573E-02}& \multicolumn{1}{r}{-2.2343E-02}\\
 0.95-CI of 0.975-VaR & \multicolumn{1}{r}{[-1.7880E-02, -1.6700E-02]} & \multicolumn{1}{r}{[-9.3516E-03, -8.2934E-03]} & \multicolumn{1}{r}{[3.8163E-02, 4.1121E-02]} & \multicolumn{1}{r}{[-2.2465E-02, -2.2228E-02]} \\
 0.975-MS & \multicolumn{1}{r}{-1.4809E-02}& \multicolumn{1}{r}{-5.3823E-03}& \multicolumn{1}{r}{5.1441E-02}& \multicolumn{1}{r}{-2.1531E-02}\\
 0.95-CI of 0.975-MS & \multicolumn{1}{r}{[-1.5073E-02, -1.4303E-02]} & \multicolumn{1}{r}{[-6.1535E-03, -4.4011E-03]} & \multicolumn{1}{r}{[4.9155E-02, 5.4390E-02]} & \multicolumn{1}{r}{[-2.1692E-02, -2.1285E-02]} \\
 0.95-CVaR & \multicolumn{1}{r}{-1.6585E-02}& \multicolumn{1}{r}{-7.1078E-03}& \multicolumn{1}{r}{4.7055E-02}& \multicolumn{1}{r}{-2.1600E-02}\\
 0.95-CI of 0.95-CVaR & \multicolumn{1}{r}{[-1.6914E-02, -1.6237E-02]} & \multicolumn{1}{r}{[-7.6653E-03, -6.5500E-03]} & \multicolumn{1}{r}{[4.5079E-02, 4.9624E-02]} & \multicolumn{1}{r}{[-2.1800E-02, -2.1369E-02]} \\
 0.975-CVaR & \multicolumn{1}{r}{-1.4122E-02}& \multicolumn{1}{r}{-3.2974E-03}& \multicolumn{1}{r}{6.0934E-02}& \multicolumn{1}{r}{-2.0440E-02}\\
 0.95-CI of 0.975-CVaR & \multicolumn{1}{r}{[-1.4509E-02, -1.3661E-02]} & \multicolumn{1}{r}{[-4.0675E-03, -2.4065E-03]} & \multicolumn{1}{r}{[5.7952E-02, 6.4633E-02]} & \multicolumn{1}{r}{[-2.0781E-02, -1.9998E-02]}  \\
		\bottomrule 
		\end{tabular}} 
		\label{table_real_put_deepout0} 
	\end{table}

	\begin{table}[!htbp]
		\centering
		\caption{\textbf{The mean, standard error, and tail risk of the final P\&L of hedging short-term deep OTM put options by the CU-RL method, the Black-Scholes, local volatility function, and SABR delta hedging method calculated from an out-of-sample test set.} The test set include all S\&P 500 short-term deep OTM put options with $K/S_{0} \in (0.8, 0.9], T \in [5, 30]$ traded on or after 10/01/2017 and expired before 01/02/2018. The p-values are calculated from a one-sided t-test for related samples that tests if the mean P\&L of our method is higher than that of the benchmark method. 0.95-CI means 95\% confidence interval. Scientific notation: $\text{1.23E-4} = 1.23\times 10^{-4}$. $\text{***}:p < 0.001$; $\text{**}:p < 0.01$; $*:p < 0.05$.}
\resizebox{\linewidth}{!}{
	\begin{tabular}{lllll}
		\toprule
		& \multicolumn{1}{r}{CU-RL} & \multicolumn{3}{c}{Traditional Model} \\ 
		\cmidrule(lr){3-5}
		& & \multicolumn{1}{r}{BS Model} & \multicolumn{1}{r}{LVF Model} & \multicolumn{1}{r}{SABR Model} \\ 
		\midrule 
		\multicolumn{5}{c}{Panel A: without transaction cost} \\ 
		\midrule
		 Mean & \multicolumn{1}{r}{0.4138}& \multicolumn{1}{r}{0.3049}& \multicolumn{1}{r}{0.1879}& \multicolumn{1}{r}{0.4132}\\
 Std Err & \multicolumn{1}{r}{2.6157E-03}& \multicolumn{1}{r}{1.8834E-03}& \multicolumn{1}{r}{1.5830E-03}& \multicolumn{1}{r}{2.5488E-03}\\
 P-Value & \multicolumn{1}{r}{}& \multicolumn{1}{r}{0.0000E+00$^{***}$}& \multicolumn{1}{r}{0.0000E+00$^{***}$}& \multicolumn{1}{r}{7.7142E-03$^{**}$}\\
 0.95-VaR & \multicolumn{1}{r}{-4.8771E-02}& \multicolumn{1}{r}{-3.8981E-02}& \multicolumn{1}{r}{6.0163E-02}& \multicolumn{1}{r}{-4.9792E-02}\\
 0.95-CI of 0.95-VaR & \multicolumn{1}{r}{[-5.0166E-02, -4.8090E-02]} & \multicolumn{1}{r}{[-4.0157E-02, -3.7644E-02]} & \multicolumn{1}{r}{[5.6466E-02, 6.3566E-02]} & \multicolumn{1}{r}{[-5.0382E-02, -4.9470E-02]} \\
 0.975-VaR & \multicolumn{1}{r}{-4.1106E-02}& \multicolumn{1}{r}{-2.2767E-02}& \multicolumn{1}{r}{0.1024}& \multicolumn{1}{r}{-4.4632E-02}\\
 0.95-CI of 0.975-VaR & \multicolumn{1}{r}{[-4.1961E-02, -3.9940E-02]} & \multicolumn{1}{r}{[-2.4812E-02, -2.1465E-02]} & \multicolumn{1}{r}{[9.6918E-02, 0.1083]} & \multicolumn{1}{r}{[-4.5362E-02, -4.3686E-02]} \\
 0.975-MS & \multicolumn{1}{r}{-2.8747E-02}& \multicolumn{1}{r}{-1.0257E-02}& \multicolumn{1}{r}{0.1574}& \multicolumn{1}{r}{-2.8819E-02}\\
 0.95-CI of 0.975-MS & \multicolumn{1}{r}{[-3.0709E-02, -2.5949E-02]} & \multicolumn{1}{r}{[-1.3455E-02, -6.4147E-03]} & \multicolumn{1}{r}{[0.1490, 0.1759]} & \multicolumn{1}{r}{[-3.4076E-02, -2.6308E-02]} \\
 0.95-CVaR & \multicolumn{1}{r}{-3.6316E-02}& \multicolumn{1}{r}{-1.8801E-02}& \multicolumn{1}{r}{0.1465}& \multicolumn{1}{r}{-3.6995E-02}\\
 0.95-CI of 0.95-CVaR & \multicolumn{1}{r}{[-3.7417E-02, -3.5206E-02]} & \multicolumn{1}{r}{[-2.0315E-02, -1.7107E-02]} & \multicolumn{1}{r}{[0.1373, 0.1565]} & \multicolumn{1}{r}{[-3.8371E-02, -3.5450E-02]} \\
 0.975-CVaR & \multicolumn{1}{r}{-2.7437E-02}& \multicolumn{1}{r}{-5.8987E-03}& \multicolumn{1}{r}{0.2149}& \multicolumn{1}{r}{-2.6365E-02}\\
 0.95-CI of 0.975-CVaR & \multicolumn{1}{r}{[-2.8966E-02, -2.6000E-02]} & \multicolumn{1}{r}{[-7.8539E-03, -3.7092E-03]} & \multicolumn{1}{r}{[0.2001, 0.2310]} & \multicolumn{1}{r}{[-2.8647E-02, -2.3516E-02]}  \\
		\midrule 
		\multicolumn{5}{c}{Panel B: with $0.1\%$ transaction cost} \\ 
		\midrule
		 Mean & \multicolumn{1}{r}{0.4006}& \multicolumn{1}{r}{0.2559}& \multicolumn{1}{r}{0.1013}& \multicolumn{1}{r}{0.3916}\\
 Std Err & \multicolumn{1}{r}{2.6129E-03}& \multicolumn{1}{r}{1.6459E-03}& \multicolumn{1}{r}{1.3821E-03}& \multicolumn{1}{r}{2.4645E-03}\\
 P-Value & \multicolumn{1}{r}{}& \multicolumn{1}{r}{0.0000E+00$^{***}$}& \multicolumn{1}{r}{0.0000E+00$^{***}$}& \multicolumn{1}{r}{1.7061E-109$^{***}$}\\
 0.95-VaR & \multicolumn{1}{r}{-4.4230E-02}& \multicolumn{1}{r}{-2.6262E-02}& \multicolumn{1}{r}{0.1520}& \multicolumn{1}{r}{-4.5208E-02}\\
 0.95-CI of 0.95-VaR & \multicolumn{1}{r}{[-4.5117E-02, -4.2671E-02]} & \multicolumn{1}{r}{[-2.7969E-02, -2.4354E-02]} & \multicolumn{1}{r}{[0.1462, 0.1568]} & \multicolumn{1}{r}{[-4.6047E-02, -4.4291E-02]} \\
 0.975-VaR & \multicolumn{1}{r}{-3.1203E-02}& \multicolumn{1}{r}{-1.1862E-02}& \multicolumn{1}{r}{0.2186}& \multicolumn{1}{r}{-2.9030E-02}\\
 0.95-CI of 0.975-VaR & \multicolumn{1}{r}{[-3.3209E-02, -2.8028E-02]} & \multicolumn{1}{r}{[-1.3490E-02, -9.5395E-03]} & \multicolumn{1}{r}{[0.2063, 0.2286]} & \multicolumn{1}{r}{[-3.3061E-02, -2.5331E-02]} \\
 0.975-MS & \multicolumn{1}{r}{-1.7395E-02}& \multicolumn{1}{r}{5.4970E-03}& \multicolumn{1}{r}{0.3108}& \multicolumn{1}{r}{-1.5243E-02}\\
 0.95-CI of 0.975-MS & \multicolumn{1}{r}{[-1.8426E-02, -1.5570E-02]} & \multicolumn{1}{r}{[1.9718E-03, 8.5926E-03]} & \multicolumn{1}{r}{[0.2930, 0.3278]} & \multicolumn{1}{r}{[-1.8050E-02, -7.8001E-03]} \\
 0.95-CVaR & \multicolumn{1}{r}{-2.6874E-02}& \multicolumn{1}{r}{-4.4017E-03}& \multicolumn{1}{r}{0.2777}& \multicolumn{1}{r}{-1.3720E-02}\\
 0.95-CI of 0.95-CVaR & \multicolumn{1}{r}{[-2.8295E-02, -2.5459E-02]} & \multicolumn{1}{r}{[-6.0792E-03, -2.3641E-03]} & \multicolumn{1}{r}{[0.2652, 0.2912]} & \multicolumn{1}{r}{[-1.7014E-02, -9.8549E-03]} \\
 0.975-CVaR & \multicolumn{1}{r}{-1.5173E-02}& \multicolumn{1}{r}{1.0600E-02}& \multicolumn{1}{r}{0.3766}& \multicolumn{1}{r}{1.2644E-02}\\
 0.95-CI of 0.975-CVaR & \multicolumn{1}{r}{[-1.6950E-02, -1.3417E-02]} & \multicolumn{1}{r}{[7.9409E-03, 1.3339E-02]} & \multicolumn{1}{r}{[0.3583, 0.3977]} & \multicolumn{1}{r}{[6.9763E-03, 1.9468E-02]}  \\
		\bottomrule 
		\end{tabular}} 
		\label{table_real_put_deepout1} 
	\end{table} 
	
	For short-term deep OTM put groups with $K/S_{0} \in (0.1, 0.8]$ and $T \in [5, 30]$, the panel A of Table~\ref{table_real_put_deepout0} shows that the SABR model obtains the significantly highest mean and lowest risk of final P\&L when there is no transaction costs.
	The panel B of Table~\ref{table_real_put_deepout0} shows the results when transaction costs are considered: (i) CU-RL obtains the significantly highest mean P\&L; (ii) gaps between risk of CU-RL and SABR model has shrunk compared with the non-transaction costs case; (iii) CU-RL obtains significantly higher mean and lower risk of final P\&L than BS and LVF models.	
	
	For short-term deep OTM put groups with $K/S_{0} \in (0.8, 0.9]$ and $T \in [5, 30]$, the panel A of Table~\ref{table_real_put_deepout1} shows that: (i) CU-RL obtains the significantly highest mean final P\&L among all the models; (ii) CU-RL obtains the lowest CVaR at 0.975 level, which is the designate risk measure in the objective; (iii) CU-RL obtains the second lowest tail risk measured by other tail risk measures, and higher risk than those obtained by the SABR model. However, the confidence intervals of tail risk of SABR model overlap with those of the CU-RL method. 
	The panel B of Table~\ref{table_real_put_deepout1} shows the results when transaction costs are considered: (i) CU-RL obtains the significantly highest mean P\&L and lowest risk except 0.95-VaR among all the models; (ii) CU-RL obtains 
	the second lowest VaR at 0.95 level, which is higher than that obtained by the SABR model. However, the confidence interval of VaR at 0.95 level of SABR model overlap with that of the CU-RL method. 

	It is clear from Table~\ref{table_real_put_short_asym}, Table~\ref{table_real_put_deepout0}, and Table~\ref{table_real_put_deepout1} that CU-RL performs very well for all short-term put options, fairly well for short-term deep OTM put groups with $K/S_{0} \in (0.8, 0.9]$ and $T \in [5, 30]$, and not so well for short-term deep OTM put groups with $K/S_{0} \in (0.1, 0.8]$ and $T \in [5, 30]$. The number of training sample paths for the three groups of options are 153433, 61890, and 28790, respectively, which may explain the difference of the performance of the CU-RL method for the three groups of put options.

	\subsection{Performance of Zero Intermediate Reward}\label{subsec:cu_rl_zero_reward}
	We consider a different reward function defined in \eqref{reward_zero}, which we call the zero intermediate reward, to test whether the CU-RL method can still perform well without intermediate information about the hedging error before maturity. We look into the performance of four unified CU-RL models, which are for all short-term near-the-money call options with $T \in [5, 30]$ and $K/S_{0} \in [0.9, 1.1]$, all short-term near-the-money put options with $T \in [5, 30]$ and $K/S_{0} \in [0.9, 1.1]$,
	all call options, and all put options, 
	 respectively. 
	
	\begin{equation}
		\label{reward_zero}
		R_{t+1}=
		\begin{cases}
			0, & t=0, \cdots, T-2, \\
			-\lambda_{1}\left[\omega + \frac{1}{1-\alpha} \max (-W_{t+1}-\omega, 0)\right] + \lambda_{2} W_{t+1}, & t=T-1.
		\end{cases} 
	\end{equation} 
	
	For the short-term near-the-money call/put options, we utilize the same value of hyperparameters as those in the experiments with the asymmetric reward in \eqref{R_relu}, which is demonstrated in the previous sections. 
	The performance of the CU-RL method under zero intermediate reward for 
	the short-term near-the-money call (resp., put) options	
	is shown in Table~\ref{table_real_call_short_zero} (resp., Table~\ref{table_real_put_short_zero}). 
	
	\begin{table}[!htbp]
		\centering
		\caption{\textbf{The mean, standard error, and tail risk of the final P\&L of hedging short-term call options by the CU-RL method under zero intermediate reward, the Black-Scholes, local volatility function, and SABR delta hedging method calculated from an out-of-sample test set.} The test set include all S\&P 500 short-term call options with $T \in [5,30], K/S_{0} \in[0.9,1.1]$ traded on or after 10/01/2017 and expired before 01/02/2018. The p-values are calculated from a one-sided t-test for related samples that tests if the mean P\&L of our method is higher than that of the benchmark method. 0.95-CI means 95\% confidence interval. Scientific notation: $\text{1.23E-4} = 1.23\times 10^{-4}$. $\text{***}:p < 0.001$; $\text{**}:p < 0.01$; $*:p < 0.05$.}
\resizebox{\linewidth}{!}{
	\begin{tabular}{lllll}
		\toprule
		& \multicolumn{1}{r}{CU-RL} & \multicolumn{3}{c}{Traditional Model} \\ 
		\cmidrule(lr){3-5}
		& & \multicolumn{1}{r}{BS Model} & \multicolumn{1}{r}{LVF Model} & \multicolumn{1}{r}{SABR Model} \\ 
		\midrule 
		\multicolumn{5}{c}{Panel A: without transaction cost} \\ 
		\midrule
		 Mean & \multicolumn{1}{r}{2.2823}& \multicolumn{1}{r}{-0.3079}& \multicolumn{1}{r}{-1.0298}& \multicolumn{1}{r}{0.7770}\\
 Std Err & \multicolumn{1}{r}{2.1255E-02}& \multicolumn{1}{r}{1.1242E-02}& \multicolumn{1}{r}{1.4010E-02}& \multicolumn{1}{r}{1.8853E-02}\\
 P-Value & \multicolumn{1}{r}{}& \multicolumn{1}{r}{0.0000E+00$^{***}$}& \multicolumn{1}{r}{0.0000E+00$^{***}$}& \multicolumn{1}{r}{0.0000E+00$^{***}$}\\
 0.95-VaR & \multicolumn{1}{r}{1.5612}& \multicolumn{1}{r}{4.0753}& \multicolumn{1}{r}{5.9742}& \multicolumn{1}{r}{4.4919}\\
 0.95-CI of 0.95-VaR & \multicolumn{1}{r}{[1.5339, 1.5876]} & \multicolumn{1}{r}{[4.0187, 4.1357]} & \multicolumn{1}{r}{[5.8757, 6.0803]} & \multicolumn{1}{r}{[4.3136, 4.6473]} \\
 0.975-VaR & \multicolumn{1}{r}{2.0183}& \multicolumn{1}{r}{4.9829}& \multicolumn{1}{r}{7.5718}& \multicolumn{1}{r}{6.5233}\\
 0.95-CI of 0.975-VaR & \multicolumn{1}{r}{[1.9709, 2.0557]} & \multicolumn{1}{r}{[4.9061, 5.0849]} & \multicolumn{1}{r}{[7.4473, 7.6647]} & \multicolumn{1}{r}{[6.3532, 6.7066]} \\
 0.975-MS & \multicolumn{1}{r}{2.4958}& \multicolumn{1}{r}{6.5480}& \multicolumn{1}{r}{9.2724}& \multicolumn{1}{r}{8.8325}\\
 0.95-CI of 0.975-MS & \multicolumn{1}{r}{[2.4380, 2.5615]} & \multicolumn{1}{r}{[6.3322, 6.7633]} & \multicolumn{1}{r}{[9.0487, 9.5064]} & \multicolumn{1}{r}{[8.5744, 9.0669]} \\
 0.95-CVaR & \multicolumn{1}{r}{2.3147}& \multicolumn{1}{r}{5.6511}& \multicolumn{1}{r}{8.2637}& \multicolumn{1}{r}{7.7263}\\
 0.95-CI of 0.95-CVaR & \multicolumn{1}{r}{[2.2593, 2.3713]} & \multicolumn{1}{r}{[5.5673, 5.7563]} & \multicolumn{1}{r}{[8.1390, 8.3906]} & \multicolumn{1}{r}{[7.5117, 7.9199]} \\
 0.975-CVaR & \multicolumn{1}{r}{2.8765}& \multicolumn{1}{r}{6.8327}& \multicolumn{1}{r}{9.8537}& \multicolumn{1}{r}{10.0016}\\
 0.95-CI of 0.975-CVaR & \multicolumn{1}{r}{[2.8021, 2.9604]} & \multicolumn{1}{r}{[6.7043, 6.9722]} & \multicolumn{1}{r}{[9.6701, 10.0390]} & \multicolumn{1}{r}{[9.7116, 10.3233]}  \\
		\midrule 
		\multicolumn{5}{c}{Panel B: with $0.1\%$ transaction cost} \\ 
		\midrule
		 Mean & \multicolumn{1}{r}{0.6826}& \multicolumn{1}{r}{-2.3988}& \multicolumn{1}{r}{-3.2881}& \multicolumn{1}{r}{-2.1633}\\
 Std Err & \multicolumn{1}{r}{2.4415E-02}& \multicolumn{1}{r}{1.4498E-02}& \multicolumn{1}{r}{1.8475E-02}& \multicolumn{1}{r}{2.1290E-02}\\
 P-Value & \multicolumn{1}{r}{}& \multicolumn{1}{r}{0.0000E+00$^{***}$}& \multicolumn{1}{r}{0.0000E+00$^{***}$}& \multicolumn{1}{r}{0.0000E+00$^{***}$}\\
 0.95-VaR & \multicolumn{1}{r}{4.4408}& \multicolumn{1}{r}{7.3841}& \multicolumn{1}{r}{9.8077}& \multicolumn{1}{r}{9.5096}\\
 0.95-CI of 0.95-VaR & \multicolumn{1}{r}{[4.4080, 4.4744]} & \multicolumn{1}{r}{[7.3077, 7.4348]} & \multicolumn{1}{r}{[9.7235, 9.9045]} & \multicolumn{1}{r}{[9.3464, 9.6470]} \\
 0.975-VaR & \multicolumn{1}{r}{4.9933}& \multicolumn{1}{r}{8.3330}& \multicolumn{1}{r}{11.2466}& \multicolumn{1}{r}{12.4072}\\
 0.95-CI of 0.975-VaR & \multicolumn{1}{r}{[4.9482, 5.0347]} & \multicolumn{1}{r}{[8.2405, 8.4352]} & \multicolumn{1}{r}{[11.1512, 11.3675]} & \multicolumn{1}{r}{[12.1395, 12.5717]} \\
 0.975-MS & \multicolumn{1}{r}{5.5183}& \multicolumn{1}{r}{9.8437}& \multicolumn{1}{r}{13.0815}& \multicolumn{1}{r}{15.0653}\\
 0.95-CI of 0.975-MS & \multicolumn{1}{r}{[5.4365, 5.5752]} & \multicolumn{1}{r}{[9.6109, 10.0783]} & \multicolumn{1}{r}{[12.7899, 13.3427]} & \multicolumn{1}{r}{[14.7583, 15.4513]} \\
 0.95-CVaR & \multicolumn{1}{r}{5.3392}& \multicolumn{1}{r}{8.9888}& \multicolumn{1}{r}{12.0782}& \multicolumn{1}{r}{13.9373}\\
 0.95-CI of 0.95-CVaR & \multicolumn{1}{r}{[5.2765, 5.4039]} & \multicolumn{1}{r}{[8.9041, 9.0956]} & \multicolumn{1}{r}{[11.9504, 12.1997]} & \multicolumn{1}{r}{[13.6502, 14.1959]} \\
 0.975-CVaR & \multicolumn{1}{r}{5.9848}& \multicolumn{1}{r}{10.1843}& \multicolumn{1}{r}{13.6890}& \multicolumn{1}{r}{17.1279}\\
 0.95-CI of 0.975-CVaR & \multicolumn{1}{r}{[5.8917, 6.0812]} & \multicolumn{1}{r}{[10.0535, 10.3302]} & \multicolumn{1}{r}{[13.4952, 13.8857]} & \multicolumn{1}{r}{[16.7323, 17.5958]}  \\
		\bottomrule 
		\end{tabular}} 
		\label{table_real_call_short_zero} 
	\end{table}
	
	\begin{table}[!htbp]
		\centering
		\caption{\textbf{The mean, standard error, and tail risk of the final P\&L of hedging short-term put options by the CU-RL method under zero intermediate reward, the Black-Scholes, local volatility function, and SABR delta hedging method calculated from an out-of-sample test set.} The test set include all S\&P 500 short-term put options with $T \in [5,30], K/S_{0} \in[0.9,1.1]$ traded on or after 10/01/2017 and expired before 01/02/2018. The p-values are calculated from a one-sided t-test for related samples that tests if the mean P\&L of our method is higher than that of the benchmark method. 0.95-CI means 95\% confidence interval. Scientific notation: $\text{1.23E-4} = 1.23\times 10^{-4}$. $\text{***}:p < 0.001$; $\text{**}:p < 0.01$; $*:p < 0.05$.}
\resizebox{\linewidth}{!}{
	\begin{tabular}{lllll}
		\toprule
		& \multicolumn{1}{r}{CU-RL} & \multicolumn{3}{c}{Traditional Model} \\ 
		\cmidrule(lr){3-5}
		& & \multicolumn{1}{r}{BS Model} & \multicolumn{1}{r}{LVF Model} & \multicolumn{1}{r}{SABR Model} \\ 
		\midrule 
		\multicolumn{5}{c}{Panel A: without transaction cost} \\ 
		\midrule
		 Mean & \multicolumn{1}{r}{3.3823}& \multicolumn{1}{r}{2.1169}& \multicolumn{1}{r}{1.2713}& \multicolumn{1}{r}{2.8813}\\
 Std Err & \multicolumn{1}{r}{1.3178E-02}& \multicolumn{1}{r}{1.1516E-02}& \multicolumn{1}{r}{1.3597E-02}& \multicolumn{1}{r}{1.5848E-02}\\
 P-Value & \multicolumn{1}{r}{}& \multicolumn{1}{r}{0.0000E+00$^{***}$}& \multicolumn{1}{r}{0.0000E+00$^{***}$}& \multicolumn{1}{r}{0.0000E+00$^{***}$}\\
 0.95-VaR & \multicolumn{1}{r}{-0.3389}& \multicolumn{1}{r}{0.2967}& \multicolumn{1}{r}{2.8337}& \multicolumn{1}{r}{0.1565}\\
 0.95-CI of 0.95-VaR & \multicolumn{1}{r}{[-0.3453, -0.3286]} & \multicolumn{1}{r}{[0.2501, 0.3555]} & \multicolumn{1}{r}{[2.7215, 2.9420]} & \multicolumn{1}{r}{[8.2989E-02, 0.2311]} \\
 0.975-VaR & \multicolumn{1}{r}{-0.2076}& \multicolumn{1}{r}{1.6854}& \multicolumn{1}{r}{5.0178}& \multicolumn{1}{r}{2.4963}\\
 0.95-CI of 0.975-VaR & \multicolumn{1}{r}{[-0.2224, -0.1994]} & \multicolumn{1}{r}{[1.5671, 1.8278]} & \multicolumn{1}{r}{[4.7887, 5.2264]} & \multicolumn{1}{r}{[2.2016, 2.7177]} \\
 0.975-MS & \multicolumn{1}{r}{0.3872}& \multicolumn{1}{r}{3.0683}& \multicolumn{1}{r}{7.4019}& \multicolumn{1}{r}{5.8128}\\
 0.95-CI of 0.975-MS & \multicolumn{1}{r}{[8.7402E-02, 0.5821]} & \multicolumn{1}{r}{[2.9316, 3.2132]} & \multicolumn{1}{r}{[7.1735, 7.6450]} & \multicolumn{1}{r}{[5.5032, 6.1951]} \\
 0.95-CVaR & \multicolumn{1}{r}{0.5382}& \multicolumn{1}{r}{2.2504}& \multicolumn{1}{r}{5.7857}& \multicolumn{1}{r}{3.9512}\\
 0.95-CI of 0.95-CVaR & \multicolumn{1}{r}{[0.4625, 0.6242]} & \multicolumn{1}{r}{[2.1449, 2.3651]} & \multicolumn{1}{r}{[5.6220, 5.9502]} & \multicolumn{1}{r}{[3.7296, 4.1748]} \\
 0.975-CVaR & \multicolumn{1}{r}{1.3534}& \multicolumn{1}{r}{3.6460}& \multicolumn{1}{r}{7.8232}& \multicolumn{1}{r}{6.8865}\\
 0.95-CI of 0.975-CVaR & \multicolumn{1}{r}{[1.2013, 1.5114]} & \multicolumn{1}{r}{[3.4992, 3.8027]} & \multicolumn{1}{r}{[7.6186, 8.0282]} & \multicolumn{1}{r}{[6.5428, 7.2275]}  \\
		\midrule 
		\multicolumn{5}{c}{Panel B: with $0.1\%$ transaction cost} \\ 
		\midrule
		 Mean & \multicolumn{1}{r}{2.2157}& \multicolumn{1}{r}{0.4693}& \multicolumn{1}{r}{-0.5903}& \multicolumn{1}{r}{1.1193}\\
 Std Err & \multicolumn{1}{r}{1.2139E-02}& \multicolumn{1}{r}{1.0408E-02}& \multicolumn{1}{r}{1.4632E-02}& \multicolumn{1}{r}{1.7650E-02}\\
 P-Value & \multicolumn{1}{r}{}& \multicolumn{1}{r}{0.0000E+00$^{***}$}& \multicolumn{1}{r}{0.0000E+00$^{***}$}& \multicolumn{1}{r}{0.0000E+00$^{***}$}\\
 0.95-VaR & \multicolumn{1}{r}{0.6401}& \multicolumn{1}{r}{3.3338}& \multicolumn{1}{r}{7.0026}& \multicolumn{1}{r}{4.8350}\\
 0.95-CI of 0.95-VaR & \multicolumn{1}{r}{[0.5642, 0.7139]} & \multicolumn{1}{r}{[3.1943, 3.4723]} & \multicolumn{1}{r}{[6.8557, 7.1900]} & \multicolumn{1}{r}{[4.6088, 5.0868]} \\
 0.975-VaR & \multicolumn{1}{r}{2.3640}& \multicolumn{1}{r}{5.6212}& \multicolumn{1}{r}{9.7243}& \multicolumn{1}{r}{8.8841}\\
 0.95-CI of 0.975-VaR & \multicolumn{1}{r}{[2.1802, 2.5743]} & \multicolumn{1}{r}{[5.4690, 5.8033]} & \multicolumn{1}{r}{[9.4652, 9.9380]} & \multicolumn{1}{r}{[8.5287, 9.1957]} \\
 0.975-MS & \multicolumn{1}{r}{5.0646}& \multicolumn{1}{r}{7.6481}& \multicolumn{1}{r}{12.2785}& \multicolumn{1}{r}{13.3470}\\
 0.95-CI of 0.975-MS & \multicolumn{1}{r}{[4.7650, 5.3493]} & \multicolumn{1}{r}{[7.3692, 7.9166]} & \multicolumn{1}{r}{[12.0182, 12.5933]} & \multicolumn{1}{r}{[12.8422, 13.7619]} \\
 0.95-CVaR & \multicolumn{1}{r}{3.4671}& \multicolumn{1}{r}{6.2739}& \multicolumn{1}{r}{10.5583}& \multicolumn{1}{r}{10.5216}\\
 0.95-CI of 0.95-CVaR & \multicolumn{1}{r}{[3.3021, 3.6385]} & \multicolumn{1}{r}{[6.1303, 6.4243]} & \multicolumn{1}{r}{[10.3660, 10.7509]} & \multicolumn{1}{r}{[10.2091, 10.8206]} \\
 0.975-CVaR & \multicolumn{1}{r}{5.6452}& \multicolumn{1}{r}{8.1932}& \multicolumn{1}{r}{12.9010}& \multicolumn{1}{r}{14.4257}\\
 0.95-CI of 0.975-CVaR & \multicolumn{1}{r}{[5.3870, 5.8671]} & \multicolumn{1}{r}{[8.0094, 8.3915]} & \multicolumn{1}{r}{[12.6551, 13.1424]} & \multicolumn{1}{r}{[14.0018, 14.8322]}  \\
		\bottomrule 
		\end{tabular}} 
		\label{table_real_put_short_zero} 
	\end{table}

	The panel A of Table~\ref{table_real_call_short_zero} and Table~\ref{table_real_put_short_zero} share some similar results: (i) CU-RL obtains the significantly highest mean final P\&L among all the methods; (ii) CU-RL obtains the lowest CVaR at 0.975 level, the designated risk measure in the total reward of CU-RL, the lowest MS at 0.975 level, and the lowest VaR at 0.975 level; (iii) CU-RL obtains the lowest VaR and CVaR at 0.95 level, suggesting it can effectively minimize the tail risk measured at an alternative level; (iv) all the results with respect to tail risk measures are statistically significant, since all the confidence intervals of CU-RL are lower and have no overlap with those of other benchmark models.
		
	The panel B of Table~\ref{table_real_call_short_zero} and Table~\ref{table_real_put_short_zero} shows that similar conclusion holds when transaction costs are considered. Although the mean of final P\&L falls and the risk measure rises due to transaction costs, CU-RL still obtains the significantly highest mean P\&L and lowest tail risk among all the models.

	We also demonstrate the performance of CU-RL under the zero intermediate reward for all call/put options. The initial learning rate is 1e-4 for the group of call options and 5e-4 for the group of put options. Other hyperparameters are the same as those in the previous experiments with asymmetric reward. 
	The performance of CU-RL for all call (resp., put) options is presented in Table~\ref{table_real_call_all_zero} (resp., Table~\ref{table_real_put_all_zero}).
	
	\begin{table}[!htbp]
		\centering
		\caption{\textbf{The mean, standard error, and tail risk of the final P\&L of hedging all call options by the CU-RL method under zero intermediate reward, the Black-Scholes, local volatility function, and SABR delta hedging method calculated from an out-of-sample test set.} The test set include all S\&P 500 call options traded on or after 10/01/2017 and expired before 01/02/2018. The p-values are calculated from a one-sided t-test for related samples that tests if the mean P\&L of our method is higher than that of the benchmark method. 0.95-CI means 95\% confidence interval. Scientific notation: $\text{1.23E-4} = 1.23\times 10^{-4}$. $\text{***}:p < 0.001$; $\text{**}:p < 0.01$; $*:p < 0.05$.}
\resizebox{\linewidth}{!}{
	\begin{tabular}{lllll}
		\toprule
		& \multicolumn{1}{r}{CU-RL} & \multicolumn{3}{c}{Traditional Model} \\ 
		\cmidrule(lr){3-5}
		& & \multicolumn{1}{r}{BS Model} & \multicolumn{1}{r}{LVF Model} & \multicolumn{1}{r}{SABR Model} \\ 
		\midrule 
		\multicolumn{5}{c}{Panel A: without transaction cost} \\ 
		\midrule
		 Mean & \multicolumn{1}{r}{1.8178}& \multicolumn{1}{r}{-1.0406}& \multicolumn{1}{r}{-2.0096}& \multicolumn{1}{r}{0.2089}\\
 Std Err & \multicolumn{1}{r}{1.5774E-02}& \multicolumn{1}{r}{7.9599E-03}& \multicolumn{1}{r}{1.0556E-02}& \multicolumn{1}{r}{1.2193E-02}\\
 P-Value & \multicolumn{1}{r}{}& \multicolumn{1}{r}{0.0000E+00$^{***}$}& \multicolumn{1}{r}{0.0000E+00$^{***}$}& \multicolumn{1}{r}{0.0000E+00$^{***}$}\\
 0.95-VaR & \multicolumn{1}{r}{4.0525}& \multicolumn{1}{r}{5.8561}& \multicolumn{1}{r}{8.7729}& \multicolumn{1}{r}{5.2496}\\
 0.95-CI of 0.95-VaR & \multicolumn{1}{r}{[4.0010, 4.1030]} & \multicolumn{1}{r}{[5.8085, 5.9103]} & \multicolumn{1}{r}{[8.6952, 8.8562]} & \multicolumn{1}{r}{[5.1924, 5.3053]} \\
 0.975-VaR & \multicolumn{1}{r}{5.4518}& \multicolumn{1}{r}{7.1287}& \multicolumn{1}{r}{11.0766}& \multicolumn{1}{r}{6.7506}\\
 0.95-CI of 0.975-VaR & \multicolumn{1}{r}{[5.3833, 5.5354]} & \multicolumn{1}{r}{[7.0642, 7.1983]} & \multicolumn{1}{r}{[10.9690, 11.1896]} & \multicolumn{1}{r}{[6.6765, 6.8438]} \\
 0.975-MS & \multicolumn{1}{r}{6.8008}& \multicolumn{1}{r}{8.2424}& \multicolumn{1}{r}{12.4982}& \multicolumn{1}{r}{8.5785}\\
 0.95-CI of 0.975-MS & \multicolumn{1}{r}{[6.7074, 6.8981]} & \multicolumn{1}{r}{[8.1687, 8.3331]} & \multicolumn{1}{r}{[12.4096, 12.5886]} & \multicolumn{1}{r}{[8.4534, 8.7365]} \\
 0.95-CVaR & \multicolumn{1}{r}{6.0031}& \multicolumn{1}{r}{7.4568}& \multicolumn{1}{r}{11.2591}& \multicolumn{1}{r}{8.1723}\\
 0.95-CI of 0.95-CVaR & \multicolumn{1}{r}{[5.9372, 6.0746]} & \multicolumn{1}{r}{[7.4043, 7.5127]} & \multicolumn{1}{r}{[11.1849, 11.3346]} & \multicolumn{1}{r}{[8.0410, 8.2998]} \\
 0.975-CVaR & \multicolumn{1}{r}{7.3161}& \multicolumn{1}{r}{8.4783}& \multicolumn{1}{r}{12.7464}& \multicolumn{1}{r}{10.4350}\\
 0.95-CI of 0.975-CVaR & \multicolumn{1}{r}{[7.2162, 7.4155]} & \multicolumn{1}{r}{[8.4106, 8.5385]} & \multicolumn{1}{r}{[12.6687, 12.8228]} & \multicolumn{1}{r}{[10.2362, 10.6633]}  \\
		\midrule 
		\multicolumn{5}{c}{Panel B: with $0.1\%$ transaction cost} \\ 
		\midrule
		 Mean & \multicolumn{1}{r}{-0.4011}& \multicolumn{1}{r}{-3.4508}& \multicolumn{1}{r}{-4.6257}& \multicolumn{1}{r}{-2.7045}\\
 Std Err & \multicolumn{1}{r}{1.8508E-02}& \multicolumn{1}{r}{1.0031E-02}& \multicolumn{1}{r}{1.3667E-02}& \multicolumn{1}{r}{1.3308E-02}\\
 P-Value & \multicolumn{1}{r}{}& \multicolumn{1}{r}{0.0000E+00$^{***}$}& \multicolumn{1}{r}{0.0000E+00$^{***}$}& \multicolumn{1}{r}{0.0000E+00$^{***}$}\\
 0.95-VaR & \multicolumn{1}{r}{7.1001}& \multicolumn{1}{r}{9.2068}& \multicolumn{1}{r}{13.0101}& \multicolumn{1}{r}{9.1282}\\
 0.95-CI of 0.95-VaR & \multicolumn{1}{r}{[7.0291, 7.1711]} & \multicolumn{1}{r}{[9.1579, 9.2491]} & \multicolumn{1}{r}{[12.9186, 13.0887]} & \multicolumn{1}{r}{[9.0536, 9.2073]} \\
 0.975-VaR & \multicolumn{1}{r}{8.9425}& \multicolumn{1}{r}{10.5362}& \multicolumn{1}{r}{15.6054}& \multicolumn{1}{r}{11.4791}\\
 0.95-CI of 0.975-VaR & \multicolumn{1}{r}{[8.8347, 9.0652]} & \multicolumn{1}{r}{[10.4645, 10.5985]} & \multicolumn{1}{r}{[15.4899, 15.7274]} & \multicolumn{1}{r}{[11.3261, 11.6428]} \\
 0.975-MS & \multicolumn{1}{r}{10.7439}& \multicolumn{1}{r}{11.6221}& \multicolumn{1}{r}{17.3305}& \multicolumn{1}{r}{14.7662}\\
 0.95-CI of 0.975-MS & \multicolumn{1}{r}{[10.6113, 10.8734]} & \multicolumn{1}{r}{[11.5416, 11.7006]} & \multicolumn{1}{r}{[17.2319, 17.4243]} & \multicolumn{1}{r}{[14.4888, 15.0669]} \\
 0.95-CVaR & \multicolumn{1}{r}{9.6051}& \multicolumn{1}{r}{10.8269}& \multicolumn{1}{r}{15.8995}& \multicolumn{1}{r}{14.0771}\\
 0.95-CI of 0.95-CVaR & \multicolumn{1}{r}{[9.5224, 9.6937]} & \multicolumn{1}{r}{[10.7770, 10.8816]} & \multicolumn{1}{r}{[15.8114, 15.9940]} & \multicolumn{1}{r}{[13.8530, 14.2852]} \\
 0.975-CVaR & \multicolumn{1}{r}{11.3110}& \multicolumn{1}{r}{11.8595}& \multicolumn{1}{r}{17.6550}& \multicolumn{1}{r}{18.0190}\\
 0.95-CI of 0.975-CVaR & \multicolumn{1}{r}{[11.2006, 11.4316]} & \multicolumn{1}{r}{[11.7940, 11.9177]} & \multicolumn{1}{r}{[17.5565, 17.7463]} & \multicolumn{1}{r}{[17.6693, 18.4025]}  \\
		\bottomrule 
		\end{tabular}} 
		\label{table_real_call_all_zero} 
	\end{table} 
	
	\begin{table}[!htbp]
		\centering
		\caption{\textbf{The mean, standard error, and tail risk of the final P\&L of hedging all put options by the CU-RL method under zero intermediate reward, the Black-Scholes, local volatility function, and SABR delta hedging method calculated from an out-of-sample test set.} The test set include all S\&P 500 put options traded on or after 10/01/2017 and expired before 01/02/2018. The p-values are calculated from a one-sided t-test for related samples that tests if the mean P\&L of our method is higher than that of the benchmark method. 0.95-CI means 95\% confidence interval. Scientific notation: $\text{1.23E-4} = 1.23\times 10^{-4}$. $\text{***}:p < 0.001$; $\text{**}:p < 0.01$; $*:p < 0.05$.}
\resizebox{\linewidth}{!}{
	\begin{tabular}{lllll}
		\toprule
		& \multicolumn{1}{r}{CU-RL} & \multicolumn{3}{c}{Traditional Model} \\ 
		\cmidrule(lr){3-5}
		& & \multicolumn{1}{r}{BS Model} & \multicolumn{1}{r}{LVF Model} & \multicolumn{1}{r}{SABR Model} \\ 
		\midrule 
		\multicolumn{5}{c}{Panel A: without transaction cost} \\ 
		\midrule
		 Mean & \multicolumn{1}{r}{2.9270}& \multicolumn{1}{r}{1.7698}& \multicolumn{1}{r}{1.0325}& \multicolumn{1}{r}{2.5043}\\
 Std Err & \multicolumn{1}{r}{1.1616E-02}& \multicolumn{1}{r}{6.6780E-03}& \multicolumn{1}{r}{6.4837E-03}& \multicolumn{1}{r}{9.9172E-03}\\
 P-Value & \multicolumn{1}{r}{}& \multicolumn{1}{r}{0.0000E+00$^{***}$}& \multicolumn{1}{r}{0.0000E+00$^{***}$}& \multicolumn{1}{r}{0.0000E+00$^{***}$}\\
 0.95-VaR & \multicolumn{1}{r}{0.3696}& \multicolumn{1}{r}{-1.4677E-02}& \multicolumn{1}{r}{1.0552}& \multicolumn{1}{r}{-2.4172E-02}\\
 0.95-CI of 0.95-VaR & \multicolumn{1}{r}{[0.3527, 0.3876]} & \multicolumn{1}{r}{[-1.5031E-02, -1.4325E-02]} & \multicolumn{1}{r}{[1.0077, 1.1090]} & \multicolumn{1}{r}{[-2.4224E-02, -2.4111E-02]} \\
 0.975-VaR & \multicolumn{1}{r}{1.0080}& \multicolumn{1}{r}{0.1280}& \multicolumn{1}{r}{2.9708}& \multicolumn{1}{r}{0.2625}\\
 0.95-CI of 0.975-VaR & \multicolumn{1}{r}{[0.9654, 1.0490]} & \multicolumn{1}{r}{[0.1032, 0.1551]} & \multicolumn{1}{r}{[2.8788, 3.0718]} & \multicolumn{1}{r}{[0.2065, 0.3192]} \\
 0.975-MS & \multicolumn{1}{r}{1.8201}& \multicolumn{1}{r}{1.1783}& \multicolumn{1}{r}{5.1877}& \multicolumn{1}{r}{2.3654}\\
 0.95-CI of 0.975-MS & \multicolumn{1}{r}{[1.7031, 2.0037]} & \multicolumn{1}{r}{[1.0880, 1.2697]} & \multicolumn{1}{r}{[5.0258, 5.3361]} & \multicolumn{1}{r}{[2.1944, 2.5420]} \\
 0.95-CVaR & \multicolumn{1}{r}{1.8967}& \multicolumn{1}{r}{0.9016}& \multicolumn{1}{r}{3.8585}& \multicolumn{1}{r}{1.8332}\\
 0.95-CI of 0.95-CVaR & \multicolumn{1}{r}{[1.8402, 1.9566]} & \multicolumn{1}{r}{[0.8550, 0.9415]} & \multicolumn{1}{r}{[3.7798, 3.9456]} & \multicolumn{1}{r}{[1.7549, 1.9240]} \\
 0.975-CVaR & \multicolumn{1}{r}{3.1783}& \multicolumn{1}{r}{1.7958}& \multicolumn{1}{r}{5.8443}& \multicolumn{1}{r}{3.6586}\\
 0.95-CI of 0.975-CVaR & \multicolumn{1}{r}{[3.0660, 3.2867]} & \multicolumn{1}{r}{[1.7203, 1.8731]} & \multicolumn{1}{r}{[5.7304, 5.9616]} & \multicolumn{1}{r}{[3.5077, 3.8106]}  \\
		\midrule 
		\multicolumn{5}{c}{Panel B: with $0.1\%$ transaction cost} \\ 
		\midrule
		 Mean & \multicolumn{1}{r}{3.5599}& \multicolumn{1}{r}{0.6398}& \multicolumn{1}{r}{-0.2402}& \multicolumn{1}{r}{1.3500}\\
 Std Err & \multicolumn{1}{r}{1.6160E-02}& \multicolumn{1}{r}{4.7636E-03}& \multicolumn{1}{r}{6.3944E-03}& \multicolumn{1}{r}{9.0985E-03}\\
 P-Value & \multicolumn{1}{r}{}& \multicolumn{1}{r}{0.0000E+00$^{***}$}& \multicolumn{1}{r}{0.0000E+00$^{***}$}& \multicolumn{1}{r}{0.0000E+00$^{***}$}\\
 0.95-VaR & \multicolumn{1}{r}{0.8111}& \multicolumn{1}{r}{1.4115}& \multicolumn{1}{r}{4.4806}& \multicolumn{1}{r}{1.6581}\\
 0.95-CI of 0.95-VaR & \multicolumn{1}{r}{[0.7839, 0.8352]} & \multicolumn{1}{r}{[1.3590, 1.4621]} & \multicolumn{1}{r}{[4.3757, 4.5848]} & \multicolumn{1}{r}{[1.5897, 1.7050]} \\
 0.975-VaR & \multicolumn{1}{r}{1.3976}& \multicolumn{1}{r}{3.0216}& \multicolumn{1}{r}{7.8318}& \multicolumn{1}{r}{5.1990}\\
 0.95-CI of 0.975-VaR & \multicolumn{1}{r}{[1.3821, 1.4144]} & \multicolumn{1}{r}{[2.9267, 3.1172]} & \multicolumn{1}{r}{[7.6875, 7.9599]} & \multicolumn{1}{r}{[5.0026, 5.3900]} \\
 0.975-MS & \multicolumn{1}{r}{1.9062}& \multicolumn{1}{r}{5.2327}& \multicolumn{1}{r}{10.6417}& \multicolumn{1}{r}{9.4210}\\
 0.95-CI of 0.975-MS & \multicolumn{1}{r}{[1.8538, 1.9473]} & \multicolumn{1}{r}{[5.0877, 5.3966]} & \multicolumn{1}{r}{[10.4548, 10.7951]} & \multicolumn{1}{r}{[9.1458, 9.7027]} \\
 0.95-CVaR & \multicolumn{1}{r}{1.8588}& \multicolumn{1}{r}{3.9472}& \multicolumn{1}{r}{8.6063}& \multicolumn{1}{r}{6.8241}\\
 0.95-CI of 0.95-CVaR & \multicolumn{1}{r}{[1.8205, 1.9002]} & \multicolumn{1}{r}{[3.8594, 4.0215]} & \multicolumn{1}{r}{[8.4893, 8.7219]} & \multicolumn{1}{r}{[6.6475, 6.9837]} \\
 0.975-CVaR & \multicolumn{1}{r}{2.6139}& \multicolumn{1}{r}{5.8264}& \multicolumn{1}{r}{11.2287}& \multicolumn{1}{r}{10.7077}\\
 0.95-CI of 0.975-CVaR & \multicolumn{1}{r}{[2.5419, 2.6846]} & \multicolumn{1}{r}{[5.7068, 5.9436]} & \multicolumn{1}{r}{[11.0766, 11.3523]} & \multicolumn{1}{r}{[10.4892, 10.9313]}  \\
		\bottomrule 
		\end{tabular}} 
		\label{table_real_put_all_zero} 
	\end{table}

	For all call options, the panel A of Table~\ref{table_real_call_all_zero} shows that: (i) CU-RL obtains the statistically significantly highest mean final P\&L than the benchmark methods; (ii) CU-RL obtains the lowest CVaR at 0.975 level, the designated risk measure in the total reward of CU-RL, the lowest MS at 0.975 level, and the lowest VaR at 0.975 level; (iii) CU-RL obtains the lowest VaR and CVaR at 0.95 level, suggesting it can effectively minimize the tail risk measured at an alternative level; (iv) all the results with respect to tail risk measures are statistically significant, since all the confidence intervals of CU-RL are lower and have no overlap with those of other benchmark models.
	The panel B of Table~\ref{table_real_call_all_zero} shows that conclusion (i)-(v) holds when transaction costs are considered. Although the mean of final P\&L falls and the risk measure rises due to transaction costs, CU-RL still obtains the significantly highest mean P\&L and lowest tail risk among all the models. 
	
	For all put options, the panel A of Table~\ref{table_real_put_all_zero} shows that: (i) CU-RL obtains the statistically significantly highest mean final P\&L than the benchmark methods; (ii) the SABR model obtains the lowest VaR at 0.95 level, and BS model obtains the lowest of other risk measures.
	The panel B of Table~\ref{table_real_put_all_zero} shows the results when transaction costs are considered. CU-RL obtains the significantly highest mean P\&L and significantly lowest tail risk among all the models.

	\subsection{Performance of Reward Divided by Initial Margin}\label{subsec:cu_rl_divided_by_margin_reward}
	If we are more concerned about the relative value of return, the reward function is correspondingly defined as a ratio. We consider a naive case, in which the reward is divided by the initial margin. At first, we define the reward as \eqref{reward_asym_margin}. 
	\begin{equation}
		\label{reward_asym_margin}
		R_{t+1}=
		\begin{cases}
			-\frac{|W_{t+1}|}{M_{0}} \cdot \mathbb{I}_{ \{ W_{t+1}<0 \} }, & t=0, \cdots, T-2, \\
			-\lambda_{1}\left[\omega + \frac{1}{1-\alpha} \max (-\frac{W_{t+1}}{M_{0}}-\omega, 0)\right] + \lambda_{2}\frac{W_{t+1}}{M_{0}}, & t=T-1,
		\end{cases} 
	\end{equation} 
	where $M_{0}$ is the initial value in the margin account. The computation of the initial margin obeys the following rule given by CBOE.
	
	\begin{itemize}
		\item Rule for initial margin:
		
		Writers of uncovered calls must deposit/maintain 100\% of the option proceeds plus 15\% of the aggregate contract value (current index level $\times$ 100) minus the amount by which the option is out-of-the-money, if any, subject to a minimum for calls of option proceeds plus 10\% of the aggregate contract value.
	\end{itemize}
	Briefly, in our experiment\footnote{We assume that one option targets one share of underlying such that the aggregate contract value is current index level $\times$ 1.},
	\begin{equation}
		\label{margin_rule}
		M_{0} =
			\max\{Z_{0} + 15\% S_{0} - \max(K - S_{0}, 0), Z_{0} + 10\% S_{0}\}.
	\end{equation}
	
	Also, we look into the performance on the short-term near-the-money call options with $T \in [5, 30]$ and $K/S_{0} \in [0.9, 1.1]$ as well as the whole dataset of call options. 
	
	For the short-term near-the-money call options, the network structure and hyperparameters in the implementation of CU-RL are the same as those in the experiments with the asymmetric reward in \eqref{R_relu}, except that the initial value of the linearly decaying learning rate is 1e-7 for short-term call options.
	The comparison of final P\&L between different models is shown in Table~\ref{table_real_call_short_asym_margin}. 
	
	In the implementation of CU-RL for all call options in margin formation, the size of buffer $\mathcal{D}_k$ is 59990, the minibatch size $n=30000$, the initial value of linearly decaying learning rate is 1e-5, and the number of training epochs for the intializer is 3000, the networks are all feedforward neural networks with 9 hidden layers of 32 neurons, and the initializers are based on SABR delta.
	The experimental results of these two groups are shown in Table~\ref{table_real_call_all_asym_margin}.
	
	\begin{table}[!htbp]
		\centering
		\caption{\textbf{The mean, standard error, and tail risk of the relative final P\&L of hedging short-term call options by the CU-RL method under asymmetric reward divided by initial margin, the Black-Scholes, local volatility function, and SABR delta hedging method calculated from an out-of-sample test set.} The test set include all S\&P 500 short-term call options with $T \in [5,30], K/S_{0} \in[0.9,1.1]$ traded on or after 10/01/2017 and expired before 01/02/2018. The p-values are calculated from a one-sided t-test for related samples that tests if the mean P\&L of our method is higher than that of the benchmark method. 0.95-CI means 95\% confidence interval. Scientific notation: $\text{1.23E-4} = 1.23\times 10^{-4}$. $\text{***}:p < 0.001$; $\text{**}:p < 0.01$; $*:p < 0.05$.}
\resizebox{\linewidth}{!}{
	\begin{tabular}{lllll}
		\toprule
		& \multicolumn{1}{r}{CU-RL} & \multicolumn{3}{c}{Traditional Model} \\ 
		\cmidrule(lr){3-5}
		& & \multicolumn{1}{r}{BS Model} & \multicolumn{1}{r}{LVF Model} & \multicolumn{1}{r}{SABR Model} \\ 
		\midrule 
		\multicolumn{5}{c}{Panel A: without transaction cost} \\ 
		\midrule
		 Mean & \multicolumn{1}{r}{3.1397E-03}& \multicolumn{1}{r}{-2.6187E-04}& \multicolumn{1}{r}{-1.8676E-03}& \multicolumn{1}{r}{2.5142E-03}\\
 Std Err & \multicolumn{1}{r}{3.1366E-05}& \multicolumn{1}{r}{2.7744E-05}& \multicolumn{1}{r}{3.4199E-05}& \multicolumn{1}{r}{5.0769E-05}\\
 P-Value & \multicolumn{1}{r}{}& \multicolumn{1}{r}{0.0000E+00$^{***}$}& \multicolumn{1}{r}{0.0000E+00$^{***}$}& \multicolumn{1}{r}{2.3747E-33$^{***}$}\\
 0.95-VaR & \multicolumn{1}{r}{3.2257E-03}& \multicolumn{1}{r}{8.6222E-03}& \multicolumn{1}{r}{1.3355E-02}& \multicolumn{1}{r}{1.0296E-02}\\
 0.95-CI of 0.95-VaR & \multicolumn{1}{r}{[3.1442E-03, 3.3243E-03]} & \multicolumn{1}{r}{[8.4544E-03, 8.7870E-03]} & \multicolumn{1}{r}{[1.3103E-02, 1.3596E-02]} & \multicolumn{1}{r}{[9.9422E-03, 1.0636E-02]} \\
 0.975-VaR & \multicolumn{1}{r}{4.9374E-03}& \multicolumn{1}{r}{1.1915E-02}& \multicolumn{1}{r}{1.8358E-02}& \multicolumn{1}{r}{1.5684E-02}\\
 0.95-CI of 0.975-VaR & \multicolumn{1}{r}{[4.7626E-03, 5.1704E-03]} & \multicolumn{1}{r}{[1.1621E-02, 1.2294E-02]} & \multicolumn{1}{r}{[1.7917E-02, 1.8788E-02]} & \multicolumn{1}{r}{[1.5288E-02, 1.6205E-02]} \\
 0.975-MS & \multicolumn{1}{r}{1.1671E-02}& \multicolumn{1}{r}{1.6796E-02}& \multicolumn{1}{r}{2.3744E-02}& \multicolumn{1}{r}{2.2342E-02}\\
 0.95-CI of 0.975-MS & \multicolumn{1}{r}{[1.0923E-02, 1.2452E-02]} & \multicolumn{1}{r}{[1.6061E-02, 1.7439E-02]} & \multicolumn{1}{r}{[2.3080E-02, 2.4471E-02]} & \multicolumn{1}{r}{[2.1643E-02, 2.2977E-02]} \\
 0.95-CVaR & \multicolumn{1}{r}{8.2889E-03}& \multicolumn{1}{r}{1.3892E-02}& \multicolumn{1}{r}{2.0444E-02}& \multicolumn{1}{r}{1.9374E-02}\\
 0.95-CI of 0.95-CVaR & \multicolumn{1}{r}{[7.9718E-03, 8.6364E-03]} & \multicolumn{1}{r}{[1.3630E-02, 1.4243E-02]} & \multicolumn{1}{r}{[2.0074E-02, 2.0816E-02]} & \multicolumn{1}{r}{[1.8764E-02, 1.9932E-02]} \\
 0.975-CVaR & \multicolumn{1}{r}{1.2679E-02}& \multicolumn{1}{r}{1.7798E-02}& \multicolumn{1}{r}{2.5396E-02}& \multicolumn{1}{r}{2.6081E-02}\\
 0.95-CI of 0.975-CVaR & \multicolumn{1}{r}{[1.2094E-02, 1.3286E-02]} & \multicolumn{1}{r}{[1.7382E-02, 1.8218E-02]} & \multicolumn{1}{r}{[2.4840E-02, 2.5936E-02]} & \multicolumn{1}{r}{[2.5234E-02, 2.7098E-02]}  \\
		\midrule 
		\multicolumn{5}{c}{Panel B: with $0.1\%$ transaction cost} \\ 
		\midrule
		 Mean & \multicolumn{1}{r}{-1.0952E-03}& \multicolumn{1}{r}{-4.9026E-03}& \multicolumn{1}{r}{-6.8922E-03}& \multicolumn{1}{r}{-4.5536E-03}\\
 Std Err & \multicolumn{1}{r}{3.2253E-05}& \multicolumn{1}{r}{3.3791E-05}& \multicolumn{1}{r}{4.3422E-05}& \multicolumn{1}{r}{5.5580E-05}\\
 P-Value & \multicolumn{1}{r}{}& \multicolumn{1}{r}{0.0000E+00$^{***}$}& \multicolumn{1}{r}{0.0000E+00$^{***}$}& \multicolumn{1}{r}{0.0000E+00$^{***}$}\\
 0.95-VaR & \multicolumn{1}{r}{9.1583E-03}& \multicolumn{1}{r}{1.6725E-02}& \multicolumn{1}{r}{2.3106E-02}& \multicolumn{1}{r}{2.2920E-02}\\
 0.95-CI of 0.95-VaR & \multicolumn{1}{r}{[8.9951E-03, 9.3405E-03]} & \multicolumn{1}{r}{[1.6476E-02, 1.6946E-02]} & \multicolumn{1}{r}{[2.2805E-02, 2.3380E-02]} & \multicolumn{1}{r}{[2.2516E-02, 2.3439E-02]} \\
 0.975-VaR & \multicolumn{1}{r}{1.5522E-02}& \multicolumn{1}{r}{2.0508E-02}& \multicolumn{1}{r}{2.8101E-02}& \multicolumn{1}{r}{3.1004E-02}\\
 0.95-CI of 0.975-VaR & \multicolumn{1}{r}{[1.4879E-02, 1.6623E-02]} & \multicolumn{1}{r}{[2.0158E-02, 2.0847E-02]} & \multicolumn{1}{r}{[2.7619E-02, 2.8581E-02]} & \multicolumn{1}{r}{[3.0262E-02, 3.1798E-02]} \\
 0.975-MS & \multicolumn{1}{r}{2.2954E-02}& \multicolumn{1}{r}{2.5189E-02}& \multicolumn{1}{r}{3.3416E-02}& \multicolumn{1}{r}{3.9520E-02}\\
 0.95-CI of 0.975-MS & \multicolumn{1}{r}{[2.2134E-02, 2.3941E-02]} & \multicolumn{1}{r}{[2.4538E-02, 2.5899E-02]} & \multicolumn{1}{r}{[3.2759E-02, 3.4164E-02]} & \multicolumn{1}{r}{[3.8745E-02, 4.0528E-02]} \\
 0.95-CVaR & \multicolumn{1}{r}{1.7806E-02}& \multicolumn{1}{r}{2.2421E-02}& \multicolumn{1}{r}{3.0336E-02}& \multicolumn{1}{r}{3.6213E-02}\\
 0.95-CI of 0.95-CVaR & \multicolumn{1}{r}{[1.7355E-02, 1.8300E-02]} & \multicolumn{1}{r}{[2.2140E-02, 2.2793E-02]} & \multicolumn{1}{r}{[2.9944E-02, 3.0731E-02]} & \multicolumn{1}{r}{[3.5335E-02, 3.7005E-02]} \\
 0.975-CVaR & \multicolumn{1}{r}{2.4329E-02}& \multicolumn{1}{r}{2.6536E-02}& \multicolumn{1}{r}{3.5446E-02}& \multicolumn{1}{r}{4.5898E-02}\\
 0.95-CI of 0.975-CVaR & \multicolumn{1}{r}{[2.3716E-02, 2.4973E-02]} & \multicolumn{1}{r}{[2.6100E-02, 2.6988E-02]} & \multicolumn{1}{r}{[3.4860E-02, 3.6033E-02]} & \multicolumn{1}{r}{[4.4740E-02, 4.7399E-02]}  \\
		\bottomrule 
		\end{tabular}} 
		\label{table_real_call_short_asym_margin} 
	\end{table} 
	The panel A of Table~\ref{table_real_call_short_asym_margin} shows that: (i) CU-RL obtains the significantly highest mean of relative final P\&L among all the methods; (ii) CU-RL obtains the lowest CVaR at 0.975 level, the designated risk measure in the total reward of CU-RL, the lowest MS at 0.975 level, and the lowest VaR at 0.975 level; (iii) CU-RL obtains the lowest VaR and CVaR at 0.95 level, suggesting it can effectively minimize the tail risk measured at an alternative level; (iv) all the results with respect to tail risk measures are statistically significant, since all the confidence intervals of CU-RL are lower and have no overlap with those of other benchmark models.
		
	The panel B of Table~\ref{table_real_call_short_asym_margin} shows that similar conclusion holds when transaction costs are considered. Although the mean of final P\&L falls and the risk measure rises due to transaction costs, CU-RL still obtains the significantly highest mean P\&L and lowest tail risk among all the models.
	\begin{table}[!htbp]
		\centering
		\caption{\textbf{The mean, standard error, and tail risk of the relative final P\&L of hedging all call options by the CU-RL method under asymmetric reward divided by initial margin, the Black-Scholes, local volatility function, and SABR delta hedging method calculated from an out-of-sample test set.} The test set include all S\&P 500 call options traded on or after 10/01/2017 and expired before 01/02/2018. The p-values are calculated from a one-sided t-test for related samples that tests if the mean P\&L of our method is higher than that of the benchmark method. 0.95-CI means 95\% confidence interval. Scientific notation: $\text{1.23E-4} = 1.23\times 10^{-4}$. $\text{***}:p < 0.001$; $\text{**}:p < 0.01$; $*:p < 0.05$.}
\resizebox{\linewidth}{!}{
	\begin{tabular}{lllll}
		\toprule
		& \multicolumn{1}{r}{CU-RL} & \multicolumn{3}{c}{Traditional Model} \\ 
		\cmidrule(lr){3-5}
		& & \multicolumn{1}{r}{BS Model} & \multicolumn{1}{r}{LVF Model} & \multicolumn{1}{r}{SABR Model} \\ 
		\midrule 
		\multicolumn{5}{c}{Panel A: without transaction cost} \\ 
		\midrule
		 Mean & \multicolumn{1}{r}{2.3970E-03}& \multicolumn{1}{r}{-1.3067E-03}& \multicolumn{1}{r}{-3.3349E-03}& \multicolumn{1}{r}{1.4441E-03}\\
 Std Err & \multicolumn{1}{r}{2.5146E-05}& \multicolumn{1}{r}{1.7018E-05}& \multicolumn{1}{r}{2.2938E-05}& \multicolumn{1}{r}{3.1437E-05}\\
 P-Value & \multicolumn{1}{r}{}& \multicolumn{1}{r}{0.0000E+00$^{***}$}& \multicolumn{1}{r}{0.0000E+00$^{***}$}& \multicolumn{1}{r}{7.3631E-263$^{***}$}\\
 0.95-VaR & \multicolumn{1}{r}{7.2071E-03}& \multicolumn{1}{r}{9.7598E-03}& \multicolumn{1}{r}{1.6684E-02}& \multicolumn{1}{r}{8.4188E-03}\\
 0.95-CI of 0.95-VaR & \multicolumn{1}{r}{[7.1301E-03, 7.2933E-03]} & \multicolumn{1}{r}{[9.6739E-03, 9.8307E-03]} & \multicolumn{1}{r}{[1.6503E-02, 1.6842E-02]} & \multicolumn{1}{r}{[8.3140E-03, 8.5129E-03]} \\
 0.975-VaR & \multicolumn{1}{r}{9.1565E-03}& \multicolumn{1}{r}{1.1937E-02}& \multicolumn{1}{r}{2.1775E-02}& \multicolumn{1}{r}{1.3115E-02}\\
 0.95-CI of 0.975-VaR & \multicolumn{1}{r}{[9.0113E-03, 9.3224E-03]} & \multicolumn{1}{r}{[1.1824E-02, 1.2038E-02]} & \multicolumn{1}{r}{[2.1483E-02, 2.2105E-02]} & \multicolumn{1}{r}{[1.2821E-02, 1.3462E-02]} \\
 0.975-MS & \multicolumn{1}{r}{1.4802E-02}& \multicolumn{1}{r}{1.4700E-02}& \multicolumn{1}{r}{2.7850E-02}& \multicolumn{1}{r}{2.0665E-02}\\
 0.95-CI of 0.975-MS & \multicolumn{1}{r}{[1.4366E-02, 1.5349E-02]} & \multicolumn{1}{r}{[1.4456E-02, 1.4964E-02]} & \multicolumn{1}{r}{[2.7488E-02, 2.8264E-02]} & \multicolumn{1}{r}{[2.0009E-02, 2.1370E-02]} \\
 0.95-CVaR & \multicolumn{1}{r}{1.2538E-02}& \multicolumn{1}{r}{1.3466E-02}& \multicolumn{1}{r}{2.3992E-02}& \multicolumn{1}{r}{1.8755E-02}\\
 0.95-CI of 0.95-CVaR & \multicolumn{1}{r}{[1.2319E-02, 1.2774E-02]} & \multicolumn{1}{r}{[1.3322E-02, 1.3614E-02]} & \multicolumn{1}{r}{[2.3754E-02, 2.4245E-02]} & \multicolumn{1}{r}{[1.8273E-02, 1.9236E-02]} \\
 0.975-CVaR & \multicolumn{1}{r}{1.7064E-02}& \multicolumn{1}{r}{1.6195E-02}& \multicolumn{1}{r}{2.9148E-02}& \multicolumn{1}{r}{2.7358E-02}\\
 0.95-CI of 0.975-CVaR & \multicolumn{1}{r}{[1.6659E-02, 1.7477E-02]} & \multicolumn{1}{r}{[1.5994E-02, 1.6422E-02]} & \multicolumn{1}{r}{[2.8804E-02, 2.9464E-02]} & \multicolumn{1}{r}{[2.6594E-02, 2.8229E-02]}  \\
		\midrule 
		\multicolumn{5}{c}{Panel B: with $0.1\%$ transaction cost} \\ 
		\midrule
		 Mean & \multicolumn{1}{r}{1.0003E-03}& \multicolumn{1}{r}{-6.0007E-03}& \multicolumn{1}{r}{-8.4565E-03}& \multicolumn{1}{r}{-4.7971E-03}\\
 Std Err & \multicolumn{1}{r}{4.1182E-05}& \multicolumn{1}{r}{2.0696E-05}& \multicolumn{1}{r}{2.9907E-05}& \multicolumn{1}{r}{3.4930E-05}\\
 P-Value & \multicolumn{1}{r}{}& \multicolumn{1}{r}{0.0000E+00$^{***}$}& \multicolumn{1}{r}{0.0000E+00$^{***}$}& \multicolumn{1}{r}{0.0000E+00$^{***}$}\\
 0.95-VaR & \multicolumn{1}{r}{1.1730E-02}& \multicolumn{1}{r}{1.7711E-02}& \multicolumn{1}{r}{2.7216E-02}& \multicolumn{1}{r}{1.9514E-02}\\
 0.95-CI of 0.95-VaR & \multicolumn{1}{r}{[1.1680E-02, 1.1793E-02]} & \multicolumn{1}{r}{[1.7612E-02, 1.7811E-02]} & \multicolumn{1}{r}{[2.6998E-02, 2.7432E-02]} & \multicolumn{1}{r}{[1.9148E-02, 1.9861E-02]} \\
 0.975-VaR & \multicolumn{1}{r}{1.4023E-02}& \multicolumn{1}{r}{2.0756E-02}& \multicolumn{1}{r}{3.3764E-02}& \multicolumn{1}{r}{2.9338E-02}\\
 0.95-CI of 0.975-VaR & \multicolumn{1}{r}{[1.3819E-02, 1.4307E-02]} & \multicolumn{1}{r}{[2.0564E-02, 2.0921E-02]} & \multicolumn{1}{r}{[3.3387E-02, 3.4175E-02]} & \multicolumn{1}{r}{[2.8814E-02, 3.0036E-02]} \\
 0.975-MS & \multicolumn{1}{r}{2.4669E-02}& \multicolumn{1}{r}{2.4463E-02}& \multicolumn{1}{r}{4.0843E-02}& \multicolumn{1}{r}{4.0211E-02}\\
 0.95-CI of 0.975-MS & \multicolumn{1}{r}{[2.3959E-02, 2.5348E-02]} & \multicolumn{1}{r}{[2.4157E-02, 2.4774E-02]} & \multicolumn{1}{r}{[4.0344E-02, 4.1297E-02]} & \multicolumn{1}{r}{[3.9400E-02, 4.1280E-02]} \\
 0.95-CVaR & \multicolumn{1}{r}{2.0243E-02}& \multicolumn{1}{r}{2.2499E-02}& \multicolumn{1}{r}{3.6241E-02}& \multicolumn{1}{r}{3.7885E-02}\\
 0.95-CI of 0.95-CVaR & \multicolumn{1}{r}{[1.9858E-02, 2.0617E-02]} & \multicolumn{1}{r}{[2.2320E-02, 2.2685E-02]} & \multicolumn{1}{r}{[3.5932E-02, 3.6550E-02]} & \multicolumn{1}{r}{[3.7120E-02, 3.8634E-02]} \\
 0.975-CVaR & \multicolumn{1}{r}{2.7901E-02}& \multicolumn{1}{r}{2.6014E-02}& \multicolumn{1}{r}{4.2498E-02}& \multicolumn{1}{r}{5.1933E-02}\\
 0.95-CI of 0.975-CVaR & \multicolumn{1}{r}{[2.7268E-02, 2.8635E-02]} & \multicolumn{1}{r}{[2.5765E-02, 2.6268E-02]} & \multicolumn{1}{r}{[4.2083E-02, 4.2897E-02]} & \multicolumn{1}{r}{[5.0711E-02, 5.3279E-02]}  \\
		\bottomrule 
		\end{tabular}} 
		\label{table_real_call_all_asym_margin} 
	\end{table} 
	
	The panel A of Table~\ref{table_real_call_all_asym_margin} shows that: (i) CU-RL obtains the  significantly highest mean of relative final P\&L among all the methods; (ii)  CU-RL obtains the lowest VaR at 0.95 and 0.975 level, and CVaR at 0.95 level, while BS model obtains the lowest MS and CVaR at 0.975 level; (iii) CU-RL obtains these lowest risk measures with statistically significance at 95\% level.
	The panel B of Table~\ref{table_real_call_all_asym_margin} shows the results (i)-(iii) still holds when transaction costs are considered.
	
	We also define a relative reward in \eqref{reward_zero_margin}, which always keeps zero unless the option matures. 
	
	\begin{equation}
		\label{reward_zero_margin}
		R_{t+1}=
		\begin{cases}
			0, & t=0, \cdots, T-2, \\
			-\lambda_{1}\left[\omega + \frac{1}{1-\alpha} \max (-\frac{W_{t+1}}{M_{0}}-\omega, 0)\right] + \lambda_{2}\frac{W_{t+1}}{M_{0}}, & t=T-1,
		\end{cases} 
	\end{equation} 
	where $M_{0}$ is the initial value in the margin account. The initial margin is still computed following the rule in \eqref{margin_rule}.
	
	In the implementation of CU-RL for short-term near-the-money call options in margin formation with zero intermediate reward in \eqref{reward_zero_margin}, the size of buffer $\mathcal{D}_k$ is 29988, the minibatch size $n=20000$, and the initial value of linearly decaying learning rate is 1e-6 for the non-transaction cost case and 1e-5 for the proportional transaction cost case, the networks are all feedforward neural networks with 9 hidden layers of 32 neurons, and the initializers are based on SABR delta. The numbers of training epochs for all initializers are 2000.
	The comparison of final P\&L between different models is shown in Table~\ref{table_real_call_short_zero_margin}.

	In the implementation of CU-RL for all call options in margin formation with zero intermediate reward in \eqref{reward_zero_margin}, the size of buffer $\mathcal{D}_k$ is 59990, the minibatch size $n=60000$, the numbers of training epochs for the initializers are 2000, and the initial value of linearly decaying learning rate is 1e-9 for the non-transaction cost case and 1e-8 for the proportional transaction cost case. The networks are all feedforward neural networks with 9 hidden layers of 32 neurons, and the initializers are based on Black-Scholes delta.
	The experimental results of these two groups are shown in Table~\ref{table_real_call_all_zero_margin}.
	
	\begin{table}[!htbp]
		\centering
		\caption{\textbf{The mean, standard error, and tail risk of the relative final P\&L of hedging short-term call options by the CU-RL method under zero intermediate reward divided by initial margin, the Black-Scholes, local volatility function, and SABR delta hedging method calculated from an out-of-sample test set.} The test set include all S\&P 500 short-term call options with $T \in [5,30], K/S_{0} \in[0.9,1.1]$ traded on or after 10/01/2017 and expired before 01/02/2018. The p-values are calculated from a one-sided t-test for related samples that tests if the mean P\&L of our method is higher than that of the benchmark method. 0.95-CI means 95\% confidence interval. Scientific notation: $\text{1.23E-4} = 1.23\times 10^{-4}$. $\text{***}:p < 0.001$; $\text{**}:p < 0.01$; $*:p < 0.05$.}
\resizebox{\linewidth}{!}{
	\begin{tabular}{lllll}
		\toprule
		& \multicolumn{1}{r}{CU-RL} & \multicolumn{3}{c}{Traditional Model} \\ 
		\cmidrule(lr){3-5}
		& & \multicolumn{1}{r}{BS Model} & \multicolumn{1}{r}{LVF Model} & \multicolumn{1}{r}{SABR Model} \\ 
		\midrule 
		\multicolumn{5}{c}{Panel A: without transaction cost} \\ 
		\midrule
		 Mean & \multicolumn{1}{r}{7.3005E-03}& \multicolumn{1}{r}{-2.6187E-04}& \multicolumn{1}{r}{-1.8676E-03}& \multicolumn{1}{r}{2.5142E-03}\\
 Std Err & \multicolumn{1}{r}{6.7308E-05}& \multicolumn{1}{r}{2.7744E-05}& \multicolumn{1}{r}{3.4199E-05}& \multicolumn{1}{r}{5.0769E-05}\\
 P-Value & \multicolumn{1}{r}{}& \multicolumn{1}{r}{0.0000E+00$^{***}$}& \multicolumn{1}{r}{0.0000E+00$^{***}$}& \multicolumn{1}{r}{0.0000E+00$^{***}$}\\
 0.95-VaR & \multicolumn{1}{r}{2.8653E-03}& \multicolumn{1}{r}{8.6222E-03}& \multicolumn{1}{r}{1.3355E-02}& \multicolumn{1}{r}{1.0296E-02}\\
 0.95-CI of 0.95-VaR & \multicolumn{1}{r}{[2.8131E-03, 2.9220E-03]} & \multicolumn{1}{r}{[8.4544E-03, 8.7870E-03]} & \multicolumn{1}{r}{[1.3103E-02, 1.3596E-02]} & \multicolumn{1}{r}{[9.9422E-03, 1.0636E-02]} \\
 0.975-VaR & \multicolumn{1}{r}{3.9160E-03}& \multicolumn{1}{r}{1.1915E-02}& \multicolumn{1}{r}{1.8358E-02}& \multicolumn{1}{r}{1.5684E-02}\\
 0.95-CI of 0.975-VaR & \multicolumn{1}{r}{[3.8367E-03, 3.9954E-03]} & \multicolumn{1}{r}{[1.1621E-02, 1.2294E-02]} & \multicolumn{1}{r}{[1.7917E-02, 1.8788E-02]} & \multicolumn{1}{r}{[1.5288E-02, 1.6205E-02]} \\
 0.975-MS & \multicolumn{1}{r}{5.1041E-03}& \multicolumn{1}{r}{1.6796E-02}& \multicolumn{1}{r}{2.3744E-02}& \multicolumn{1}{r}{2.2342E-02}\\
 0.95-CI of 0.975-MS & \multicolumn{1}{r}{[4.9795E-03, 5.2856E-03]} & \multicolumn{1}{r}{[1.6061E-02, 1.7439E-02]} & \multicolumn{1}{r}{[2.3080E-02, 2.4471E-02]} & \multicolumn{1}{r}{[2.1643E-02, 2.2977E-02]} \\
 0.95-CVaR & \multicolumn{1}{r}{5.2750E-03}& \multicolumn{1}{r}{1.3892E-02}& \multicolumn{1}{r}{2.0444E-02}& \multicolumn{1}{r}{1.9374E-02}\\
 0.95-CI of 0.95-CVaR & \multicolumn{1}{r}{[5.0914E-03, 5.4721E-03]} & \multicolumn{1}{r}{[1.3630E-02, 1.4243E-02]} & \multicolumn{1}{r}{[2.0074E-02, 2.0816E-02]} & \multicolumn{1}{r}{[1.8764E-02, 1.9932E-02]} \\
 0.975-CVaR & \multicolumn{1}{r}{7.2227E-03}& \multicolumn{1}{r}{1.7798E-02}& \multicolumn{1}{r}{2.5396E-02}& \multicolumn{1}{r}{2.6081E-02}\\
 0.95-CI of 0.975-CVaR & \multicolumn{1}{r}{[6.8975E-03, 7.5695E-03]} & \multicolumn{1}{r}{[1.7382E-02, 1.8218E-02]} & \multicolumn{1}{r}{[2.4840E-02, 2.5936E-02]} & \multicolumn{1}{r}{[2.5234E-02, 2.7098E-02]}  \\
		\midrule 
		\multicolumn{5}{c}{Panel B: with $0.1\%$ transaction cost} \\ 
		\midrule
		 Mean & \multicolumn{1}{r}{3.0086E-03}& \multicolumn{1}{r}{-4.9026E-03}& \multicolumn{1}{r}{-6.8922E-03}& \multicolumn{1}{r}{-4.5536E-03}\\
 Std Err & \multicolumn{1}{r}{7.0405E-05}& \multicolumn{1}{r}{3.3791E-05}& \multicolumn{1}{r}{4.3422E-05}& \multicolumn{1}{r}{5.5580E-05}\\
 P-Value & \multicolumn{1}{r}{}& \multicolumn{1}{r}{0.0000E+00$^{***}$}& \multicolumn{1}{r}{0.0000E+00$^{***}$}& \multicolumn{1}{r}{0.0000E+00$^{***}$}\\
 0.95-VaR & \multicolumn{1}{r}{8.9019E-03}& \multicolumn{1}{r}{1.6725E-02}& \multicolumn{1}{r}{2.3106E-02}& \multicolumn{1}{r}{2.2920E-02}\\
 0.95-CI of 0.95-VaR & \multicolumn{1}{r}{[8.8369E-03, 8.9762E-03]} & \multicolumn{1}{r}{[1.6476E-02, 1.6946E-02]} & \multicolumn{1}{r}{[2.2805E-02, 2.3380E-02]} & \multicolumn{1}{r}{[2.2516E-02, 2.3439E-02]} \\
 0.975-VaR & \multicolumn{1}{r}{1.0468E-02}& \multicolumn{1}{r}{2.0508E-02}& \multicolumn{1}{r}{2.8101E-02}& \multicolumn{1}{r}{3.1004E-02}\\
 0.95-CI of 0.975-VaR & \multicolumn{1}{r}{[1.0263E-02, 1.0687E-02]} & \multicolumn{1}{r}{[2.0158E-02, 2.0847E-02]} & \multicolumn{1}{r}{[2.7619E-02, 2.8581E-02]} & \multicolumn{1}{r}{[3.0262E-02, 3.1798E-02]} \\
 0.975-MS & \multicolumn{1}{r}{1.5126E-02}& \multicolumn{1}{r}{2.5189E-02}& \multicolumn{1}{r}{3.3416E-02}& \multicolumn{1}{r}{3.9520E-02}\\
 0.95-CI of 0.975-MS & \multicolumn{1}{r}{[1.4462E-02, 1.5855E-02]} & \multicolumn{1}{r}{[2.4538E-02, 2.5899E-02]} & \multicolumn{1}{r}{[3.2759E-02, 3.4164E-02]} & \multicolumn{1}{r}{[3.8745E-02, 4.0528E-02]} \\
 0.95-CVaR & \multicolumn{1}{r}{1.3572E-02}& \multicolumn{1}{r}{2.2421E-02}& \multicolumn{1}{r}{3.0336E-02}& \multicolumn{1}{r}{3.6213E-02}\\
 0.95-CI of 0.95-CVaR & \multicolumn{1}{r}{[1.3235E-02, 1.3909E-02]} & \multicolumn{1}{r}{[2.2140E-02, 2.2793E-02]} & \multicolumn{1}{r}{[2.9944E-02, 3.0731E-02]} & \multicolumn{1}{r}{[3.5335E-02, 3.7005E-02]} \\
 0.975-CVaR & \multicolumn{1}{r}{1.7635E-02}& \multicolumn{1}{r}{2.6536E-02}& \multicolumn{1}{r}{3.5446E-02}& \multicolumn{1}{r}{4.5898E-02}\\
 0.95-CI of 0.975-CVaR & \multicolumn{1}{r}{[1.6995E-02, 1.8240E-02]} & \multicolumn{1}{r}{[2.6100E-02, 2.6988E-02]} & \multicolumn{1}{r}{[3.4860E-02, 3.6033E-02]} & \multicolumn{1}{r}{[4.4740E-02, 4.7399E-02]}  \\
		\bottomrule 
		\end{tabular}}  
		\label{table_real_call_short_zero_margin} 
	\end{table}
	
	The panel A of Table~\ref{table_real_call_short_zero_margin} shows that: (i) CU-RL obtains the significantly highest mean of relative final P\&L among all the methods; (ii) CU-RL obtains the lowest CVaR at 0.975 level, the designated risk measure in the total reward of CU-RL, the lowest MS at 0.975 level, and the lowest VaR at 0.975 level; (iii) CU-RL obtains the lowest VaR and CVaR at 0.95 level, suggesting it can effectively minimize the tail risk measured at an alternative level; (iv) all the results with respect to tail risk measures are statistically significant, since all the confidence intervals of CU-RL are lower and have no overlap with those of other benchmark models.
		
	The panel B of Table~\ref{table_real_call_short_zero_margin} shows that similar conclusion holds when transaction costs are considered. Although the mean of final P\&L falls and the risk measure rises due to transaction costs, CU-RL still obtains the significantly highest mean P\&L and lowest tail risk among all the models.	

	\begin{table}[!htbp]
		\centering
		\caption{\textbf{The mean, standard error, and tail risk of the relative final P\&L of hedging all call options by the CU-RL method under zero intermediate reward divided by initial margin, the Black-Scholes, local volatility function, and SABR delta hedging method calculated from an out-of-sample test set.} The test set include all S\&P 500 call options traded on or after 10/01/2017 and expired before 01/02/2018. The p-values are calculated from a one-sided t-test for related samples that tests if the mean P\&L of our method is higher than that of the benchmark method. 0.95-CI means 95\% confidence interval. Scientific notation: $\text{1.23E-4} = 1.23\times 10^{-4}$. $\text{***}:p < 0.001$; $\text{**}:p < 0.01$; $*:p < 0.05$.}
\resizebox{\linewidth}{!}{
	\begin{tabular}{lllll}
		\toprule
		& \multicolumn{1}{r}{CU-RL} & \multicolumn{3}{c}{Traditional Model} \\ 
		\cmidrule(lr){3-5}
		& & \multicolumn{1}{r}{BS Model} & \multicolumn{1}{r}{LVF Model} & \multicolumn{1}{r}{SABR Model} \\ 
		\midrule 
		\multicolumn{5}{c}{Panel A: without transaction cost} \\ 
		\midrule
		 Mean & \multicolumn{1}{r}{-1.2085E-04}& \multicolumn{1}{r}{-1.3067E-03}& \multicolumn{1}{r}{-3.3349E-03}& \multicolumn{1}{r}{1.4441E-03}\\
 Std Err & \multicolumn{1}{r}{1.8511E-05}& \multicolumn{1}{r}{1.7018E-05}& \multicolumn{1}{r}{2.2938E-05}& \multicolumn{1}{r}{3.1437E-05}\\
 P-Value & \multicolumn{1}{r}{}& \multicolumn{1}{r}{0.0000E+00$^{***}$}& \multicolumn{1}{r}{0.0000E+00$^{***}$}& \multicolumn{1}{r}{1.0000$^{}$}\\
 0.95-VaR & \multicolumn{1}{r}{8.4498E-03}& \multicolumn{1}{r}{9.7598E-03}& \multicolumn{1}{r}{1.6684E-02}& \multicolumn{1}{r}{8.4188E-03}\\
 0.95-CI of 0.95-VaR & \multicolumn{1}{r}{[8.3976E-03, 8.5019E-03]} & \multicolumn{1}{r}{[9.6739E-03, 9.8307E-03]} & \multicolumn{1}{r}{[1.6503E-02, 1.6842E-02]} & \multicolumn{1}{r}{[8.3140E-03, 8.5129E-03]} \\
 0.975-VaR & \multicolumn{1}{r}{9.8070E-03}& \multicolumn{1}{r}{1.1937E-02}& \multicolumn{1}{r}{2.1775E-02}& \multicolumn{1}{r}{1.3115E-02}\\
 0.95-CI of 0.975-VaR & \multicolumn{1}{r}{[9.7235E-03, 9.8718E-03]} & \multicolumn{1}{r}{[1.1824E-02, 1.2038E-02]} & \multicolumn{1}{r}{[2.1483E-02, 2.2105E-02]} & \multicolumn{1}{r}{[1.2821E-02, 1.3462E-02]} \\
 0.975-MS & \multicolumn{1}{r}{1.1330E-02}& \multicolumn{1}{r}{1.4700E-02}& \multicolumn{1}{r}{2.7850E-02}& \multicolumn{1}{r}{2.0665E-02}\\
 0.95-CI of 0.975-MS & \multicolumn{1}{r}{[1.1205E-02, 1.1449E-02]} & \multicolumn{1}{r}{[1.4456E-02, 1.4964E-02]} & \multicolumn{1}{r}{[2.7488E-02, 2.8264E-02]} & \multicolumn{1}{r}{[2.0009E-02, 2.1370E-02]} \\
 0.95-CVaR & \multicolumn{1}{r}{1.0501E-02}& \multicolumn{1}{r}{1.3466E-02}& \multicolumn{1}{r}{2.3992E-02}& \multicolumn{1}{r}{1.8755E-02}\\
 0.95-CI of 0.95-CVaR & \multicolumn{1}{r}{[1.0420E-02, 1.0573E-02]} & \multicolumn{1}{r}{[1.3322E-02, 1.3614E-02]} & \multicolumn{1}{r}{[2.3754E-02, 2.4245E-02]} & \multicolumn{1}{r}{[1.8273E-02, 1.9236E-02]} \\
 0.975-CVaR & \multicolumn{1}{r}{1.1959E-02}& \multicolumn{1}{r}{1.6195E-02}& \multicolumn{1}{r}{2.9148E-02}& \multicolumn{1}{r}{2.7358E-02}\\
 0.95-CI of 0.975-CVaR & \multicolumn{1}{r}{[1.1850E-02, 1.2077E-02]} & \multicolumn{1}{r}{[1.5994E-02, 1.6422E-02]} & \multicolumn{1}{r}{[2.8804E-02, 2.9464E-02]} & \multicolumn{1}{r}{[2.6594E-02, 2.8229E-02]}  \\
		\midrule 
		\multicolumn{5}{c}{Panel B: with $0.1\%$ transaction cost} \\ 
		\midrule
		 Mean & \multicolumn{1}{r}{-4.7287E-03}& \multicolumn{1}{r}{-6.0007E-03}& \multicolumn{1}{r}{-8.4565E-03}& \multicolumn{1}{r}{-4.7971E-03}\\
 Std Err & \multicolumn{1}{r}{1.9512E-05}& \multicolumn{1}{r}{2.0696E-05}& \multicolumn{1}{r}{2.9907E-05}& \multicolumn{1}{r}{3.4930E-05}\\
 P-Value & \multicolumn{1}{r}{}& \multicolumn{1}{r}{0.0000E+00$^{***}$}& \multicolumn{1}{r}{0.0000E+00$^{***}$}& \multicolumn{1}{r}{2.8414E-02$^{*}$}\\
 0.95-VaR & \multicolumn{1}{r}{1.5313E-02}& \multicolumn{1}{r}{1.7711E-02}& \multicolumn{1}{r}{2.7216E-02}& \multicolumn{1}{r}{1.9514E-02}\\
 0.95-CI of 0.95-VaR & \multicolumn{1}{r}{[1.5243E-02, 1.5385E-02]} & \multicolumn{1}{r}{[1.7612E-02, 1.7811E-02]} & \multicolumn{1}{r}{[2.6998E-02, 2.7432E-02]} & \multicolumn{1}{r}{[1.9148E-02, 1.9861E-02]} \\
 0.975-VaR & \multicolumn{1}{r}{1.7085E-02}& \multicolumn{1}{r}{2.0756E-02}& \multicolumn{1}{r}{3.3764E-02}& \multicolumn{1}{r}{2.9338E-02}\\
 0.95-CI of 0.975-VaR & \multicolumn{1}{r}{[1.6993E-02, 1.7180E-02]} & \multicolumn{1}{r}{[2.0564E-02, 2.0921E-02]} & \multicolumn{1}{r}{[3.3387E-02, 3.4175E-02]} & \multicolumn{1}{r}{[2.8814E-02, 3.0036E-02]} \\
 0.975-MS & \multicolumn{1}{r}{1.8970E-02}& \multicolumn{1}{r}{2.4463E-02}& \multicolumn{1}{r}{4.0843E-02}& \multicolumn{1}{r}{4.0211E-02}\\
 0.95-CI of 0.975-MS & \multicolumn{1}{r}{[1.8838E-02, 1.9086E-02]} & \multicolumn{1}{r}{[2.4157E-02, 2.4774E-02]} & \multicolumn{1}{r}{[4.0344E-02, 4.1297E-02]} & \multicolumn{1}{r}{[3.9400E-02, 4.1280E-02]} \\
 0.95-CVaR & \multicolumn{1}{r}{1.8034E-02}& \multicolumn{1}{r}{2.2499E-02}& \multicolumn{1}{r}{3.6241E-02}& \multicolumn{1}{r}{3.7885E-02}\\
 0.95-CI of 0.95-CVaR & \multicolumn{1}{r}{[1.7929E-02, 1.8133E-02]} & \multicolumn{1}{r}{[2.2320E-02, 2.2685E-02]} & \multicolumn{1}{r}{[3.5932E-02, 3.6550E-02]} & \multicolumn{1}{r}{[3.7120E-02, 3.8634E-02]} \\
 0.975-CVaR & \multicolumn{1}{r}{1.9948E-02}& \multicolumn{1}{r}{2.6014E-02}& \multicolumn{1}{r}{4.2498E-02}& \multicolumn{1}{r}{5.1933E-02}\\
 0.95-CI of 0.975-CVaR & \multicolumn{1}{r}{[1.9804E-02, 2.0109E-02]} & \multicolumn{1}{r}{[2.5765E-02, 2.6268E-02]} & \multicolumn{1}{r}{[4.2083E-02, 4.2897E-02]} & \multicolumn{1}{r}{[5.0711E-02, 5.3279E-02]}  \\
		\bottomrule 
		\end{tabular}} 
		\label{table_real_call_all_zero_margin} 
	\end{table} 
	
	The panel A of Table~\ref{table_real_call_all_zero_margin} shows that: (i) CU-RL's mean of relative final P\&L is higher than BS and LVF model but lower than SABR model; (ii) the SABR model obtains the lowest VaR at 0.95 level while CU-RL obtains the significantly lowest tail risk measured by other risk measures.
	The panel B of Table~\ref{table_real_call_all_zero_margin} shows the results when transaction costs are considered. CU-RL has the significantly highest mean of final relative P\&L while obtaining the significantly lowest tail risk. 

	\section{Conclusion}\label{sec:conclusion}

	In conclusion, we propose a contract-unified reinforcement learning (CU-RL) algorithm for option hedging. It is contract-unified in the sense that it provides a unified hedging strategy for hedging all options contracts with different initial states such as initial stock prices and different contract parameters such as maturities and strikes. 

We formulate the hedging problem as a reinforcement learning problem that maximizes a weighted sum of the negative CVaR and expectation of the final P\&L of an option contract. The VaR and CVaR of final P\&L for an option contract depend on the initial state of the option. The CU-RL algorithm solves the reinforcement learning problem and the estimation of the VaR of the final P\&L simultaneously by extending the PPO algorithm. A key innovation of the CU-RL algorithm is to represent the VaR of the final P\&L of an option contract by a neural network with the initial state of the option contract as input. The neural network allows the CU-RL algorithm to train a unified hedging function that applies to an   option contract with any initial state. 

For the first time in the literature, we train a risk sensitive contract-unified hedging model using only the historical market price of S\&P 500 index and index options. Comprehensive empirical results demonstrate the effectiveness and robustness of the CU-RL model. Compared with the BS and SABR delta hedging methods and a minimum variance hedging method, the CU-RL approach achieves significantly higher mean of P\&L and lower tail risk in most empirical experiments. The performance of CU-RL remains stable in the presence of transaction costs and alternative definitions of reward in the problem formulation. 

Although we demonstrate the CU-RL approach by European stock option, the approach can be extended in a straightforward way for hedging general European style derivatives with possibly multiple underlying assets. 

\theendnotes

\section*{Acknowledgement}
Xianhua Peng is partially supported by the Natural Science Foundation of Shenzhen (Grant No. JCYJ20190813104607549).

\bibliographystyle{dcu} 
\bibliography{ref}

\end{document}